\documentclass[english,10pt,aps,prd,a4paper,preprintnumbers,floatfix,nofootinbib,showpacs,superscriptaddress,notitlepage,hyperref]{revtex4-2} 
\usepackage{amssymb}
\usepackage{slashed}
\usepackage{amsmath}
\usepackage{mathtools}
\usepackage{braket}
\usepackage{float}
\usepackage[utf8]{inputenc}
\usepackage{graphicx}
\usepackage{subfigure}
\usepackage{latexsym}
\usepackage{epic}
\usepackage{eepic}
\usepackage{epsfig}
\usepackage{color}
\usepackage{cancel}
\usepackage{natbib}
\usepackage[colorlinks=true,citecolor=darkred,urlcolor=darkred, pdfborder={0 0 0}]{hyperref}
\usepackage[normalem]{ulem}

\newcommand{\bea}{\begin{eqnarray}}
\newcommand{\eea}{\end{eqnarray}}

\definecolor{darkred}{rgb}{0.6,0,0}

\definecolor{linkcolor}{rgb}{0,0,0.5}

\begin{document}

\title{Heavy Majorana neutrino pair production from $Z^\prime$ at hadron and lepton colliders}

\author{Arindam Das}
\email{arindamdas@oia.hokudai.ac.jp}
\affiliation{Institute for the Advancement of Higher Education, Hokkaido University, Sapporo 060-0817, Japan}
\affiliation{Department of Physics, Hokkaido University, Sapporo 060-0810, Japan}

\author{Sanjoy Mandal}
\email{smandal@kias.re.kr}
\affiliation{Korea Institute for Advanced Study, Seoul 02455, Korea}

\author{Takaaki Nomura}
\email{nomura@scu.edu.cn}
\affiliation{College of Physics, Sichuan University, Chengdu 610065, China}

\author{Sujay Shil}
\email{sujay@iopb.res.in}
\affiliation{Institute of Physics, Sachivalaya Marg, Bhubaneswar 751005, India}
\affiliation{Homi Bhabha National Institute, BARC Training School Complex, Anushakti Nagar, Mumbai 400094, India}
\affiliation{IIT Gandhinagar, Palaj Campus, Gujarat 382355, India}
\date{\today}
\preprint{EPHOU-21-016}
\bibliographystyle{unsrt} 
\begin{abstract}
A gauged U$(1)$ extension of the Standard Model (SM) is a simple and anomaly free framework where three generations of Majorana type right-handed neutrinos (RHNs) are introduced to generate light neutrino mass and flavor mixings through the seesaw mechanism. We investigate such models at different hadron and lepton colliders via $Z^\prime$ induced Majorana type RHNs pair production. We derive bounds on U$(1)$ gauge coupling $(g^\prime)$ comparing the model cross sections with experimentally observed data for different $Z^\prime$ mass $(M_{Z^\prime})$ and RHNs mass $(M_N)$. Using these limits we estimate the allowed RHN pair production cross section which can be manifested by lepton number violating signatures in association with fat-jets at the hadron colliders depending  on the mass of the RHNs. Hence we study dilepton and trilepton modes with fat-jet/s of the signal for different benchmark values of $M_{Z^\prime}$ and $M_N$.  Using fat-jet signatures and studying the signal and corresponding SM backgrounds, we estimate bounds on $M_N-M_{Z^\prime}$ plane at different center of mass energies which could be probed at different hadron colliders. In the context of the lepton colliders we consider electron positron initial states where Majorana type RHNs can be produced from $Z^\prime$ manifesting same sign dilepton plus jets signature and trilepton plus jets in association with missing energy. Studying the signal and corresponding SM backgrounds we estimate the bounds on the $M_N-M_{Z^\prime}$ plane for different center of mass energies. In the context of U$(1)$ extension of the SM there is an SM singlet BSM scalar which couples with the RHNs. We can probe the Majorana nature of RHNs via this BSM scalar production at electron positron colliders.
\end{abstract}
\vspace{-3cm}
\maketitle
\section{Introduction}
\label{intro}
The experimental observations of the SM have set it on a stable foundation, however, the observations of the light neutrino mass and the flavor mixing \cite{Patrignani:2016xqp} give a strong indication of physics beyond the Standard Model (BSM). This leads us to extend the SM. From a perspective of the low energy effective theory, one can introduce a dimension-5 operator \cite{Weinberg:1979sa} involving the Higgs and lepton doublets within the SM framework which violates lepton number by two units. After the breaking of electroweak (EW) symmetry, the neutrinos acquire tiny Majorana masses which are suppressed by the scale of the dimension-5 operator. In the context of a renormalizable theory, the dimension-5 operator can be naturally generated by introducing SM singlet heavy Majorana RHNs. This is the well known type-I seesaw mechanism \cite{Minkowski:1977sc,Mohapatra:1979ia,Schechter:1980gr,Yanagida:1979as,GellMann:1980vs,Glashow:1979nm,Mohapatra:1979ia}.

RHNs in the TeV scale or lighter can be produced at the Large Hadron Collider (LHC) or other hadron colliders with a distinctive same-sign di-lepton plus jets final state which manifest the lepton number violation. The RHNs are SM-singlet therefore they can only be produced at the colliders through the mixings with the light neutrinos. The estimated light-heavy neutrino mixing becomes naturally small ($\sim 10^{-6}$) when the TeV scale RHNs reproduce the observed light neutrino mass around $0.1$ eV. This mixing parameter can be comparatively large when the Dirac mass matrix is generally parametrized \cite{Casas:2001sr} in order to satisfy the neutrino oscillation data, electroweak precision measurements and the lepton flavor violating processes \cite{Das:2017nvm}. The study of the heavy neutrinos have been a point of interest for a long period of time at different high energy colliders from a variety of production modes to study different final states to estimate bounds on light-heavy neutrino mixing as a function of heavy neutrino mass \cite{Asaka:2005pn,Han:2006ip,Gorbunov:2007ak,delAguila:2008cj,Mitra:2011qr,BhupalDev:2012zg,Dev:2013wba,Das:2014jxa,Das:2015toa,Antusch:2015gjw,Das:2016hof,Das:2017gke,Bhardwaj:2018lma,Antusch:2018bgr,Das:2017pvt,Arganda:2015ija,Dib:2016wge,Dib:2017vux,Chakraborty:2018khw,Das:2018usr,Liu:2020vur,Liu:2019ayx,Mondal:2016kof,Helo:2018qej,Drewes:2019fou,Drewes:2019vjy,Hirsch:2020klk,Gao:2021one,Deppisch:2019kvs,Cvetic:2020lyh,Cvetic:2014nla,Tastet:2021vwp,Tastet:2020tzh,Mekala:2022cmm,Liu:2022kid,Abada:2018sfh,Choudhury:2020cpm,Deka:2021koh,Jana:2018rdf,Abada:2021yot,FileviezPerez:2020cgn}.

Apart from the canonical seesaw scenario there is an interesting aspect, namely B$-$L, which incorporates the seesaw mechanism for the neutrino mass generation mechanism 
\cite{Davidson:1979wr, Mohapatra:1980qe, Marshak:1979fm,Davidson:1978pm,Davidson:1987mh, Wetterich:1981bx, Masiero:1982fi, Mohapatra:1982xz, Buchmuller:1991ce,Amrith:2018yfb} after the global B$-$L symmetry breaking. The U$(1)$ extension introduces a neutral BSM gauge boson, $Z^\prime$ which acquires mass after the B$-$L symmetry breaking. In addition to the RHN productions through the light-heavy neutrino mixing \cite{Basso:2008iv,FileviezPerez:2009hdc,Deppisch:2013cya,Kang:2015uoc,Cox:2017eme,Accomando:2017qcs}, this model provides a new mechanism for the RHN production in pair through $Z^\prime$ at the colliders. Once produced these RHNs can decay into the SM particles through the usual light-heavy neutrino mixings. 

In this paper we consider a general U$(1)_X$ scenario which includes three generations of Majorana type RHNs to cancel the gauge and mixed gauge gravity anomalies. After the U$(1)_X$ symmetry breaking, the light neutrino masses are generated by the type-I seesaw mechanism. The $U(1)_X$ symmetry can be identified as the linear combination of the $U(1)_Y$ in SM and the $U(1)_{B-L}$ gauge groups, hence the $U(1)_X$ scenario is the generalization of the $U(1)_{B-L}$ extension of the SM. A suitable choice of the U$(1)_X$ charges can even enhance the RHN pair production cross section from the $Z^\prime$ compare to $U(1)_{B-L}$~\cite{Appelquist:2002mw,Das:2017flq,Das:2018tbd}, which can further increase the discovery potential of the RHNs. In addition to that, there is an alternative $U(1)_X$ scenario where two of the RHNs have U$(1)_X$ charge as $-4$ and the third one has the U$(1)_X$ charge as $+5$ \cite{Das:2017deo}. Note that, although the RHNs production is light-heavy neutrino mixing independent, RHN decays to SM final sates such as $\ell W, \nu Z$ and $\nu h$ through the mixing. For small enough values of the mixing, the heavy neutrinos can be long-lived, leading to displaced decays. The displaced decays of the RHNs in this model have been studied in \cite{Das:2019fee, Chiang:2019ajm}.                                                                                                                                                                                                                                                                                                                                                                                                                                                                                                                                                                                                                                                                                                                                                                                                                                                                                                                                                                                                                                                                                                                                                                                                                                                                                                                                                                                                                                                                                                                                                                                                                                                                                                                                                                                                                                                                                                                                                                                                                                                                                                                                                                                                                                                                                                                           

If the $Z^\prime$ and Majorana type RHN masses reside in TeV scale, they can be produced at high energy colliders. The U$(1)_X$ coupling can be constrained from the existing and prospective experimental results. Using the allowed parameters, pair production of RHNs can be possible from $Z^\prime$ at hadron colliders. Subsequently, each of the RHNs will dominantly decay into a charged lepton and SM $W$ boson. The $W$ can decay dominantly into quarks $(q\overline{q^\prime})$ or sub-dominantly into leptons $(\ell \nu_\ell)$. Depending on the mass of the RHNs, the $W$ boson produced can be sufficiently boosted so that the hadrons produced from the $W$ can form a fat-jet. Hence from each RHN we obtain $(\ell^\pm)$ and a fat-jet $(J)$. As a result each event contains two same sign charged leptons and two fat-jets $(\ell^\pm \ell^\pm+2J)$ due to the Majorana nature of the RHNs. On the other hand there is another possibility where one of the $W$ bosons can decay leptonically which shows a trilepton plus one fat-jet accompanied by missing momentum $(\ell^\pm\ell^\pm\ell^\mp+J+p_T^{\rm miss})$. In this paper, analyzing these signals we investigate the discovery potential of the RHNs. Majorana type RHN can be produced at the electron positron colliders from $Z^\prime$ which can also produce $\ell W$ in pair. Hence same sign dilepton signature can be produced in association with four jets $(\ell^\pm \ell^\pm+4j)$. We can also study the trilepton mode in association with two jets and missing energy when one of the $W$ from RHN decays hadronically $(\ell^\pm \ell^\pm \ell^\mp+2j+\rm MET)$. General U$(1)$ extension of the SM contains SM singlet scalars which has Yukawa interaction withe Majorana type RHNs. In this article we consider production of the BSM scalar in $e^-e^+$ collision by $Z$ association and vector boson fusion process. The BSM scalar can decay into a pair of RHNs which can further decay into same sign dilepton mode in association with jets and missing energy. 

We arrange this paper in the following way. In Sec.~\ref{Sec2} we describe the different gauged $U(1)$ extensions of the the SM and study the constraints on the U$(1)$ coupling as a function of $M_{Z^\prime}$. We give the relation on neutrino mass, mixing and partial widths in Sec.~\ref{HNL}. The bounds on $g^\prime-M_{Z^\prime}$ plane has been discussed for two cases~(general $U(1)_X$ and alternative $U(1)_X$) in Sec.~\ref{bounds} where we compare with the existing and prospective limits. We estimate the production cross sections of the heavy neutrino pair using the experimental bounds and propose theoretically estimated density plots on $M_N-M_{Z^\prime}$ plane in Sec.~\ref{RHN} at hadron colliders. In this section we choosing different benchmarks values of $M_{Z^\prime}$ and $M_N$ to study the same sign dilepton and trilepton modes in association with fat-jets. Simulating the signal, backgrounds and applying kinematic cuts we postulate a 2$-\sigma$ exclusion contour on $M_N-M_{Z^\prime}$ plane which could be probed at different hadron colliders. In Sec.~\ref{RHNemep} we study the RHN pair production at the electron positron collider and study multilpeton modes in association with jets and missing energy. We study the pair production of the RHNs from the BSM scalar at the electron positron colliders in Sec~\ref{RHNhiggs}. Finally we conclude the article in Sec.~\ref{Sec8}. 
\section{Gauged U$(1)$ extension of the Standard Model}
\label{Sec2}
We consider a gauged U$(1)_X$ of the SM where three generations of the RHNs are introduced to cancel all the gauge and mixed gauge-gravitational anomalies. The U$(1)$ extension of the SM introduces neutral BSM gauge boson which directly interact with the RHNs.
In this paper we study two type of $U(1)_X$ extensions which are described as below.
\subsection{Case-I}
\begin{table}[t]
\begin{center}
\begin{tabular}{|c|c|c|c|c|}
\hline\hline
      &  $SU(3)_c$  & $SU(2)_L$ & $U(1)_Y$ & $U(1)_X$  \\ 
\hline\hline
$q_{L_i}$ & {\bf 3 }    &  {\bf 2}         & $ 1/6$       & $(1/6) x_{H} + (1/3) x_{\Phi}$    \\
$u_{R_i}$ & {\bf 3 }    &  {\bf 1}         & $ 2/3$       & $(2/3) x_{H} + (1/3) x_{\Phi}$  \\
$d_{R_i}$ & {\bf 3 }    &  {\bf 1}         & $-1/3$       & $-(1/3) x_{H} + (1/3) x_{\Phi}$  \\
\hline
\hline
$\ell_{L_i}$ & {\bf 1 }    &  {\bf 2}         & $-1/2$       & $(-1/2) x_{H} - x_{\Phi}$  \\
$e_{R_i}$    & {\bf 1 }    &  {\bf 1}         & $-1$                   & $-x_{H} - x_{\Phi}$  \\
\hline
\hline
$N_{R_i}$    & {\bf 1 }    &  {\bf 1}         &$0$                    & $- x_{\Phi}$   \\
\hline
\hline
$H$            & {\bf 1 }    &  {\bf 2}         & $- 1/2$       & $(-1/2) x_{H}$  \\  
$\Phi$            & {\bf 1 }       &  {\bf 1}       &$ 0$                  & $ + 2x_{\Phi}$  \\ 
\hline\hline
\end{tabular}
\end{center}
\caption{
Particle content of the minimal $U(1)_X$ model where $i$ is the generation index.}
\label{tab1}
\end{table}  

The minimal particle content of $U(1)_X$ extension has been listed in Tab.~\ref{tab1}. There are three RHNs ($N_{R_i}$) which are introduced 
 to cancel the gauge and the mixed gauge-gravitational anomalies. In this model we introduce a new, SM singlet  
scalar field $\Phi$. This scalar field is introduced to break the U$(1)_X$ gauge symmetry by its vacuum expectation value (VEV) which further generates the Majorana mass term for the RHNs.
The Yukawa Lagrangian for the RHNs sector is given by
\bea
-\mathcal{L}_{\rm int}\ \supset \ \sum_{i,j=1}^{3} Y_1^{ij} \overline{\ell_{L_i}} H N_{R_j}
  +\frac{1}{2} \sum_{i,j=1}^{3} Y_{2}^{ij} \overline{N_{R_i}^{C}} \Phi N_{R_j}+ \rm{H. c.} \, , 
\label{U1xy}
\eea 
 where $C$ stands for the charge-conjugation. The Higgs potential of this model is given by
  \begin{align}
  V \ = \ m_H^2(H^\dag H)+\lambda_h(H^\dag H)^2+m_\Phi^2 (\Phi^\dag \Phi)+\lambda_\Phi(\Phi^\dag \Phi)^2+\lambda^\prime (H^\dag H)(\Phi^\dag \Phi) \, .
  \label{pot}
  \end{align}
In the limit where $\lambda^\prime$ is small we can analyze separately the Higgs potential for $H$ and $\Phi$ as a good approximation. To break the electroweak and the $U(1)_X$ gauge symmetries
we consider the parameters of the potential for the scalar fields $H$ and $\Phi$ to develop their VEVs 
  \begin{align}
  \langle H \rangle \ = \ \frac{1}{\sqrt{2}}\begin{pmatrix} v_h\\0 
  \end{pmatrix} \, , \quad {\rm and}\quad 
 \langle \Phi \rangle \ =\  \frac{v_\Phi}{\sqrt{2}} \, ,
  \end{align}
 at the potential minimum where $v_h\simeq 246$ GeV is the electroweak scale and $v_\Phi$ is a free parameter.  
 After the symmetry breaking, the mass term of the $U(1)_X$ gauge boson $(Z^\prime)$ is generated
 \bea
 M_{Z^\prime} \ & = & \ g^\prime \sqrt{4 v_\Phi^2+  \frac{1}{4}x_H^2 v_h^2} \ \simeq \ 2 g^\prime v_\Phi.
\eea 
The process of $U(1)_X$ symmetry breaking also induces the Majorana mass term for the RHNs from the second term of the Eq.~\ref{U1xy}
 \bea
    M_{N_i} \ & = & \ \frac{Y^i_{2}}{\sqrt{2}} v_\Phi, 
    \label{mNI}   
    \eea    
 and followed by the electroweak symmetry breaking the neutrino Dirac mass term is generated 
\bea
    M_{D}^{ij} \ & = & \ \frac{Y_{1}^{ij}}{\sqrt{2}} v_h \label{mDI}.
\eea  
In this model $x_H$ and $x_\Phi$ are real parameters and U$(1)_X$ coupling $g^\prime$ is a free parameter. Without the loss of generality we consider the basis where $Y_2$ is a diagonal matrix. With the Majorana and Dirac neutrino mass terms in Eqs.~\ref{mNI} and \ref{mDI} respectively the seesaw mechanism becomes accountable for the generation of the tiny Majorana masses of the light neutrino mass eigenstates.
\subsection{Case-II}
\begin{table}[t]
\begin{center}
\begin{tabular}{|c|c|c|c|c|c|}
\hline\hline
      &  $SU(3)_c$  & $SU(2)_L$ & $U(1)_Y$ & $U(1)_X$ \\ 
\hline
$q_{L_i}$ & {\bf 3 }    &  {\bf 2}         & $ 1/6$       &  $ (1/6) x_{H} + (1/3)$ \\
$u_{R_i}$ & {\bf 3 }    &  {\bf 1}         & $ 2/3$       & $(2/3) x_{H} + (1/3) $ \\
$d_{R_i}$ & {\bf 3 }    &  {\bf 1}         & $-1/3$       & $-(1/3) x_{H} + (1/3) $\\
\hline
\hline
$\ell_{L_i}$ & {\bf 1 }    &  {\bf 2}         & $-1/2$       & $(-1/2) x_{H} - 1 $ \\
$e_{R_i}$    & {\bf 1 }    &  {\bf 1}         & $-1$         & $-x_{H} - 1 $ \\
\hline
\hline
$N_{R_{1,2}}$    & {\bf 1 }    &  {\bf 1}         &$0$                    & $- 4 $ \\ 
$N_{R_3}$    & {\bf 1 }    &  {\bf 1}         &$0$                           & $+ 5 $   \\
\hline
\hline
$H_1$            & {\bf 1 }    &  {\bf 2}         & $- 1/2$       & $(-1/2) x_{H}$ \\  
$H_2$            & {\bf 1 }       &  {\bf 2}       &$ -1/2$                  & $(-1/2) x_{H}+3 $  \\ 
$\Phi_1$            & {\bf 1 }       &  {\bf 1}       &$ 0$                  & $ +8  $  \\ 
$\Phi_2$            & {\bf 1 }       &  {\bf 1}       &$ 0$                  & $ -10 $  \\ 
$\Phi_3$          & {\bf 1 }       &  {\bf 1}       &$ 0$                  & $ -3 $  \\
\hline\hline
\end{tabular}
\end{center}
\caption{
Particle content of the alternative $U(1)_X$ extension of the SM where $i$ denotes the generation index. 
}
\label{tab2}
\end{table}   

There is another interesting $U(1)_X$ extension of the SM whose minimal particle content is shown in Tab.~\ref{tab2}. We call it an alternative U$(1)_X$ scenario.  The U$(1)_X$ charge $x_H$ is a real parameter and the U$(1)_X$ coupling $g^\prime$ is a free parameter. The RHNs in this model are differently charged under the U$(1)_X$.
In this model first two generations of RHNs have charge $-4$ whereas the third one has a charge $+5$. This non-universal charge assignment is a unique choice in order to cancel all the anomalies \cite{Montero:2007cd}.
In this model we introduce two Higgs doublets $(H_1, H_2)$ and three additional SM-singlet scalars $(\Phi_{1,2,3})$.
The Higgs doublet $H_2$ is responsible for the generation of the Dirac mass term for $N_{R_{1,2}}$.
The SM-singlet scalar $\Phi_{1}$ is responsible for the generation of the Majorana mass term of $N_{R_{1,2}}$ after the U$(1)_X$ breaking.
The Majorana mass term of $N_{R_{3}}$ is generated from the VEV of $\Phi_2$, however, there is no Dirac mass term for $N_{R_{3}}$ due to the preservation of U$(1)_X$ symmetry.
Hence $N_{R_3}$ does not participate in the neutrino mass generation mechanism. The relevant part of the interaction Lagrangian of the RHNs is given by
\bea
-\mathcal{L} _{\rm int}& \ \supset \ & \sum_{i=1}^{3} \sum_{j=1}^{2} Y_{1}^{ij} \overline{\ell_{L_i}} H_2 N_{R_j}+\frac{1}{2} \sum_{k=1}^{2} Y_{2}^{k}  \overline{N_{R_k}^{C}}\Phi_1 N_{R_k} 
+\frac{1}{2} Y_{3} \overline{N_{R_3}^{C}} \Phi_2  N_{R_3}+ \rm{H. c.} \, ,
\label{ExoticYukawa}
\eea 
where we have assumed a basis in which $Y_2$ is diagonal, without the loss of generality. The scalar potential is given by 
\bea
  V&\ =\ &
m_{H_1}^2 (H_1^\dagger H_1) + \lambda_{H_1}  (H_1^\dagger H_1)^2 + m_{H_2}^2 (H_2^\dagger H_2) + \lambda_{H_2}  (H_2^\dagger H_2)^2 \nonumber \\
&& + m_{\Phi_1}^2 (\Phi_1^\dagger \Phi_1) + \lambda_1  (\Phi_1^\dagger \Phi_1)^2 
+ m_{\Phi_2}^2 (\Phi_2^\dagger \Phi_2) + \lambda_2   (\Phi_2^\dagger \Phi_2)^2 \nonumber \\
&&+ m_{\Phi_3}^2 (\Phi_3^\dagger \Phi_3) + \lambda_3   (\Phi_3^\dagger \Phi_3)^2 
+ ( \mu \Phi_3 (H_1^\dagger H_2) + {\rm H.c.} )  \nonumber \\
&&+ \lambda_4 (H_1^\dagger H_1) (H_2^\dagger H_2)+ \lambda_5 (H_1^\dagger H_2) (H_2^\dagger H_1) +\lambda_6 (H_1^\dagger H_1) (\Phi_1^\dagger \Phi_1)\nonumber \\
&&+ \lambda_7 (H_1^\dagger H_1) (\Phi_2^\dagger \Phi_2)+ \lambda_8 (H_1^\dagger H_2) (\Phi_3^\dagger \Phi_3) +\lambda_9 (H_2^\dagger H_2) (\Phi_1^\dagger \Phi_1)  \nonumber \\
&&+ \lambda_{10} (H_1^\dagger H_1) (\Phi_2^\dagger \Phi_2)+ \lambda_{11} (H_1^\dagger H_2) (\Phi_3^\dagger \Phi_3)+  \lambda_{12} (\Phi_1^\dagger \Phi_1) (\Phi_2^\dagger \Phi_2) \nonumber \\
&&+ \lambda_{13} (\Phi_2^\dagger \Phi_2) (\Phi_3^\dagger \Phi_3)+ \lambda_{14} (\Phi_3^\dagger \Phi_3) (\Phi_1^\dagger \Phi_1).
\label{HiggsPotential-2}
\eea
We choose suitable parameters for the Higgs fields to develop their respective VEVs: 
\bea
  \langle H_1 \rangle \ = \  \frac{1}{\sqrt 2}\left(  \begin{array}{c}  
    v_{h_1} \\
    0 \end{array}
\right),   \; 
\langle H_2 \rangle \ = \   \frac{1}{\sqrt{2}} \left(  \begin{array}{c}  
    v_{h_2}\\
    0 \end{array}
\right),  
\langle \Phi_1 \rangle \ = \  \frac{v_{1}}{\sqrt{2}},  \; 
\langle \Phi_2 \rangle \ = \  \frac{v_{2}}{\sqrt{2}},  \; 
\langle \Phi_3 \rangle \ = \  \frac{v_{3}}{\sqrt{2}},~~~~ 
\eea   
with the condition, $v_{h_1}^2 + v_{h_2}^2 = (246 \,  {\rm GeV})^2$. 
We consider negligibly small mixed-quartic couplings between the Higgs doublets 
  and the SM singlets for simplicity, so that the Higgs singlet sector is effectively separated from
  the Higgs doublets. It ensures that any higher-order mixing effect between the three generations of the RHNs after $U(1)_X$ symmetry-breaking will be extremely suppressed.
  The singlet and doublet Higgs sectors communicate only through the triple coupling $\Phi_3 (H_1^\dagger H_2)+{\rm H.c.}$ 
Taking the collider constraints into account $v_1^2 + v_2^2+ v_3^2 \gg v_{h_1}^2 + v_{h_2}^2$, 
  the triple coupling has no significant effect on determining the VEVs $(v_{1, 2, 3})$ of the SM-singlet scalars $(\Phi_1, \Phi_2, \Phi_3)$
  when we arrange the parameters in the scalar potential to have the VEVs of the SM-singlet scalars almost same $(v_1 \sim v_2 \sim v_3)$ 
  and $\mu < v_1$. The third SM-singlet scalar, $\Phi_3$,  can be used as a spurion for the Higgs doublet sector  
  which generates the mixing between $H_1$ and $H_2$ through the term $\mu \Phi_3 (H_1^\dagger H_2)+{\rm H.c.}$ Using $\langle \Phi_3 \rangle =\frac{v_3}{\sqrt{2}}$ the mixing mass term becomes $m_{\rm{mix}}^2=\frac{\mu v_3}{\sqrt{2}}$.
  Hence the Higgs doublet sector potential effectively becomes the Higgs potential of the two Higgs doublet model.  
There is no mixing mass term among $\Phi_{1,2,3}$ due to the U$(1)_X$ symmetry. As a result there are two physical Nambu-Goldstone (NG) modes present in our model. 
These NG modes are originated from the SM singlet scalars and they are not phenomenologically dangerous. 
We consider the SM-singlet scalars heavier than $Z^\prime$, so that $Z^\prime$ cannot decay into the NG modes. 
After the $U(1)_X$ symmetry is broken the $Z^\prime$ boson acquires the mass term as
\bea
 M_{Z^\prime} = g' \sqrt{64 v_{1}^2+ 100 v_{2}^2+ 9v_3^2 +\frac{1}{4} x_H^2 v_{h_{1}}^2 + \left(-\frac{1}{2} x_H +3\right)^2  v_{h_{2}}^2}
\simeq g' \sqrt{64 v_{1}^2+ 100 v_{2}^2+ 9 v_{3}^2}~~~~~~~.
\label{masses-Alt}   
\eea 
and the Majorana masses of the RHNs are generated as
\bea
 M_{N_{1,2}}=\frac{Y_2^{1,2}}{\sqrt{2}} v_1,\;\; M_{N_3} = \frac{Y_3^{3}}{\sqrt{2}} v_2 
\label{mN3II} 
\eea
using the collider constraints to set $(v_1^2 + v_2^2+ v_3^2) \gg (v_{h_{1}}^2 + v_{h_{2}}^2)$.  
The Dirac mass terms of the neutrinos are generated by $\langle H_2 \rangle$: 
\bea
M_{D}^{ij} \ = \ \frac{Y_{1}^{ij}}{\sqrt{2}} \, v_{h_{2}} \, , \label{mDII}
\eea
after which the seesaw mechanism is implemented.
Because of the $U(1)_X$ charges, only two RHNs~($N_{R_{1,2}}$) are involved in the minimal seesaw mechanism 
  \cite{Smirnov:1993af,King:1999mb,Frampton:2002qc,Ibarra:2003up}    
   while the third RHN ($N_{R_3}$) has no direct interaction with the SM sector. 
   Hence it can be a potential Dark Matter (DM) candidate.
  Due to the $U(1)_X$ symmetry the Higgs doublet $H_1$ has no coupling with the RHNs and
the neutrino Dirac masses are generated by the VEV of $H_2$ as mentioned in Eq.~\ref{mDII}. This structure can be considered as a type of the 
neutrinophilic two Higgs Doublet Model (2HDM) \cite{Ma:2000cc,Wang:2006jy,Gabriel:2006ns,Davidson:2009ha,Haba:2010zi}. 
In Eq.~(\ref{HiggsPotential-2}) we may consider $0 < m_{\rm mix}^2 = \frac{\mu v_3}{\sqrt{2}} \ll m_{\Phi_3}^2$  
  which leads to $v_{h_{2}} \sim m_{\rm mix}^2 v_{h_{1}}/m_{\Phi_{3}}^2 \ll v_{h_{1}}$~\cite{Ma:2000cc}.
\section{Heavy neutrino interactions}
\label{HNL}
After the breaking of the electroweak and $U(1)_X$ symmetry, the neutrino mass matrix can be generated as
\bea
m_{\nu} \ = \ \begin{pmatrix}
0&&M_{D}\\
M_{D}^{T}&&M_{N}
\end{pmatrix}.  
\label{typeInu}
\eea
without the loss of generality we consider that $M_N$ is a diagonal matrix. The eigenvalues of this mass matrix $M_N$ have been written in the Eq.~\ref{mNI} for Case-I and in the Eq.~\ref{mN3II} for Case-II respectively.
Similarly the Dirac mass matrices for the Case-I  and Case-II are written in Eqs.~\ref{mDI} and  \ref{mDII} respectively. Diagonalizing Eq.~\ref{typeInu} we get the light neutrino mass eigenvalue as $m_\nu \simeq -M_D M_N^{-1} M_D^T$ which allows us to express the light neutrino flavor eigenstate $(\nu_\alpha)$ in terms of the light $(\nu_i)$ and heavy $(N_i)$ mass eigenstates
\bea 
\nu_\alpha \simeq  U_{\alpha i}\nu_i  + V_{\alpha i} N_i,  
\eea 
where $\alpha$ and $i$ are the generation indices, $U_{\alpha i}$ is the $3\times 3$ light neutrino mixing matrix and can be written as $U_{\alpha i} = (1-\frac{\epsilon}{2})U_{\rm PMNS}$ with $\epsilon= V^\ast V^T$, the non-unitary parameter. 
Here
\bea
V_{\alpha i} \simeq M_D M_N^{-1}
\label{eq:mixing}
\eea
is the mixing between the SM neutrinos and the heavy neutrinos which is assumed to be much less than $1$. Hence the SM gauge singlet heavy neutrinos interact with the $W$ and $Z$ bosons of the SM via mixing. 
$U_{\rm PMNS}$ is the $3\times3$ light neutrino mass matrix which diagonalizes the light neutrino mass matrix as 
\bea
U_{\rm PMNS}^T~m_\nu~U_{\rm PMNS} = diag(m_1, m_2, m_3)
\eea 
in the presence of $\epsilon$ the mixing matrix $U$ is non-unitary. For the Case-I, three generations of the RHNs  are involved the seesaw mechanism, however, in Case-II, only two generations are involved in the seesaw mechanism.
For simplicity, we assume that the heavy neutrinos are degenerate in mass. Hence the light neutrino mass matrix can be given by~\cite{Casas:2001sr}
\bea 
   m_\nu = \frac{1}{M_N} M_D M_D^T 
 = U_{\rm{PMNS}}^\ast D_{\rm{NH/IH}} U_{\rm{PMNS}}^\dagger ,  
\eea  
for the Normal Hierarchy (NH)/ Inverted Hierarchy (IH) cases of the light neutrino mass eigenvalues.

The charged-current interactions can be expressed in terms of the  neutrino mass eigenstates 
\bea 
{\mathcal{L}_{CC} \supset 
 -\frac{g}{\sqrt{2}} W_{\mu}
  \bar{e} \gamma^{\mu} P_L   V_{\alpha i} N_i  + \rm{h.c}.}, 
\label{CC}
\eea
where $e$ represents the three generations of the charged leptons, and $P_L =\frac{1}{2} (1- \gamma_5)$ is the projection operator. Similarly, in terms of the mass eigenstates, the neutral-current interactions are written as
\bea 
{\mathcal{L}_{NC} \supset 
 -\frac{g}{2 c_w}  Z_{\mu} 
\left[ 
  \overline{N}_m \gamma^{\mu} P_L  (V^{\dagger} V)_{mi} N_i
+ \left\{ 
  \overline{\nu}_m \gamma^{\mu} P_L (U^{\dagger}V)_{mi}  N_i
  + \rm{h.c.} \right\} 
\right] , }
\label{NC}
\eea
 where $c_w \equiv \cos \theta_w$ with $\theta_w$ being the weak mixing angle.

From the Eqs.~\ref{CC} and \ref{NC} we see that the heavy neutrinos $(N)$ decay into $\ell W$, $\nu Z$ and $\nu h$ respectively. Here $h$ is the SM Higgs boson. 
For sterile neutrinos heavier than $W$, $Z$ and $h$, the decays are on-shell, i.e., tow body followed by the further decays of the SM bosons otherwise  
there will be three-body decays of the heavy neutrinos, with the off-shell SM bosons. For simplicity we assume that the $U(1)_X$ Higgs bosons in Cases-I and II are heavier than the RHNs.
In these models we consider heavy $Z^\prime$ in the TeV scale. As a result, RHNs decay through the off-shell $Z^\prime$ will be negligibly small due to the $Z^\prime$ mass suppression.
The coupling between the $Z^\prime$, light neutrino and RHN arises after the $U(1)_X$ breaking, however, this is proportional to the light-heavy mixing and suppressed.
Therefore we neglect all the $Z^\prime$ mediated off-shell decays of the RHNs into the SM fermions. The allowed decay modes of the RHNs are written in the following.
When we consider the RHNs are heavier than the SM bosons so that they can decay into $\ell W$, $\nu_{\ell} Z$, and $\nu_{\ell} h$ on-shell modes. The corresponding partial decay widths are 
\bea
\Gamma(N_i \rightarrow \ell_{\alpha} W)
 &=& \frac{|V_{\alpha i}|^{2}}{16 \pi} 
\frac{ (M_{N_i}^2 - M_W^2)^2 (M_{N_i}^2+2 M_W^2)}{M_{N_i}^3 v_h^2} ,
\nonumber \\
\Gamma(N_i \rightarrow \nu_\alpha Z)
&=& \frac{|V_{\alpha i}|^{2}}{32 \pi} 
\frac{ (M_{N_i}^2 - M_Z^2)^2 (M_{N_i}^2+2 M_Z^2)}{M_{N_i}^3 v_h^2} ,
\nonumber \\
\Gamma(N_i \rightarrow \nu_\alpha h)
 &=& \frac{|V_{\alpha i}|^{2}}{32 \pi}\frac{(M_{N_i}^2-M_h^2)^2}{M_{N_i} v_h^2}.
\label{eq:dwofshell}
\eea 
When RHNs are lighter than the SM bosons they decay into three body modes where partial decay widths of $N_i$ are approximately given by
\bea
 \Gamma(N_i \to \ell_\alpha^- \ell_\beta^+  \nu_{\ell_\beta}) & = \Gamma(N_i \to \ell_\alpha^+ \ell_\beta^- \bar \nu_{\ell_\beta}) \simeq  |V_{\alpha i}|^2 \frac{G_F^2}{192 \pi^3} M_{N_i}^5 \quad (\alpha \neq \beta), \\
 \Gamma(N_i \to \ell_\beta^- \ell_\beta^+  \nu_{\ell_\alpha}) & = \Gamma(N_i \to \ell_\beta^+ \ell_\beta^- \bar \nu_{\ell_\alpha}) \nonumber \\ 
& \simeq  |V_{ \alpha i}|^2 \frac{G_F^2}{192 \pi^3} M_{N_i}^5 \left( \frac14 \cos^2 2 \theta_W + \sin^4 \theta_W \right) \quad (\alpha \neq \beta), \\ 
\label{int}
 \Gamma(N_i \to \ell_\alpha^- \ell_\alpha^+  \nu_{\ell_\alpha}) & = \Gamma(N_i \to \ell_\alpha^+ \ell_\alpha^- \bar \nu_{\ell_\alpha}) \nonumber \\
& \simeq  |V_{ \alpha i}|^2 \frac{G_F^2}{192 \pi^3} M_{N_i}^5 \left( \frac14 \cos^2 2 \theta_W + \cos 2 \theta_W + \sin^4 \theta_W \right), \\
\Gamma(N_i \to \nu_\beta \bar \nu_\beta  \nu_{\ell_\alpha}) & = \Gamma(N_i \to \nu_\beta \bar \nu_\beta \bar \nu_{\ell_\alpha}) \simeq   |V_{\alpha i}|^2 \frac{1}{4} \frac{G_F^2}{192 \pi^3} M_{N_i}^5  , \\
 \Gamma(N_i \to \ell_\alpha^-  q_a \bar q_b)  &= \Gamma(N_i \to \ell_\alpha^+  \bar q_a  q_b) \simeq  N_c |V_{ \alpha i}|^2 |V_{CKM}^{ab} |^2 \frac{G_F^2}{192 \pi^3} M_{N_i}^5, \\
\Gamma(N_i \to q_a \bar q_a  \nu_{\ell_\alpha}) & = \Gamma(N_i \to q_a \bar q_a \bar \nu_{\ell_\alpha}) \simeq N_c  |V_{ \alpha i}|^2 \frac{G_F^2}{192 \pi^3} M_{N_i}^5 2 \left( |g_V^q|^2 +  |g_A^q|^2 \right) , 
\eea
where
\bea
g_V^u=\frac{1}{2} -\frac{4}{3} \sin^2 \theta_W, \, \, g_A^u= -\frac{1}{2} \nonumber \\
g_V^d = -\frac{1}{2}+\frac{2}{3} \sin^2 \theta_W, \, \, g_A^u= \frac{1}{2} 
\eea
respectively which come from the $Z$ boson interaction with the quarks and $N_c=3$ is the color factor. Eq.~\ref{int} is the interference between the $Z$ and $W$ mediated channels where all flavors are same. We estimate the partial decay widths in a limit neglecting the SM fermion masses.  
\section{$Z^\prime$ production and bounds on the U$(1)_X$ gauge coupling}
\label{bounds}
In the U$(1)_X$ models, the RHNs have non-zero U$(1)_X$ charges like all SM fermions. It ensures the production of the RHNs through the $Z^\prime$ at the hadron collider.
First $Z^\prime$ boson is resonantly produced and subsequently decay into a pair of RHNs, if kinematically allowed. The production cross section depends on the choice of the parameter which is consistent with the dilepton \cite{Aad:2019fac,CMS:2021ctt,CERN-LHCC-2017-018}, dijet \cite{ATLAS:2019bov,Sirunyan:2018xlo} searches from the LHC and constraints obtained from LEP \cite{LEP:2003aa,Carena:2004xs,Schael:2013ita,Das:2021esm}. 
To estimate the cross section we calculate the partial decay widths into a pair of SM fermions neglecting their masses and a pair of Majorna RHNs as  
\bea
\Gamma(Z^\prime \to \overline{f^{L(R)}} f^{L(R)}) &=&  N_c \frac{{g^\prime}^2}{24 \pi} {Q_{f}^{L(R)}}^2 M_{Z^\prime}, \,\, \, \,\,\,
\Gamma(Z^\prime \to N_i N_i)= \frac{{g^\prime}^2}{24 \pi} {Q_N}^2 M_{Z^\prime} \Big(1-4\frac{M_{N_i}^2}{M_{Z^\prime}^2}\Big)^{\frac{3}{2}}~~~~~~
\label{dec}
\eea
respectively where $N_c=3 (1)$ for SM quarks (leptons), $Q_{f_L(R)}$ are the U$(1)_X$ charges of the left (right) handed SM fermions and $Q_N$ is the U$(1)_X$ charge of the heavy neutrino.

\begin{figure}[h]
\centering
\includegraphics[width=0.47\textwidth,angle=0]{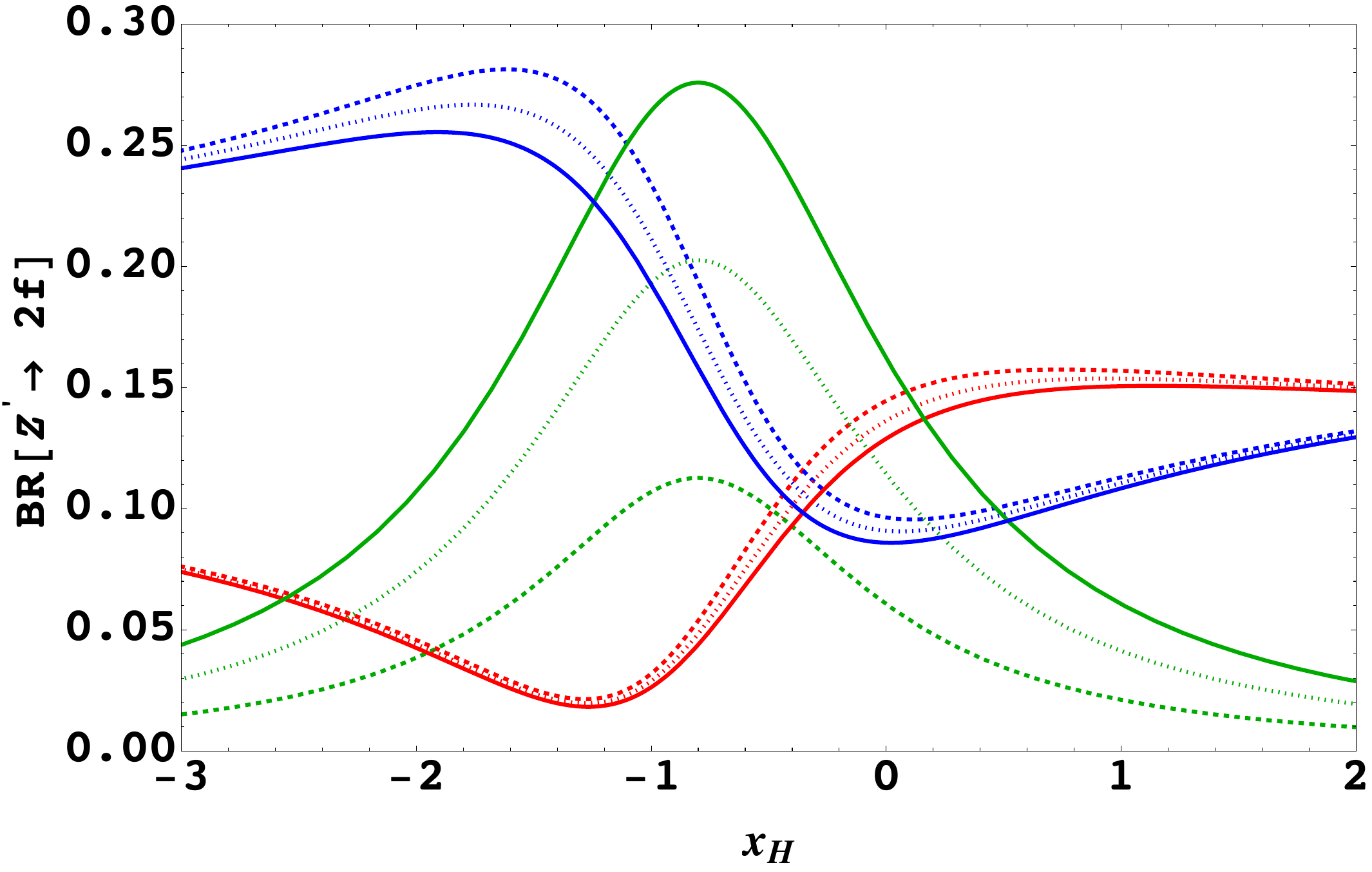}
\includegraphics[width=0.47\textwidth,angle=0]{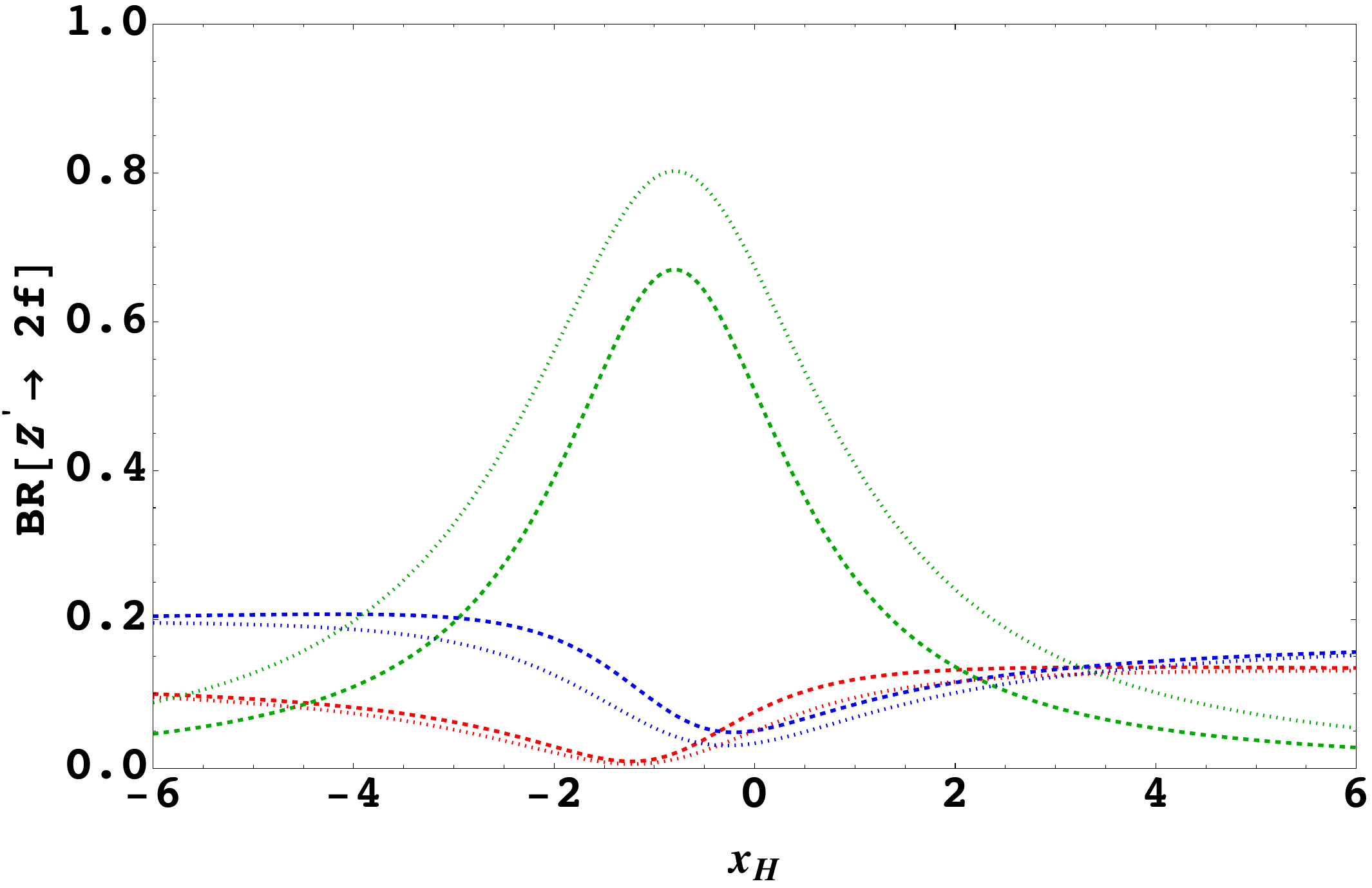}\\
\includegraphics[width=0.48\textwidth,angle=0]{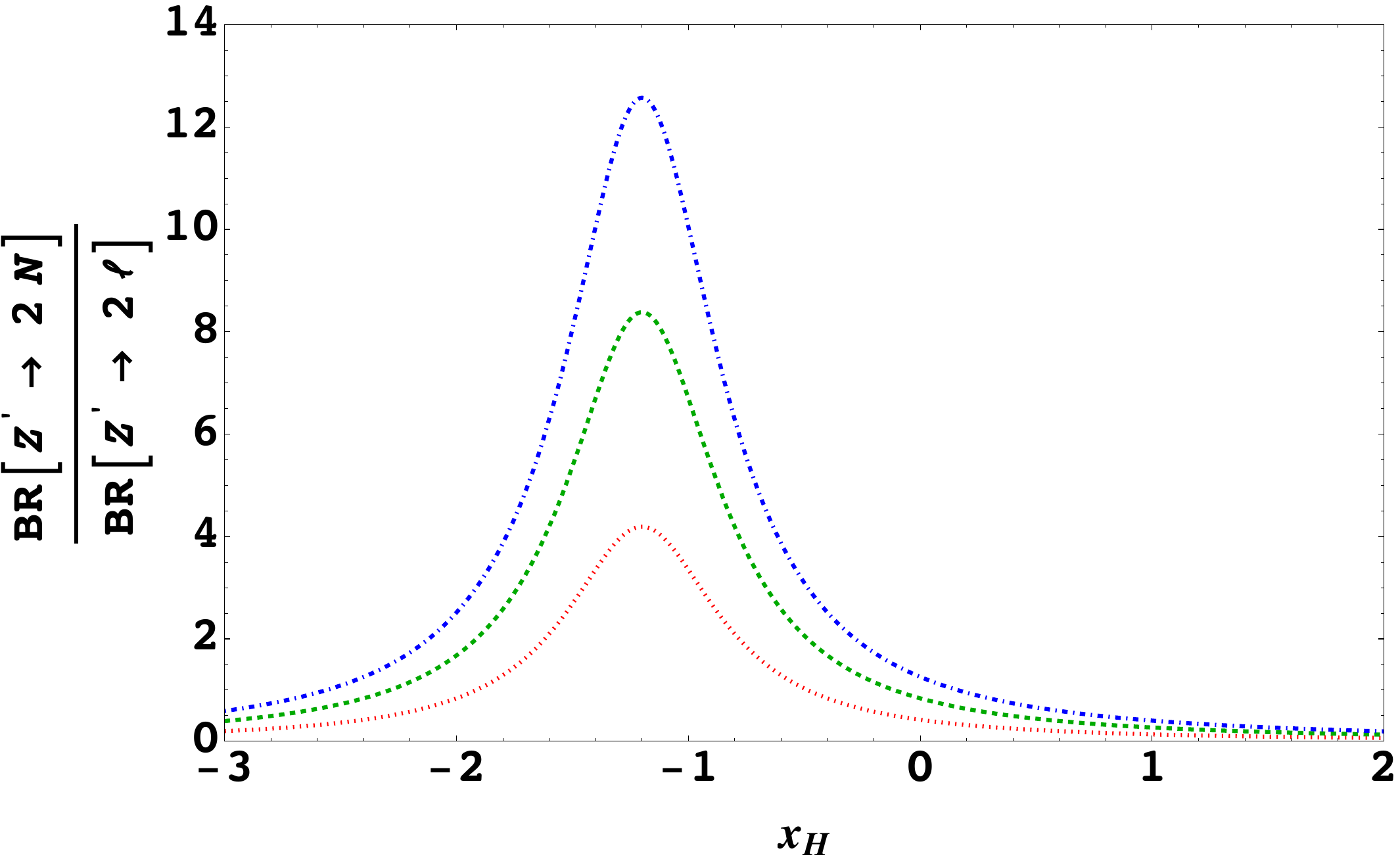}
\includegraphics[width=0.50\textwidth,angle=0]{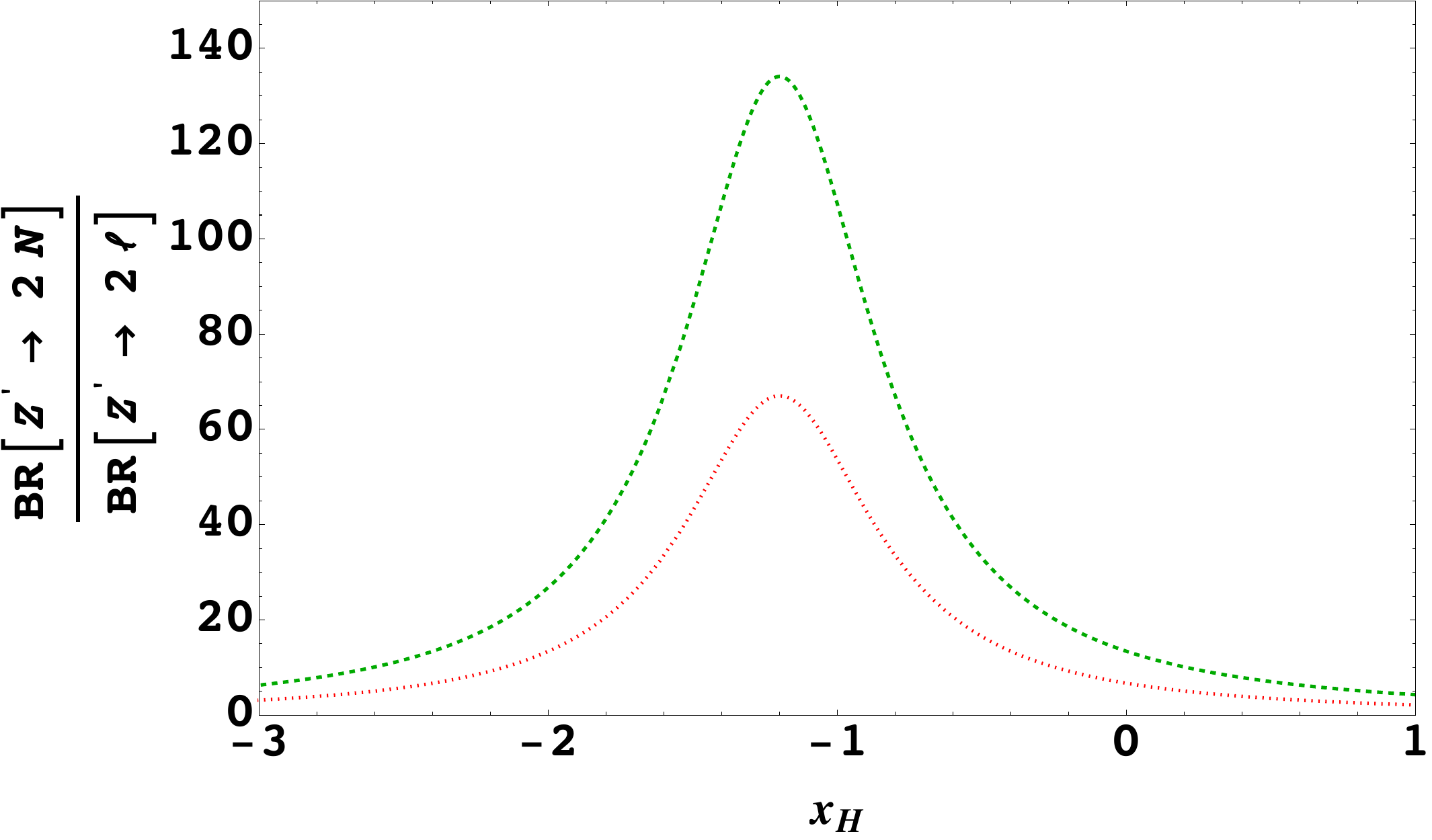}
\caption{The branching ratio of the $Z^\prime$ (top panel) into the charged lepton (red), quarks (blue) and RHNs (green) for Case-I (II) in the left (right) panel as a function of $x_H$.
The quantity $\frac{\Gamma(Z^\prime \to NN)}{\Gamma(Z^\prime \to \ell^-\ell^+)}$ (bottom panel) shows the enhancement in the RHN decay over the dilepton for Case-I (II) in the left (right) panel. 
We have considered $M_{Z^\prime}=3$ TeV (see text).}
\label{Zp-decay}
\end{figure}
The branching ratios of $Z^\prime$ into a pair of charged fermions and the heavy Majorana neutrinos for Case-I (II) are given in the left (right) side of the top panel of Fig.~\ref{Zp-decay} for $M_{Z^\prime}=3$ TeV as a function of $x_H$. The $Z^\prime$ decays into a pair of RHNs when $M_N < \frac{M_{Z^\prime}}{2}$. The solid lines correspond to the scenario when $M_{N_{1,2,3}} = 500$ GeV, the dotted line corresponds to the scenario when $M_{N_{1,2}}= 500$ GeV but $M_{N_{3}} > \frac{M_{Z^\prime}}{2}$ and the dashed line corresponds to the scenario when $M_{N_{1}}= 500$ GeV but $M_{N_{2, 3}} > \frac{M_{Z^\prime}}{2}$ from top to bottom. The lines in blue, red and green represent the decay of $Z^\prime$ into $q\overline{q}$, $\ell^-\ell^+$ and RHN pair respectively. 

In the bottom panel of Fig.~\ref{Zp-decay} we plot the ratio of the branching ratios of $Z^\prime \to NN$ to $Z^\prime \to \ell^-\ell^+$ which is defined as 
\bea
\frac{\Gamma(Z^\prime \to NN)}{\Gamma(Z^\prime \to \ell^-\ell^+)}=\frac{Q_N^2}{2+3 x_H+\frac{5}{4} x_H^2} \Big(1-4\frac{M_N^2}{M_{Z^\prime}^2}\Big)^{\frac{3}{2}}
\eea 
where $Q_N$ is the U$(1)_X$ charge of the RHNs under consideration from the Cases-I and II respectively. The blue, green and red curves represent the cases $M_{N_{1,2,3}} =500$ GeV, $M_{N_{1,2}} =500$ GeV but $M_{N_{3}} > \frac{M_{Z^\prime}}{2}$ and $M_{N_{1}}= 500$ GeV but $M_{N_{2, 3}} > \frac{M_{Z^\prime}}{2}$ respectively. In this analysis we have considered the benchmarks as $M_{Z^\prime}= 3$ TeV and $M_N= 500$ GeV when $M_{N_{i}} <  \frac{M_Z^\prime}{2}$ in the bottom left panel of Fig.~\ref{Zp-decay} for Case-I. We study the Case-II in the bottom right panel where two generations of the RHNs $(N_1, N_2)$ participate in the neutrino mass generation mechanism. The dashed green line represents $M_{N_{1,2}}=500$ GeV and the dotted red line represents the case $M_{N_{1}}=500$ GeV but $M_{N_2} > \frac{M_{Z^\prime}}{2}$. We find that $\frac{\Gamma(Z^\prime \to NN)}{\Gamma(Z^\prime \to \ell^-\ell^+)}$ becomes maximum at $x_H=-1.2$ for $x_\Phi=1$.

From the bottom left panel of Fig~\ref{Zp-decay} considering one generation of RHNs in Case-I, for example with $M_{N_1} = 500$ GeV, $\Gamma(Z^\prime \to NN) \simeq 4.22$ $\Gamma(Z^\prime \to \ell^-\ell^+)$ which can be estimated from the red dotted line. For two generations of RHNs with $M_{N_{1, 2}} =500$ GeV the situation is improved roughly by another factor of $2$ compared to the one generation case. In case of three generations of RHNs with $M_{N_{1, 2, 3}}= 500$ GeV, $\Gamma(Z^\prime \to NN)$ is nearly $13$ times greater than $\Gamma(Z^\prime \to \ell^-\ell^+)$. In Case-I we obtain the so-called B$-$L scenario for $x_H=0$ and $x_\Phi=1$ where for $M_{N_{1, 2,3}} = 500$ GeV $\Gamma(Z^\prime \to NN) \simeq \Gamma(Z^\prime \to \ell^-\ell^+)$ whereas one generation and two generation cases are suppressed compared to the dilepton mode. In Case-II, $\Gamma(Z^\prime \to NN)$ is greater than $\Gamma (Z^\prime \to \ell^+ \ell^-)$ for the choices $M_{N_{1,2}}=500$ GeV and $M_{N_{1}}=500$ GeV, $M_{N_{2}} > \frac{M_{Z^\prime}}{2}$ at $x_H=-1.2$. This property is valid for any RHN mass below $\frac{M_{Z^\prime}}{2}$ for Cases-I and II.

\begin{figure}[t]
\centering
\includegraphics[width=0.495\textwidth,angle=0]{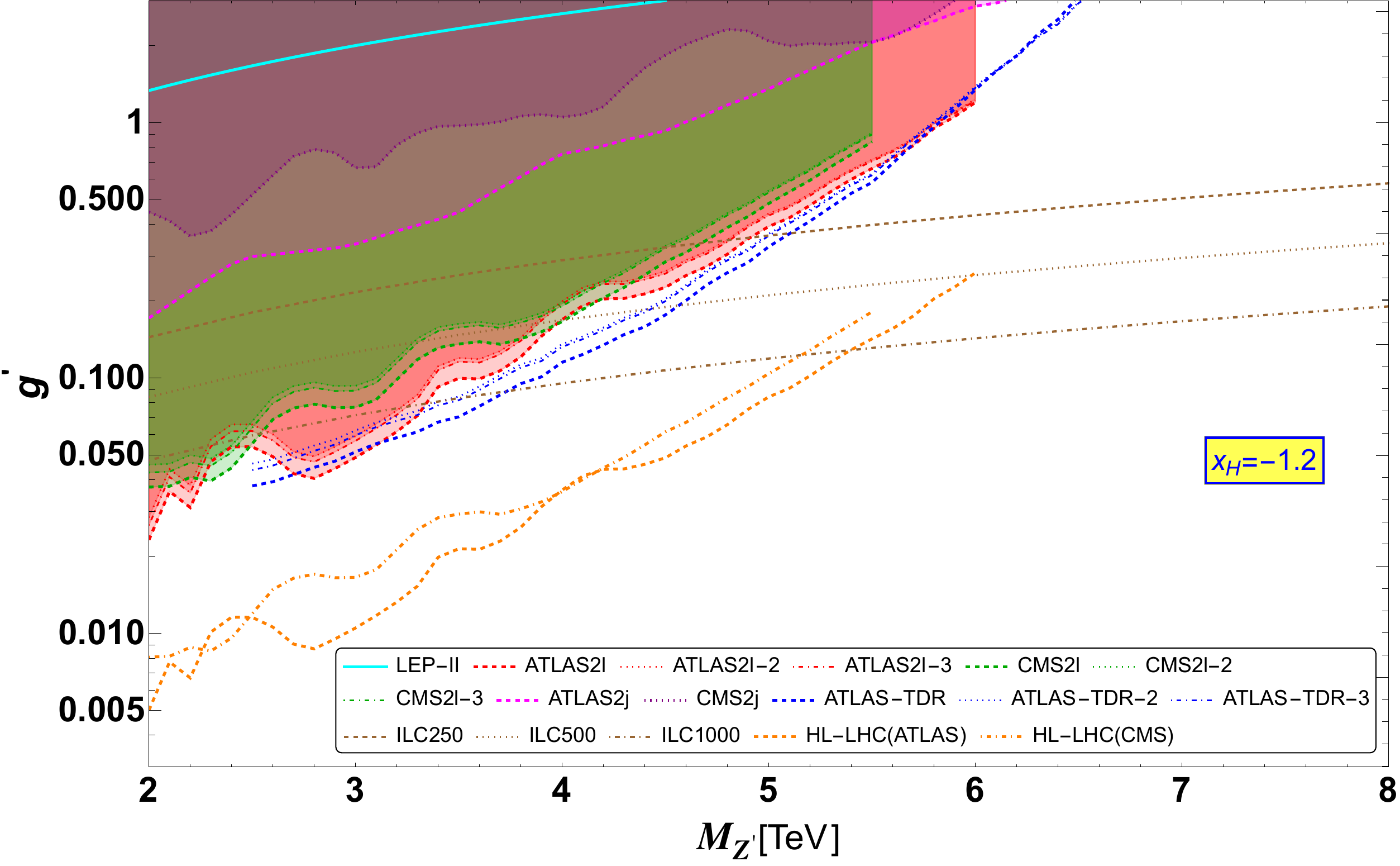}
\includegraphics[width=0.495\textwidth,angle=0]{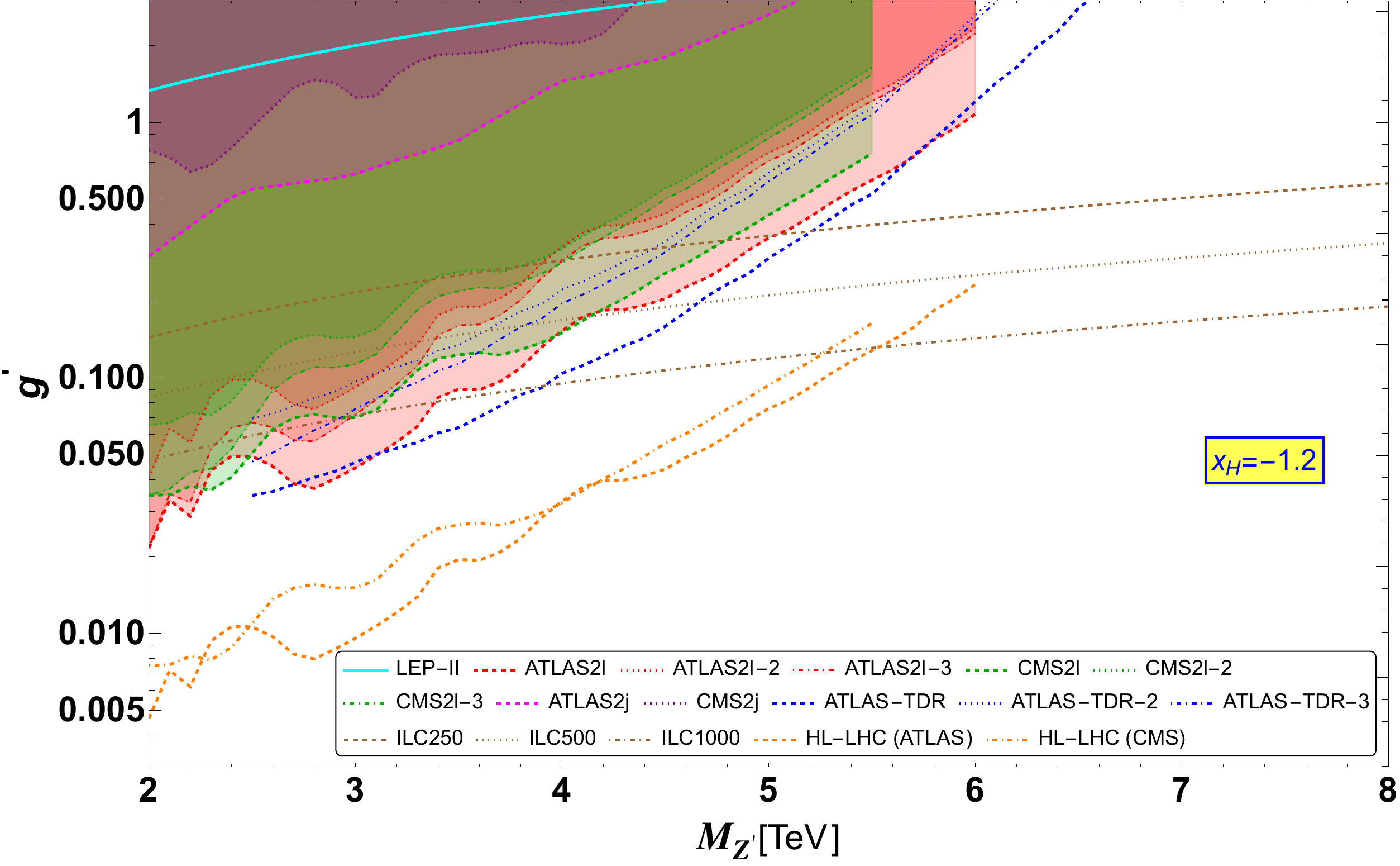}
\caption{Upper bounds on the $g^\prime$ vs $M_{Z^\prime}$ plane for Case-I (II) in the left (right) panel with  $x_H=-1.2$. The bounds are calculated from the dilepton channels at the ATLAS data \cite{Aad:2019fac}, CMS \cite{CMS:2021ctt,Sirunyan:2021khd}, ATLAS-TDR \cite{CERN-LHCC-2017-018}. The bounds from the LEP-II data have been calculated from \cite{Schael:2013ita}. The bounds from the dijet have been estimated from the ATLAS \cite{ATLAS:2019bov} and CMS \cite{Sirunyan:2018xlo} respectively. The shaded region is ruled out by the current experimental data.}
\label{gp-MZp-1}
\end{figure}

To estimate the bounds on the U$(1)_X$ gauge coupling we calculate the dilepton production cross section from the $Z^\prime$ at the $13$ TeV LHC. 
We compare our cross section with the observed dilepton cross sections from the LHC where sequential standard model (SSM) \cite{Langacker:2008yv} was studied assumming $\frac{\Gamma}{m}=3\%$. 
Using the following master equation 
\bea
g^\prime = \sqrt{g_{\rm{Model}}^2 \Big(\frac{\sigma_{\rm{ATLAS}}^{\rm{Observed}}}{\sigma_{\rm{Model}}}\Big)}
\label{gp}
\eea
we find the bounds on the U$(1)_X$ gauge coupling. The $Z^\prime$ production cross sections can be calculated using narrow width approximation as
\bea
\sigma(pp \to Z^\prime) = 2 \sum_{q, \overline{q}} \int dx \int dy q(x, Q) \overline{q}(y, Q) \hat{\sigma}(\hat{s})
\label{Xsec-1}
\eea
where $q (x, Q)$ and $\overline{q} (x, Q)$ are the parton distribution functions of the quark and antiquark respectively and $\hat{s}= xys$ is the invariant mass squared of the colliding quark at the center of mass energy $\sqrt{s}$. Due to the narrow width approximation the cross section of the colliding quarks to produce $Z^\prime$ boson is 
\bea
\hat{\sigma}=\frac{4\pi^2}{3} \frac{\Gamma(Z^\prime \to q\overline{q})}{M_{Z^\prime}}  \delta(\hat{s}-M_{Z^\prime})
\label{Xsec-2}
\eea
With a factorization scale set at $Q=M_{Z^\prime}$ and employing CTEQ6L \cite{Pumplin:2002vw}, one can estimate $\sigma_{\rm{Model}}=$ $\sigma(pp\to Z^\prime)$ BR$(Z^\prime \to 2\ell)$ or $\sigma(pp\to Z^\prime)$ BR$(Z^\prime \to 2j)$ for dilepton or dijet final states. For the dijet production cross section we use $58\%$ and $70\%$ acceptance respectively to compare with the LHC results at the ATLAS and CMS respectively. Finally the bounds are calculated from the Eq.~\ref{gp} for Case-I (II) are shown in the left (right) panel of  Fig.~\ref{gp-MZp-1}. 

To estimate these bounds from the LHC we have considered three scenarios: (i) when all the RHNs are heavier than $\frac{M_{Z^\prime}}{2}$ which are shown by the dashed lines (Red/ Green/ Blue) to represented the LHC dilepton (ATLAS2l/ CMS2l/ ATLAS-TDR) bounds. Due to this fact the $Z^\prime$ can not decay into the RHNs. Hence the other decay modes of $Z^\prime$ will be enhanced providing the strongest bound on $g^\prime$; (ii) when two generations of the RHN masses are below $\frac{M_{Z^\prime}}{2}$ leading to the decay of the $Z^\prime$ into the kinematically allowed RHN pairs. We have chosen two benchmark scenarios in this case for at least two generations of the degenerate RHNs with $M_{N_{1,2}}=500$ GeV which are shown by the dotted lines (Red/ Green/ Blue) to represent the LHC dilepton (ATLAS2l-2/ CMS2l-2/ ATLAS-TDR-2) bounds; (iii) similar bounds for $M_{N_{1,2}}=1$ TeV are represented by dot-dashed lines (Red/ Green/ Blue) to represent the LHC dilepton (ATLAS2l-3/ CMS2l-3/ ATLAS-TDR-3) bounds. The di-jet bounds are shown by the magenta lines for the case (i). It is important to notice that the dilepton bound is the strongest one with respect to the di-jet. Therefore we did not extract the dijet bounds considering the RHN masses at $500$ GeV and $1$ TeV. The lines with different RHN masses match with the case where RHNs are heavier than $\frac{M_{Z^\prime}}{2}$ for the heavier $M_{Z^\prime}$ mass because of factor $(1-4 \frac{M_{N}^2}{M_{Z^\prime}^2})$ factor in the $Z^\prime \to NN$ partial mode. It becomes almost $1$ when $M_{Z^\prime} >> M_{N}$. The bounds on the $g^\prime-M_{Z^\prime}$ plane for Case-II are shown in the right panel of Fig.~\ref{gp-MZp-1}, the branching ratio of $Z^\prime \to NN$ is more than one order of magnitude higher than the Case-I due the U$(1)_X$ charges. Such a behavior has been reflected in the nature of the constraints on $g^\prime$ for different $M_{Z^\prime}$.  The bounds from the LHC are estimated using narrow width approximation at 139(140) $\rm fb^{-1}$ luminosity at the LHC where the $Z^\prime$ production cross section is proportional to ${g^\prime}^2$. Hence we naively estimate the future or prospective upper bound on $g^\prime$ scaling the luminosity following $g^\prime \simeq g^\prime_{\rm current} \sqrt{\frac{139(140) {\rm fb}^{-1}}{\mathcal{L_{\rm future} {\rm fb^{-1}}}}}$ at a future luminosity of $\mathcal{L_{\rm future}}$. 
At the High Luminosity LHC (HL-LHC) of 3000 fb$^{-1}$ we scale the limits from ATLAS (CMS) considering $M_{N} > \frac{M_{Z^\prime}}{2}$ and obtain that the limits get approximately 0.215 (0.216) times stronger and it is uniformly applicable. Similar scenario could be obtained for the other choices of the heavy neutrino masses, however, those bounds will be weaker than the case $M_{N} > \frac{M_{Z^\prime}}{2}$. These prospective limits could be verified in the near future.

LEP-II bounds have been estimated from \cite{LEP:2003aa,Carena:2004xs,Schael:2013ita,Das:2021esm} for $x_H=-1.2$ and has been shown by the solid cyan line in Fig.~\ref{gp-MZp-1}. We estimated bounds on the $g^\prime-M_{Z^\prime}$ plane from ILC at $\sqrt{s}=$250 GeV, 500 GeV and 1 TeV respectively from~\cite{Das:2021esm} considering $M_{Z^\prime} > \sqrt{s}$. Hence from \cite{Das:2021esm} for $x_H=-1.2$ we find, the limit is $\frac{M_{Z^\prime}}{g^\prime} > 2.68$ TeV from LEP-II and the prospective limits on this quantity are 13.85 TeV, 23.74 TeV and 41.96 TeV respectively from the ILC at $\sqrt{s}=$ 250 GeV, 500 GeV and 1 TeV respectively. Hence we estimate the upper bounds on $g^\prime$-$M_{Z^\prime}$ plane from $e^-e^+$ colliders which are represented by LEP-II (cyan), ILC250 (brown, dashed), ILC500 (brown, dotted) and ILC1000 (brown, dot-dashed) respectively in Fig.~\ref{gp-MZp-1} for Case-I (II) in the left (right) panel. The LEP-II limit is weaker compared to the current LHC bounds, however,  the prospective bounds obtained from the ILC are stronger than the existing LHC limits for heavier $Z^\prime$. In Tab.~\ref{tab3} we show some allowed benchmark values of $M_{Z^\prime}$, $M_{N}$ and $g^\prime$ which will be used in the RHN pair production from $Z^\prime$.
\begin{table}[h]
\begin{center}
\begin{tabular}{|c|c|c|c|}
\hline\hline
     $U(1)_X$ coupling & $M_N > \frac{M_{Z^\prime}}{2}$ & $M_{N_{1,2}}=1000$ GeV& $M_{N_{1,2}}=500$ GeV  \\ 
\hline
\hline
$g^\prime$(Case-I)&0.051&0.061&0.064\\
\hline
\hline
$g^\prime$(Case-II)&0.046&0.072&0.095\\
\hline
\end{tabular}
\end{center}
\caption{Allowed values of the $U(1)_X$ couplings considering all the necessary constraints for $M_{Z^\prime} =3$ TeV with different benchmark scenarios of the RHN masses for $x_H=-1.2$.}
\label{tab3}
\end{table}   

BR$(Z^\prime \to NN)$ plays an important role to study RHN pair production from $Z^\prime$ at the hadron colliders.
From Fig.~\ref{Zp-decay} we find that it is maximum for $x_H=-1.2$ over any other decay modes including the otherwise stronger dilepton modes whereas for other values of $x_H$, BR$(Z^\prime \to NN)$ is subdominant. Therefore we choose $x_H=-1.2$ for RHN pair production at the hadron colliders. Also note that different values of $x_H$ for $x_\Phi=1$ have implications in the theory which has been studied in \cite{Das:2021esm} in detail manifesting the chiral nature of the model. We have studied the low energy aspect of the model for light $Z^\prime$ which could be tested at the neutrino experiments like DUNE to probe the chiral nature of such a scenario in \cite{Chakraborty:2021apc}. On the other hand variation of $x_\Phi$ can manifest the periodic nature of the model which is beyond the scope of this article, however, will be tested in future to study different aspects of this model. On the other hand we will use the aspect $M_{Z^\prime} >> \sqrt{s}$ for the $e^-e^+$ colliders to produce the RHN pair from $Z^\prime$ where we will consider different values of $x_H$ other than $-1.2$ which will also manifest the chiral nature of the $Z^\prime$ for RHN pair production. 
\section{Heavy neutrino pair production at $pp$ colliders}
\label{RHN}

The RHNs can be produced in pair at $\sqrt{s}=$ 14 TeV, 27 TeV and 100 TeV proton proton colliders respectively through the $Z'$ production for $x_H=-1.2$, $\sigma(pp \to Z^\prime)$ BR$(Z^\prime \to N N)$. After the RHNs are pair produced, they dominantly decay into $\ell W$ mode followed by the hadronic decay of the $W$ boson. Hence due to the pair production of the Majorana RHNs, 
same sign dilepton (SSDL) plus four jet signal is produced. We show 
\begin{figure}[t]
\centering
\includegraphics[width=0.32\textwidth,angle=0]{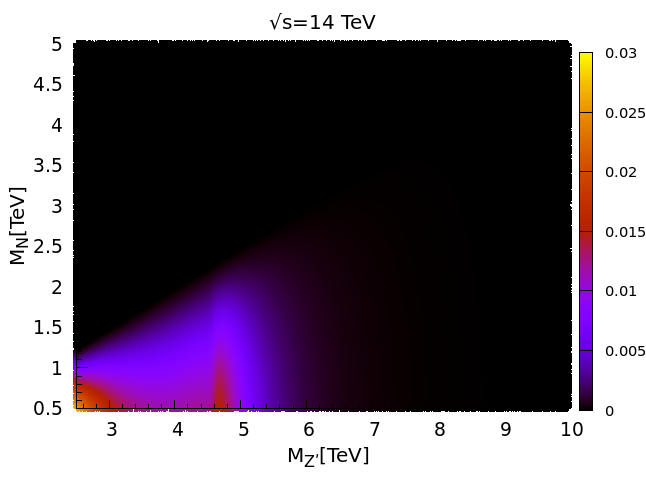}
\includegraphics[width=0.32\textwidth,angle=0]{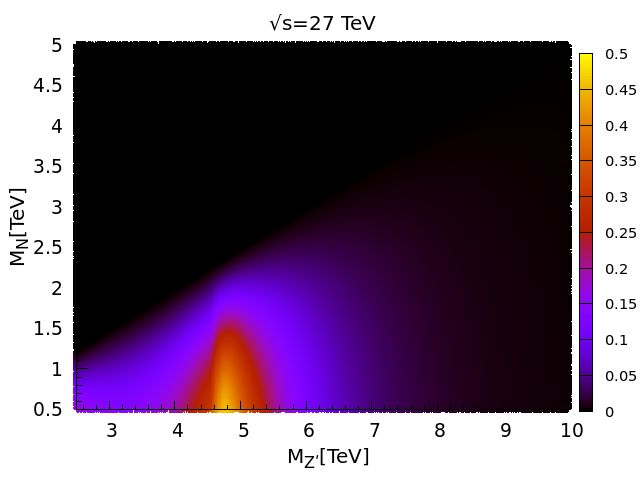}
\includegraphics[width=0.32\textwidth,angle=0]{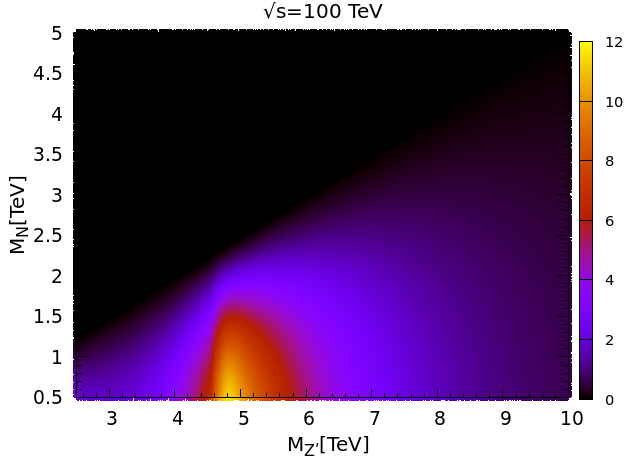}\\
\includegraphics[width=0.32\textwidth,angle=0]{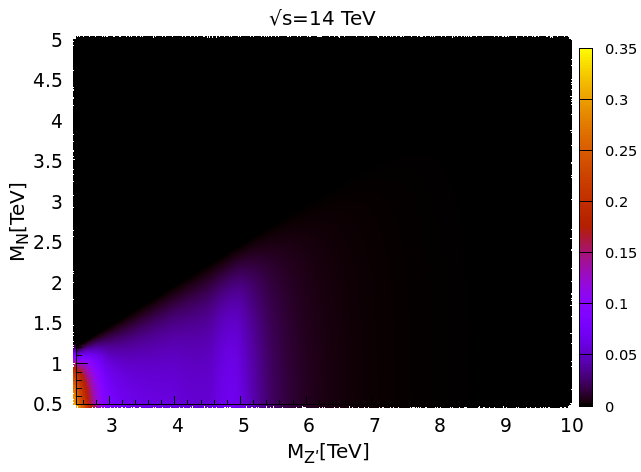}
\includegraphics[width=0.32\textwidth,angle=0]{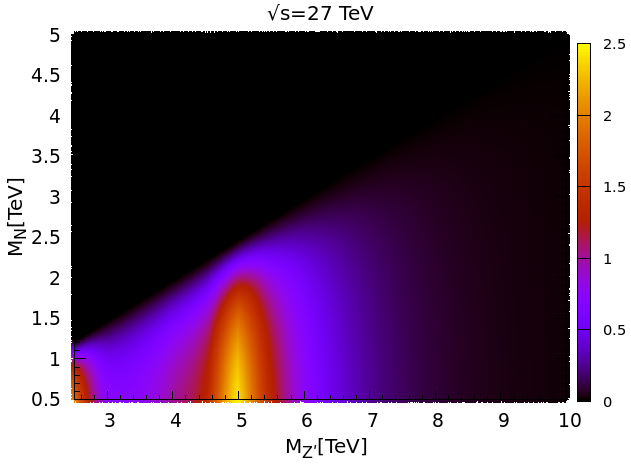}
\includegraphics[width=0.32\textwidth,angle=0]{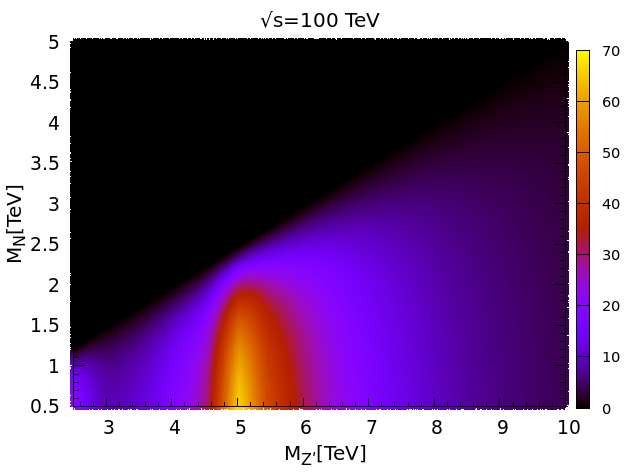}
\caption{The density plot representing SSDL plus four jet cross section on $M_N-M_{Z^\prime}$ plane for Case-I (II) in the upper (lower) panel considering $x_H=-1.2$ at different center of mass energies. The bar chart represents the cross sections in fb.}
\label{MN-MZp-I-1}
\end{figure}
the density plots on the $M_N-M_{Z^\prime}$ plane for $\sqrt{s}=$14 TeV, 27 TeV and 100 TeV in the upper (lower) panel of Fig~\ref{MN-MZp-I-1} for Case-I (II) satisfying the strongest upper limits on $g^\prime$ from the dilepton results at the LHC for $M_N > \frac{M_{Z^\prime}}{2}$. The bar charts in the density plots represent the cross sections from low to high (bottom to top) values in fb. In this case we consider the $M_{Z^\prime}$ from 2 TeV to 10 TeV. For the Case-I at $14$ TeV we did not obtain any points for $M_{Z^\prime} > 6$ TeV, at 27 TeV did not obtain any point beyond $M_{Z^\prime} > 8$ TeV. However, at the 100 TeV we obtain larger cross sections for the SSDL case from the RHN pair production. We show the density plot for the Case-II in the lower panel of Fig.~\ref{MN-MZp-I-1} for $\sqrt{s}=$14 TeV, 27 TeV and 100 TeV respectively. In this case we find the allowed points for $M_{Z^\prime}$ up to 6 TeV at 14 TeV collider, 8 TeV at 27 TeV collider and 10 TeV at 100 TeV collider.

\begin{figure}[h]
\centering
\includegraphics[width=0.32\textwidth,angle=0]{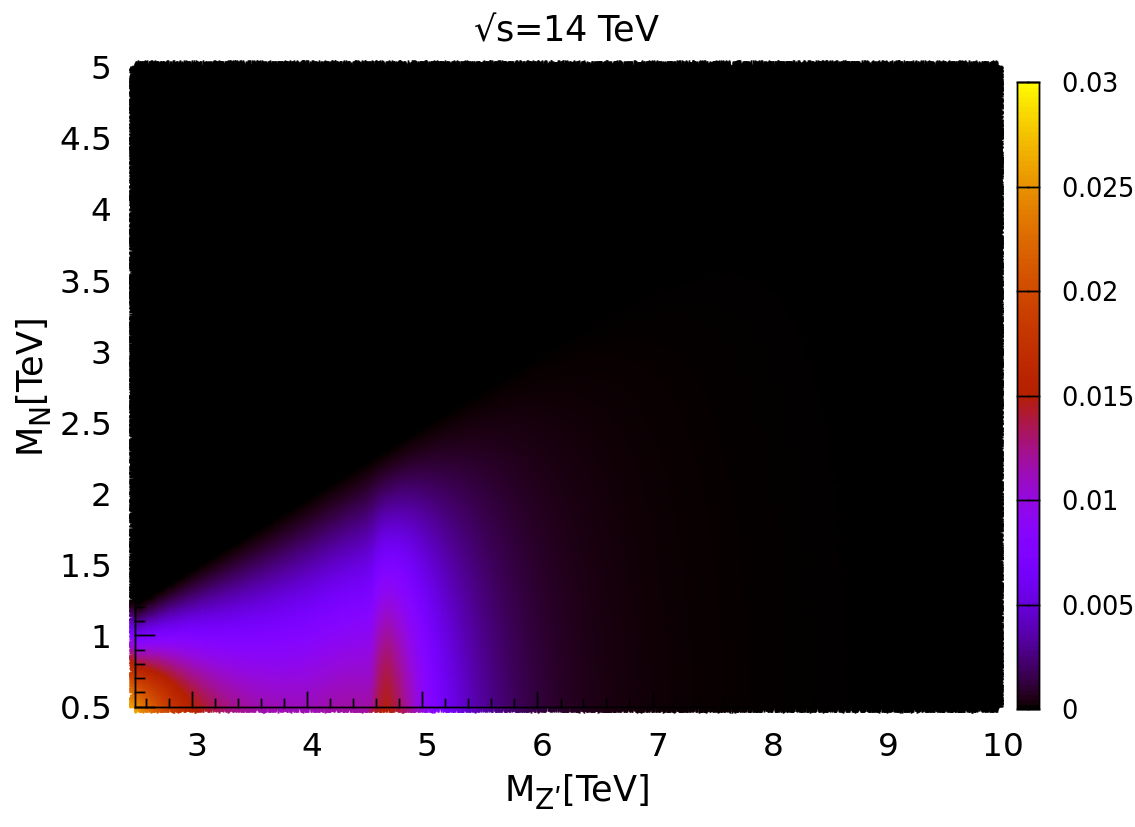}
\includegraphics[width=0.32\textwidth,angle=0]{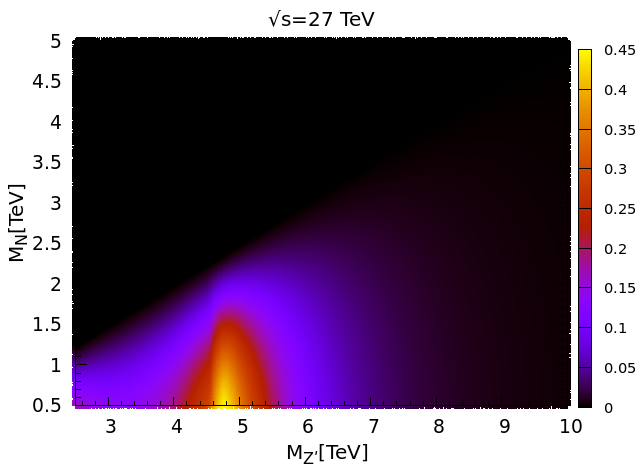}
\includegraphics[width=0.32\textwidth,angle=0]{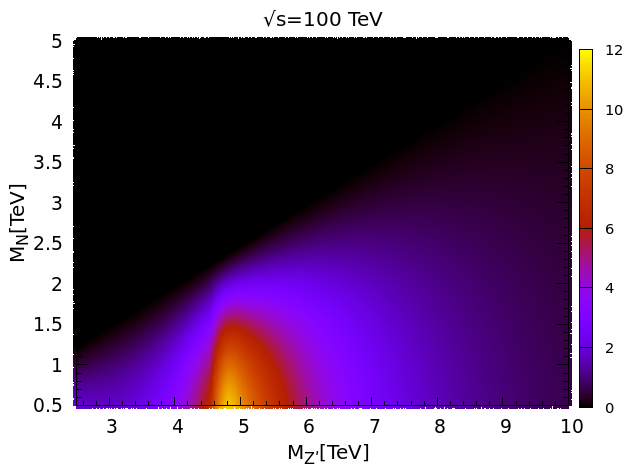}\\
\includegraphics[width=0.32\textwidth,angle=0]{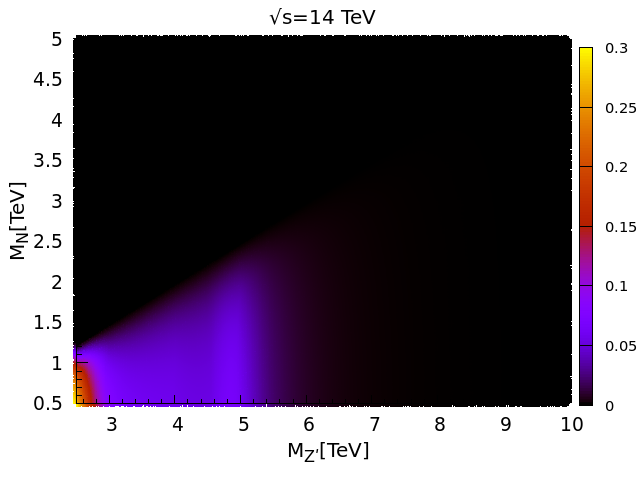}
\includegraphics[width=0.32\textwidth,angle=0]{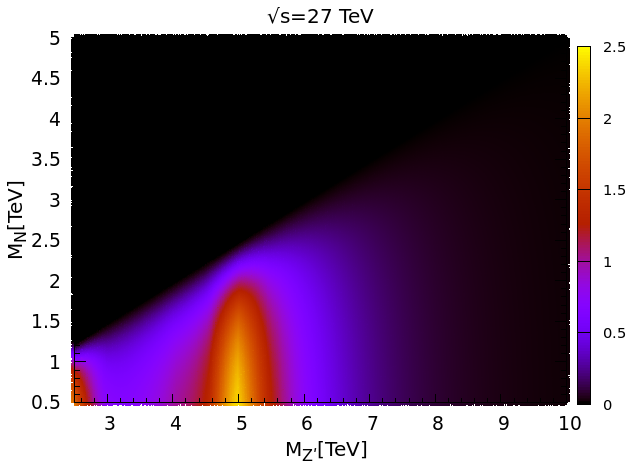}
\includegraphics[width=0.32\textwidth,angle=0]{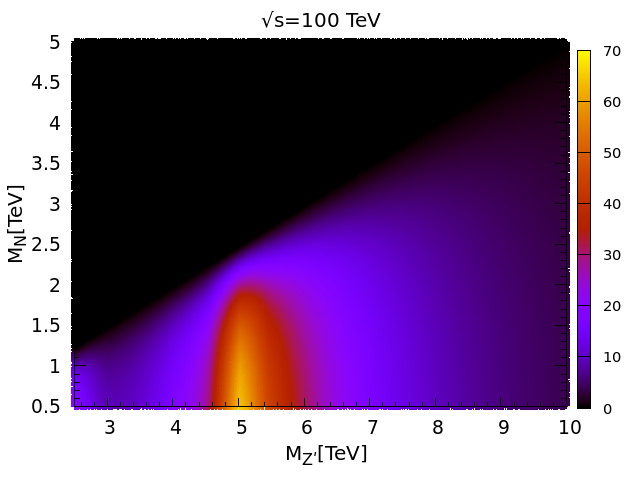}
\caption{The density plot representing trilepton plus two jet cross section in association with missing energy on $M_N-M_{Z^\prime}$ plane for Case-I (II) in the upper (lower) panel considering $x_H=-1.2$ at different center of mass energies. The bar chart represents the cross sections in fb. 
}
\label{MN-MZp-II-3L-1}
\end{figure}

Apart from the SSDL plus four jet signal from the pair production of RHNs, we can consider a trilepton final state plus two jet signature in association with missing energy at $\sqrt{s}=$ 14 TeV, 27 TeV and 100 TeV hadron colliders respectively for $x_H=-1.2$. Both the RHNs dominantly decay into $\ell W$ mode followed by the hadronic decay of one of the $W$ bosons whereas the other one decays leptonically. Using the experimentally allowed strongest upper bounds from the $g^\prime- M_{Z^\prime}$ plane for $M_N > \frac{M_{Z^\prime}}{2}$ we show   
the density plots in the $M_N-M_{Z^\prime}$ plane for 14 TeV, 27 TeV and 100 TeV are given in the upper (lower) panel of Fig~\ref{MN-MZp-II-3L-1} for Case-I (II). The bar charts in the density plots represent the cross sections from low to high (bottom to top) in fb. The colored area shows the allowed cross sections of the trilepton pus two jet in association with missing energy in $M_N-M_{Z^\prime}$ plane. In this case we consider the $Z^\prime$ mass from 2 TeV to 10 TeV. For the Case-I at $14$ TeV we did not obtain any points for $M_{Z^\prime} >$ 6 TeV, at 27 TeV did not obtain any data point for $M_{Z^\prime} > $ 8 TeV. However, at the 100 TeV we obtain larger cross sections for the trilepton case from the RHN pair production even for large $M_{Z^\prime}$. The density plots for the Case-II are shown in the lower panel of Fig.~\ref{MN-MZp-II-3L-1}. Similar behavior like the Case-I can be observed, however, the cross sections increase due to the U$(1)_X$ charge of the RHNs and the cross sections increase with center of mass energy.  

\begin{figure}[h]
\centering
\includegraphics[width=0.45\textwidth,angle=0]{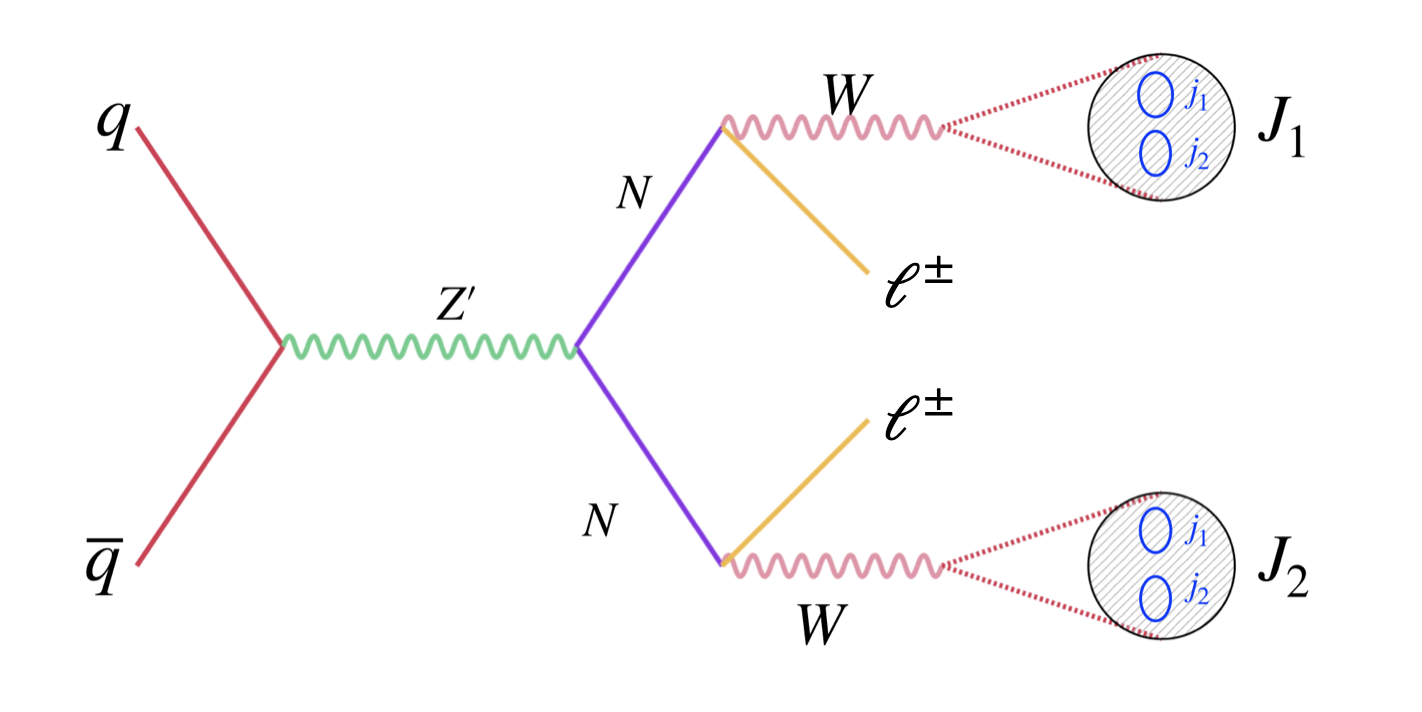}
\includegraphics[width=0.46\textwidth,angle=0]{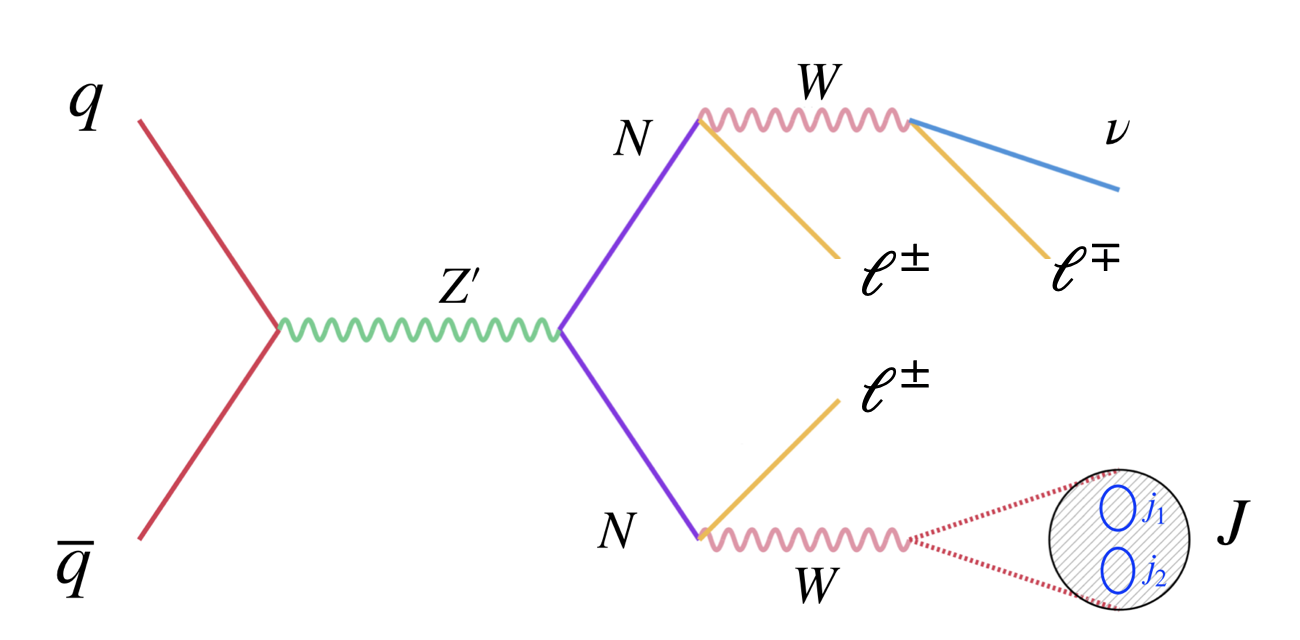}
\caption{Heavy neutrino pair production processes at the hadron collider from the $Z^\prime$. The heavy neutrinos $(N)$ decay into the SSDL plus two fat-jets (left) and trilepton plus one fat-jet in association with missing energy (right).}
\label{SSDL}
\end{figure}

Another important motivation of this paper is to study the boosted objects from the RHNs pair. After the RHNs are produced, they can decay through the dominant mode $\ell W$, followed by the hadronic decay of the $W$. The Majorana nature of the RHNs will manifest a distinct $\ell^{\pm} \ell^{\pm}$ signature along with the hadronic decay of the $W$ boson. We consider the RHNs which are sufficiently heavy, e. g., $M_N \geq 500$ GeV which allow the $W$ boson to be boosted so that the hadronic decay mode of the $W$ can be collimated to produce fat-jet. Hence SSDL plus two fat-jet and trilepton plus one fat-jet shown in Fig.~\ref{SSDL} could be interesting to study. 

\subsection{Same sign dilepton (SSDL) with fat-jets}
\label{ssdl-1}

The resonant production of the Majorana RHN pair from the $Z^\prime$ can show a distinct signature of the lepton number violation at the collider. In this case RHNs will produce same sign leptons and $W$ bosons. Due to the heavy mass of the RHNs the $W$ boson will be boosted to make a fat-jet. The production process is 
\bea
pp \to Z^\prime \to  N N, N \to \ell^{\pm} W, N \to \ell^{\pm} W, W \to j_1 j_2 (J_1), W \to j_1 j_2 (J_2)
\eea
where $J_1$ and $J_2$ are the fat-jets, shown in the left panel of Fig.~\ref{SSDL}. In our scenarios there are three RHNs out of them we consider the first two generations are degenerate. We consider two benchmark scenarios $M_{N_{1,2}}=500$ GeV and $M_{N_{1,2}}=1$ TeV respectively for the U$(1)$ scenarios stated in Cases-I and II.  

The dominant SM background in the case of SSDL scenario comes from the same sign $W$ boson production in association with jets. The same sign $W$ will decay leptonically to produce SSDL environment and the two jets will resemble the $W^\pm$ like fat-jets. Another significant background will be contributed from $W^\pm Z+$ jets where $W^\pm$ and $Z$ decay leptonically. In addition to that, another important contribution will come from $W^\pm W^\mp Z+$ jets background where one $W$ and $Z$ will decay leptonically where as the remaining $W$ will decay hadronically. In these cases there will be an SSDL pair in association with an opposite sign third lepton. A third lepton veto will reduce these backgrounds. Other significant contributions will come from $t\overline{t} W^\pm$ and $t\overline{t} Z$ channels. The SSDL events will come from either $t$ or $\overline{t}$ and the leptonic decay of $W^\pm$. A similar scenario can be observed for $t\overline{t}Z$ for the SSDL signature, however, there will be an additional lepton. In this case, the choice of the two same sign lepton and third lepton veto are very important to reduce the background. The $W$ like fat-jets can be produced from the additional jets in the event or from the remaining $t$ or $\overline{t}$. We use the veto on b-jets to reject events with b-jets. 

Implementing the model\footnote{We have written the model file in FeynRules. The FR and UFO files  can be obtained in the category of Simple extensions of the SM of the FeynRules model database \href{https://feynrules.irmp.ucl.ac.be/wiki/GeneralU1\#no1/}{General U(1) model}} in FeynRules \cite{Christensen:2008py, Alloul:2013bka} we generate the events using MadGraph \cite{Alwall:2011uj, Alwall:2014hca} and parton distribution function CTEQ6L \cite{Pumplin:2002vw} fixing the factorization scale $\mu_F$ at the default MadGraph option. The showering, fragmentation and hadronization of the signal and SM backgrounds were performed by the PYTHIA8\cite{Sjostrand:2007gs}. The detector simulation of the showered events was performed using the detector simulation package Delphes \cite{deFavereau:2013fsa} equipped with the Cambridge-Achen (C/A) jet clustering algorithm \cite{Dokshitzer:1997in,Wobisch:1998wt}.  In our analysis, we have taken jet cone radius, $R=1.0$. The jets and associated sub-jet variables are constructed using the softdrop procedure. The softdrop procedure depends on two parameters, an asymmetry cut $Z_{cut}$, and an angular exponents $\beta$ describe here \cite{Larkoski:2014wba,Dasgupta:2013ihk,Butterworth:2008iy}. For $\beta = 0$, the Softdrop algorithm to find sub-jets is the same as the modified mass-drop tagger\cite{Butterworth:2008iy}. The algorithm goes as, (i) By undoing the last step of the C/A clustering algorithm of fatjet we will get two subjets, for example, $j_1$ and $j_2$, (ii) Now, if these two subjets pass the asymmetry cut $\frac{min(P_{T}^{j_{1}},P_{T}^{j_{2}})}{(P_{T}^{j_{1}}+P_{T}^{j_{2}})} > Z_{cut}$ (here we use $Z_{cut} = 0.1$) we call the fat-jet as a softdrop jet. (iii) If the last condition is not satisfied, we consider the largest $p_{T}$ jet as a fat-jet and do the same procedure from step 1. In this algorithm, the asymmetry cut helps us to discriminate the jets created from the decays of the SM heavy resonance particles (top quark, $W$, $Z$ and Higgs bosons) with respect to the quark and gluon jets which have pure QCD origin. This asymmetry cut also removes contamination from ISR, FSR, etc. Because of that, the softdrop jet mass is close to the heavy resonance particle mass if the source of the fat-jet is a decay of heavy resonance particle. 
 
\begin{table}[h]
\begin{center}
\tiny
  \begin{tabular}{|c|c|c|c|c| } 
 \hline 
    Backgrounds processes& SSDL+jets&14 TeV& 27 TeV& 100 TeV\\
\hline
     &&(fb)&(fb)&(fb)\\
\hline
$WWjj$&$\mu^\pm \mu^\pm $&0.71&1.7&3.8\\
&$e^\pm e^\pm $&0.52&1.7&3.2\\
\hline
$WZjj$&$\mu^\pm \mu^\pm $&4.84&16.2&39.7\\
&$e^\pm e^\pm $&3.6&16.2&39.8\\
\hline
$WWZ+$jets&$\mu^\pm \mu^\pm $&0.17&0.71&3.2\\
&$e^\pm e^\pm $&0.13&0.7&3.3\\
\hline
$t\overline{t}W$&$\mu^\pm \mu^\pm $&1.12&3.6&7.2\\
&$e^\pm e^\pm $&0.82&2.5&7.0\\
\hline 
$t\overline{t}W$&$\mu^\pm \mu^\pm $&1.12&3.6&7.2\\
&$e^\pm e^\pm $&0.82&2.5&7.0\\
\hline 
$t\overline{t}Z$&$\mu^\pm \mu^\pm $&1.28&6.7&29.2\\
&$e^\pm e^\pm $&0.95&6.7&29.0\\
\hline
\end{tabular}
\end{center}
\caption{SM backgrounds after imposing the basic cuts at different pp colliders with $\sqrt{s}=14$ TeV, $27$ TeV and $100$ TeV for SSDL plus jets signature.}
\label{XsecBG-1}
\end{table} 
 
\begin{figure}[h]
\centering
\includegraphics[width=1\textwidth,angle=0]{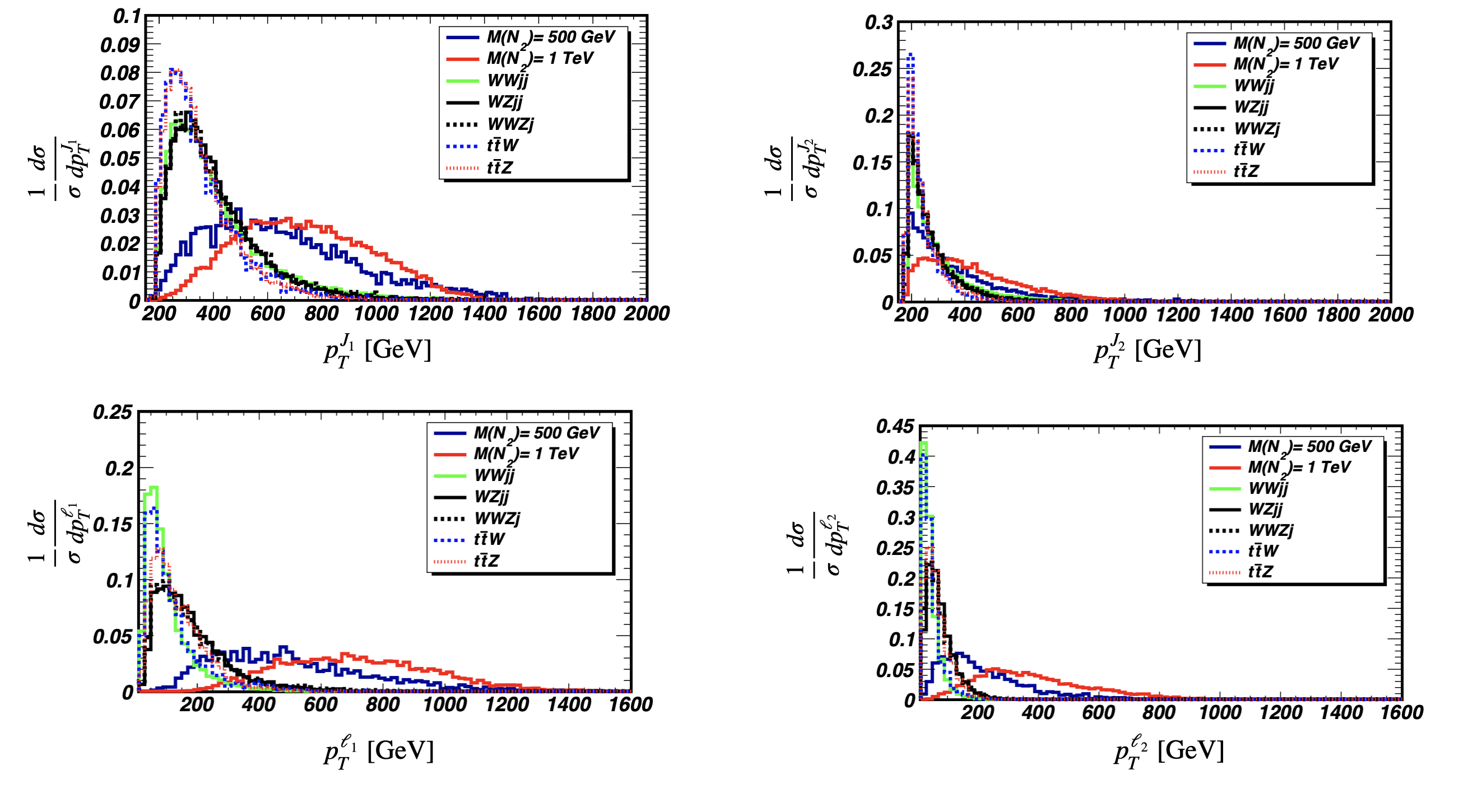}\\
\caption{Transverse momentum distribution of the leading (top, left), sub leading (top, right) fat jets and the leading (bottom, left), subleading (bottom, right) leptons for $M_N=500$ GeV (blue, solid) and $1$ TeV (red, solid) respectively with the corresponding SM backgrounds at $\sqrt{s}=$ 14 TeV.}
\label{2j2l-14-1}
\end{figure}

\begin{figure}[h]
\centering
\includegraphics[width=1\textwidth,angle=0]{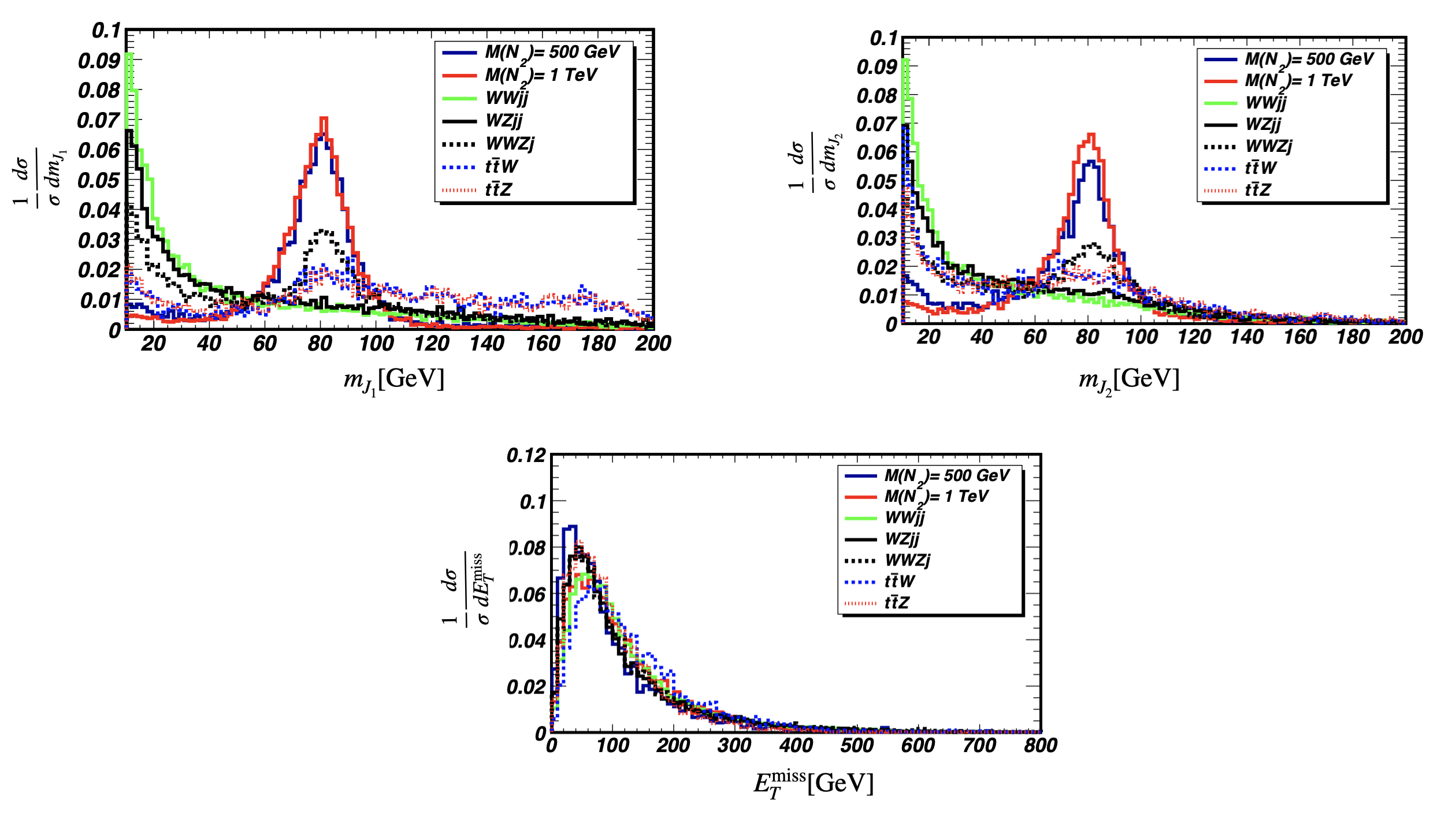}\\
\caption{Jet mass distribution of the leading (top, left), subleading (top, right) fat jets and the missing energy distribution (bottom) for $M_N=500$ GeV (blue, solid) and $1$ TeV (red, solid) respectively with the corresponding SM backgrounds at $\sqrt{s}=$ 14 TeV.}
\label{2j2l-14-2}
\end{figure}

Generating the $\mu^{\pm}\mu^{\pm}+2J$ and $e^{\pm} e^{\pm}+2J$ signals at $14$ TeV we study Cases-I and II. The corresponding SM backgrounds are $WWjj$, $WZjj$, $WWZ+$jets, $t\overline{t}W$ and $t\overline{t}Z$ respectively. The backgrounds are generated using the selection cuts $H_T > 300$ GeV where $H_T$ is the scalar sum of the transverse momentum of the jets $(p_T^j)$. The jets are selected with the transverse momentum $p_T^j > 10$ GeV, pseudorapidity of jets $|\eta^j| < 2.5$. The transverse momentum of the leptons $p_T^\ell > 10$ GeV, pseudorapidity of the leptons $|\eta^\ell| < 2.5$, separation between the leptons in the $\eta-\phi$ plane $\Delta R_{\ell \ell} > 0.4$ and lepton and jets $\Delta R_{\ell j} > 0.4$ are among them. The partonic cross sections after the basic cuts are given in Tab.~\ref{XsecBG-1} for $\sqrt{s}=14$ TeV.
To generate the $WWZ+\rm jets$, $t\bar{t}W$, and $t\bar{t}Z$ backgrounds, we slightly change our generation level cuts. Here we applied the same $H_{T}$ cut as above for all jets, including jets coming from standard model heavy resonance decay. However we did not use any $p_{T}$, $\eta$, and $\Delta_{R}$ cuts to leptons and jets coming from standard model resonance decay. After the primary cuts the $p_T$ distributions of the fat-jets and the leptons are shown in Fig.~\ref{2j2l-14-1}. The distributions of the jet masses and missing energy are shown in Fig.~\ref{2j2l-14-2}.  After selecting the signal events we use advanced cuts for the signals and SM backgrounds. To study the SM background from $WWZ+$ jets process we consider the leptonic decay of one $W$ and the hadronic decay of the other whereas the $Z$ boson decays into charged leptons. Such leptonic events allow us to use a third lepton veto in the advanced cuts. A similar procedure has been adopted for $t\overline{t} Z$. To study SM backgrounds from $t\overline{t}Z$ process, we consider the leptonic decay of one $W$ from the top quark and the hadronic decay of the other from the remaining top quark, whereas the $Z$ boson decays leptonically.



From Fig.~\ref{2j2l-14-1} we find the peak of the $p_T^{J_1}$ distribution situated in the high $p_T$ region whereas the backgrounds are distributed mainly in the low $p_T$ region. We consider this as the leading fat-jet and the remaining fat-jet $(J_2)$ as the sub-leading one. The sub-leading fat-jet has a peak coinciding with the SM backgrounds. For the heavier RHNs, these jets are more energetic, and their distribution shifted to the higher $p_T$ region. The leptons from the RHNs are more energetic than those from SM backgrounds. Hence leptons' $p_T$ will be another important discriminator along with the $p_T$ of the fat-jets. From the distribution of the jet masses in the upper panel of Fig.~\ref{2j2l-14-2}, we find peaks from the signal and some SM backgrounds like $WWZ$, $t\overline{t} W$ and $t\overline{t} Z$ around the $W$ boson mass; however, the backgrounds peaks are not that prominent as the signal peak. These allow us to choose a window of $15$ GeV around the $W$ boson mass to suppress the SM backgrounds further and get rid of other low energy hadronic effects below $60$ GeV. From the missing energy distributions shown in the lower panel of Fig.~\ref{2j2l-14-2} we find a conservative cut below 150 GeV. We have noticed that many trilepton events coming from the SM processes $WWZ$, $t\overline{t} W$ and $t\overline{t} Z$ where we use a criteria of at least two leptons having the same sign. In addition to that we use a third lepton veto which reduces such SM background events further. The backgrounds coming from the top quark induced events contain b-jets. In the final selection we use a b-veto to reject such events.

Selecting the signal and backgrounds with basic cuts we employ the post selection cuts or advanced cuts. The transverse momentum of each fat jet has been considered to be $p_T^{J} > 180$ GeV (C-I). We select the events with SSDL with transverse momenta $p_T^{\ell_1}, p_T^{\ell_2} > 100$ GeV (C-II). Both of the jets have sub-jet number greater than 1 (C-III). We have considered that the invariant mass of the two leading sub-jet is within a window of $\pm15$ GeV around the $W$ boson mass for the leading fat-jets (C-IV). Finally we consider that both jets are not b-jets (C-V). We apply a conservative cut on the missing energy as $E_T^{\rm miss} < 150$ GeV (C-VI). The cut flow has been shown in Tab.~\ref{tabI-0-1} with the corresponding efficiencies for the signals and corresponding backgrounds for $\mu^\pm \mu^\pm$ and $e^\pm e^\pm$ samples in the upper and lower halves of the table respectively. The upper part of each row represents the $\ell^+\ell^+$ signal and the lower part represents the $\ell^- \ell^-$ signal respectively. The combined significance of the SSDL mode has been calculated using $\frac{S}{\sqrt{S+B}}$ for signal $(S)$ and background $(B)$ events. Solving for a particular significance we estimate the required luminosity for Cases-I and II. We find the significance is below 1$-\sigma$ up to the total luminosity of the LHC for both of the benchmark RHN masses whereas an achievable 5$-\sigma$ significance can be reached for Case-II around $1.88$ ab$^{-1}$ for $M_N=1$ TeV and $1.52$ ab$^{-1}$ luminosities for $M_N=500$ GeV respectively. 
\begin{table}[h]
\begin{center}
\tiny
 \begin{tabular}{ |c|c|c|c|c|c|c|c|c|c| } 
 \hline
 $\mu^\pm \mu^\pm+2J$   & $W^{\pm}W^{\pm}jj$ & $W^{\pm}Zjj$ & $W^{\pm}W^{\mp}Zj$ & $t\bar{t}W^{\pm}$ & $t\bar{t}Z$ & sig($500$ GeV) & sig($1$ TeV)& sig($500$ GeV) &sig($1$ TeV)\\
   &&&&&&Case-I&Case-I&Case-II&Case-II\\
   &(fb)&(fb)&(fb)&(fb)&(fb)&(fb)&(fb)&(fb)&(fb)\\
 \hline
  C-I   & 0.1398    & 0.147     & 0.00562   & 0.0347    & 0.0224    & 0.00086   & 0.0012    & 0.028     & 0.025\\ 
        & $[100\%]$ & $[100\%]$ & $[100\%]$ & $[100\%]$ & $[100\%]$ & $[100\%]$ & $[100\%]$ & $[100\%]$ & $[100\%]$\\
   \hline
  C-II  & 0.01        & 0.0224  & 0.00093   & 0.0055 & 0.00114  & 0.00070 & 0.0011 & 0.021 & 0.024\\ 
        & $[7.15\%]$  & $[15.23\%]$ & $[16.55\%]$ & $[15.85\%]$ & $[6.25\%]$ & $[81.39\%]$ & $[91.60\%]$ & $[75\%]$ & $[96\%]$\\
     \hline
         C-III & 0.0098 & 0.0221  & 0.00091 & 0.002 & 0.00112 & 0.00065  & 0.001 & 0.022 & 0.023\\ 
       & $[7.01\%]$ & $[15.03\%]$ & $[16.19\%]$ & $[5.76\%]$ & $[5\%]$ & $[75.58\%]$ & $[83.30\%]$ & $[78.57\%]$ & $[92\%]$\\   
   \hline
C-IV & 0.00067 & 0.00206 & 0.00021 & 0.000354 & 0.00017 & 0.00043  & 0.00075 & 0.014 & 0.016\\   
     & $[0.48\%]$ & $[1.77\%]$ & $[3.74\%]$ & $[1.02\%]$ & $[0.76\%]$ & $[50\%]$ & $[62.50\%]$ & $[50\%]$ & $[64\%]$\\
     \hline
C-V & 0.00061 & 0.00107 & 0.00019 & 0.000195 & 0.000063 & 0.00036  & 0.00062 & 0.012 & 0.013\\
    & $[0.44\%]$ & $[0.73\%]$ & $[3.38\%]$ & $[0.56\%]$ & $[0.28\%]$ & $[38.37\%]$ & $[51.60\%]$ & $[42.85\%]$ & $[52\%]$\\
      \hline
 C-VI & 0.00043  & 0.00073 & 0.00014 & 0.000164 & 0.000063  & 0.00033 & 0.00048 & 0.011 & 0.01\\
      & $[0.31\%]$ & $[0.50\%]$ & $[2.49\%]$ & $[0.47\%]$ & $[0.28\%]$ & $[38.37\%]$ & $[40\%]$ & $[39.28\%]$ & $[40\%]$\\
   \hline
  \hline
  $e^\pm e^\pm+2J$ & $W^{\pm}W^{\pm}jj$ & $W^{\pm}Zjj$ & $W^{\pm}W^{\mp}Zj$ & $t\bar{t}W^{\pm}$ & $t\bar{t}Z$ & sig($500$ GeV) & sig($1$ TeV)& sig($500$ GeV) &sig($1$ TeV)\\
   &&&&&&Case-I&Case-I&Case-II&Case-II\\
   &(fb)&(fb)&(fb)&(fb)&(fb)&(fb)&(fb)&(fb)&(fb)\\
 \hline
 C-I    &0.103&0.11 &0.0041  & 0.026 & 0.0017  & 0.00061&0.0009&0.0204&0.018\\ 
        & $[100\%]$ & $[100\%]$ & $[100\%]$ & $[100\%]$ & $[100\%]$ & $[100\%]$ & $[100\%]$ & $[100\%]$ & $[100\%]$\\
   \hline
   C-II    & 0.0073 & 0.017 & 0.0007 & 0.0017 & 0.00090  & 0.00048  & 0.00082 & 0.0162 & 0.0173\\ 
           & $[7.08\%]$ & $[15.45\%]$ & $[17.07\%]$ & $[6.54\%]$ & $[52.94\%]$ & $[78.69\%]$ & $[91.10\%]$ & $[79.41\%]$ & $[96.10\%]$\\
  \hline
 C-III     & 0.00721  & 0.016  & 0.00066 & 0.00168 & 0.00082 & 0.00047  & 0.00081 & 0.016 & 0.015\\ 
           & $[7\%]$ & $[14.54\%]$ & $[16.10\%]$ & $[6.46\%]$ & $[48.23\%]$ & $[77.05\%]$ & $[90\%]$ & $[78.43\%]$ & $[83.33\%]$\\ 
   \hline
  C-IV     & 0.00050 & 0.0015  & 0.00016 & 0.000345 & 0.00013 & 0.00031  & 0.0006 & 0.01 & 0.012\\   
           & $[0.48\%]$ & $[1.36\%]$ & $[3.90\%]$ & $[1.33\%]$ & $[7.65\%]$ & $[50.82\%]$ & $[66.60\%]$ & $[49.02\%]$ & $[66.60\%]$\\
    \hline
C-V        & 0.00045 & 0.0014  & 0.00007 & 0.000234  & 0.000048 & 0.00025  & 0.0005 & 0.0084 & 0.01\\
           & $[0.44\%]$ & $[1.27\%]$ & $[1.71\%]$ & $[0.90\%]$ & $[2.82\%]$ & $[40.98\%]$ & $[55.50\%]$ & $[41.18\%]$ & $[55.50\%]$\\
     \hline
C-VI       & 0.00032 & 0.001 & 0.00005  & 0.00018  & 0.000046   & 0.00023  & 0.00035 & 0.0080 & 0.007\\
           & $[0.31\%]$ & $[0.90\%]$ & $[1.22\%]$ & $[0.69\%]$ & $[2.70\%]$ & $37.70[\%]$ & $[38.80\%]$ & $[39.21\%]$ & $[38.80\%]$\\
 \hline
 \end{tabular}
 \end{center}
\caption{Cut flow for SSDL plus two fat-jet signal. Upper block represents the muon and lower block represents the electron signals. We consider $M_{Z^\prime}=3$ TeV at $\sqrt{s}=14$ TeV.}
\label{tabI-0-1}
\end{table}

We analyze the SSDL signal with two fat jets consisting $\mu^{\pm}\mu^{\pm}+2J$ and $e^{\pm} e^{\pm}+2J$ signals at $\sqrt{s}=27$ TeV. After that we generate the corresponding SM backgrund processes $WWjj$, $WZjj$, $WWZ+$jets, $t\overline{t}W$ and $t\overline{t}Z$ to compare the signals for $M_{N_{1,2}}=500$ GeV and $M_{N_{1,2}}=1$ TeV both for Case-I and II. To generate the backgrounds we impose the selection cuts as $H_T > 300$ GeV, $p_T^j > 10$ GeV, $|\eta^j| < 2.5$, $p_T^\ell > 10$ GeV, $|\eta^\ell| < 2.5$, $\Delta R_{\ell \ell} > 0.4$ and $\Delta R_{\ell j} > 0.4$ respectively. The parton level cross sections of the SM backgrounds after the selection cuts are given in Tab.~\ref{XsecBG-1}. By studying the signal and the backgrounds we finalize the advanced cuts for the signals and backgrounds. We do not show the histograms, apart from the higher energy reaches of the kinematic variables, their nature will  remain almost the same as the previous case. The advanced cuts are used selecting at least two fat-jets $(J)$ with transverse momentum $p_T^J > 180$ GeV (C-I). Only two same sign leptons with $p_T^\ell > 100$ GeV are selected (C-II) vetoing the third lepton. Both of the jets have sub-jet number $>1$ (C-III). The Invariant mass of the leading two sub-jets is within $\pm15$ GeV window around the $W$ boson mass (C-IV). We ensure that both jets are not b-jets (C-V). The missing energy is conservatively considered as $\slashed{E_T} < 150$ GeV (C-VI). The cut flow and the corresponding efficiencies for $\mu^\pm \mu^\pm+2J$ and $e^\pm e^\pm+2J$ signals and SM backgrounds are given in the top, right panel of Tab.~\ref{tabI-1}. 

\begin{table}[h]
\begin{center}
\tiny
 \begin{tabular}{ |c|c|c|c|c|c|c|c|c|c| } 
 \hline
 $\mu^\pm \mu^\pm+2J$   & $W^{\pm}W^{\pm}jj$ & $W^{\pm}Zjj$ & $W^{\pm}W^{\mp}Zj$ & $t\bar{t}W^{\pm}$ & $t\bar{t}Z$ & sig($500$ GeV) & sig($1$ TeV)& sig($500$ GeV) &sig($1$ TeV)\\
   &&&&&&Case-I&Case-I&Case-II&Case-II\\
   &(fb)&(fb)&(fb)&(fb)&(fb)&(fb)&(fb)&(fb)&(fb)\\
 \hline
C-I     & $0.340$ & $0.40$ & $0.026$ & $0.125$ & $0.123$  & $0.008$ & $0.01$&0.37&0.31\\ 
        & $[100\%]$ & $[100\%]$ & $[100\%]$ & $[100\%]$ & $[100\%]$ & $[100\%]$ & $[100\%]$ & $[100\%]$ & $[100\%]$\\
   \hline
  C-II     & $0.037$ & $0.12$ & $0.0061$ & $0.01$ & $0.011$ & $0.006$ & $0.0097$&0.3&0.29\\ 
           & $[10.88\%]$ & $[30\%]$ & $[23.46\%]$ & $[8\%]$ & $[8.94\%]$ & $[75\%]$ & $[97\%]$ & $[81.08\%]$ & $[93.55\%]$\\
  \hline
  C-III   & $0.036$ & $0.117$ & $0.0051$ & $0.0093$ & $0.0103$ & $0.0056$ & $0.0093$&0.28&0.285\\ 
          & $[10.59\%]$ & $[29.25\%]$ & $[19.61\%]$ & $[7.44\%]$ & $[8.37\%]$ & $[70\%]$ & $[93\%]$ & $[75.67\%]$ & $[91.93\%]$\\ 
   \hline
   C-IV       & $0.0028$ & $0.01$ & $0.0011$ & $0.002$ & $0.0012$ & $0.0034$ & $0.0062$&0.17&0.19\\   
              & $[0.82\%]$ & $[2.5\%]$ & $[4.23\%]$ & $[1.6\%]$ & $[0.97\%]$ & $[42.5\%]$ & $[62\%]$ & $[45.94\%]$ & $[61.29\%]$\\
    \hline
 C-V      & $0.0026$ & $0.007$ & $0.001$ & $0.001$ & $0.0007$ & $0.0028$ & $0.0051$&0.14&0.16\\
          & $[0.76\%]$ & $[1.75\%]$ & $[3.85\%]$ & $[0.8\%]$ & $[0.57\%]$ & $[35\%]$ & $[51\%]$ & $[37.84\%]$ & $[51.61\%]$\\
     \hline
C-VI       & $0.0016$ & $0.006$ & $0.0006$ & $0.00071$ &  $0.0006$ & $0.0025$ & $0.0038$ & 0.13 & 0.142\\
           & $[0.47\%]$ & $[1.5\%]$ & $[2.31\%]$ & $[0.57\%]$ & $[0.49\%]$ & $[31.25\%]$ & $[38\%]$ & $[35.13\%]$ & $[45.81\%]$\\
 \hline
  \hline
  $e^\pm e^\pm+2J$ & $W^{\pm}W^{\pm}jj$ & $W^{\pm}Zjj$ & $W^{\pm}W^{\mp}Zj$ & $t\bar{t}W^{\pm}$ & $t\bar{t}Z$ & sig($500$ GeV) & sig($1$ TeV)& sig($500$ GeV) &sig($1$ TeV)\\
   &&&&&&Case-I&Case-I&Case-II&Case-II\\
   &(fb)&(fb)&(fb)&(fb)&(fb)&(fb)&(fb)&(fb)&(fb)\\
 \hline
C-I     & $0.19$ & $0.62$ & $0.0214$ & $0.08$ & $0.098$  & $0.0046$ & $0.0070$&0.15&0.134\\ 
        & $[100\%]$ & $[100\%]$ & $[100\%]$ & $[100\%]$ & $[100\%]$ & $[100\%]$ & $[100\%]$ & $[100\%]$ & $[100\%]$\\
   \hline
C-II       & $0.084$ & $0.125$ & $0.0046$ & $0.0071$ & $0.0094$ & $0.0038$ & $0.0063$&0.12&0.125\\ 
           & $[44.21\%]$ & $[20.16\%]$ & $[21.49\%]$ & $[8.87\%]$ & $[9.59\%]$ & $[82.61\%]$ & $[90\%]$ & $[80\%]$ & $[93.28\%]$\\
  \hline
C-III         & $0.025$ & $0.122$ & $0.0045$ & $0.007$ & $0.0092$ & $0.0036$ & $0.0060$&0.11&0.121\\ 
              & $[13.16\%]$ & $[19.68\%]$ & $[21.03\%]$ & $[8.75\%]$ & $[9.39\%]$ & $[78.26\%]$ & $[85.71\%]$ & $[73.30\%]$ & $[90.30\%]$\\ 
   \hline
C-IV          & $0.002$ & $0.0114$ & $0.001$ & $0.0014$ & $0.0014$ & $0.0022$ & $0.0041$&0.07&0.083\\   
              & $[1.05\%]$ & $[1.84\%]$ & $[4.67\%]$ & $[1.75\%]$ & $[1.43\%]$ & $[47.82\%]$ & $[58.57\%]$ & $[46.60\%]$ & $[61.94\%]$\\
    \hline
C-IV       & $0.0018$ & $0.01$ & $0.0009$ & $0.00063$ & $0.00087$ & $0.002$ & $0.0036$&0.061&0.070\\
           & $[0.95\%]$ & $[1.61\%]$ & $[4.20\%]$ & $[0.79\%]$ & $[0.89\%]$ & $[43.48\%]$ & $[51.43\%]$ & $[40.60\%]$ & $[52.24\%]$\\
     \hline
C-VI       & $0.0011$ & $0.0062$ & $0.0007$ & $0.00045$ &  $0.00053$ & $0.0018$ & $0.0031$&0.054&0.061\\
           & $[0.58\%]$ & $[1\%]$ & $[3.27\%]$ & $[0.56\%]$ & $[0.54\%]$ & $[39.13\%]$ & $[44.28\%]$ & $[36\%]$ & $[45.52\%]$\\
 \hline
 \end{tabular}
 \end{center}
\caption{Cut flow for SSDL plus two fat-jet signal. Upper block represents the muon and lower block represents the electron signals. We consider $M_{Z^\prime}=3$ TeV at $\sqrt{s}=27$ TeV.}
\label{tabI-1}
\end{table}
Solving the signal and background relation for a particular significance we can estimate the luminosity to achieve that significance. Hence we calculate the significance for 1 TeV RHN from Case-I around $5-\sigma$ at 14.23 ab$^{-1}$ luminosity and that for 1 TeV reaches up to $3.5-\sigma$ at 15 ab$^{-1}$ luminosity. The results for Case-II are also improved compared to the previous case. A significance of $5-\sigma$ can be attained at $144$ fb$^{-1}$ luminosity for 500 GeV RHN whereas that can be attained at $130.5$ fb$^{-1}$ luminosity for 1 TeV RHN respectively. 

Finally, we study the SSDL signal from Cases-I and II at $\sqrt{s}=100$ TeV proton proton collider considering $M_{Z^\prime}=3$ TeV. All primary selection cuts are the same as the previous two cases except $H_T$. Here we use $H_{T} > 500$ GeV for all the background processes. The partonic cross sections of the SM backgrounds are given in Tab.~\ref{XsecBG-1}. After selecting the signal events we use the advanced cuts for the signals and SM backgrounds. The advanced cuts are used by selecting at least two fat-jets $(J)$ with transverse momentum $p_T^J > 250$ GeV (C-I). The other advanced cuts are the same as the previous two cases. In this case we have only considered two degenerate RHNs with mass $M_{N}=1$~TeV. The cut flow table for SSDL scenario is given in Tab.~\ref{tab4} with the efficiencies after each level of cuts for the $\mu^\pm \mu^\pm+2J$ and $e^\pm e^\pm+2J$ channels. We find that the $5-\sigma$ significance could be achievable for the Case-I around $534$ fb$^{-1}$ luminosity and for the Case-II, the same benchmark point can be achievable with $22.2$ fb$^{-1}$ luminosity.  


\begin{table}[h]
\begin{center}
\tiny
  \begin{tabular}{ |c|c|c|c|c|c|c|c|c| } 
 \hline 
    $\mu^\pm\mu^\pm+2J$& $W^{\pm}W^{\pm}jj$ & $W^{\pm}Zjj$ & $W^{\pm}W^{\mp}Zj$ & $t\bar{t}W^{\pm}$ & $t\bar{t}Z$ & sig($1$ TeV)&sig($1$ TeV)\\
    &(fb)&(fb)&(fb)&(fb)&(fb)&(fb)&(fb)\\
    &&&&&&Case-I&Case-II\\
 \hline
 \hline
C-I  & $0.811$ & $2.221$ & $0.122$ & $0.202$ & $0.476$  & $0.102$&2.0\\ 
     & $[100\%]$ & $[100\%]$ & $[100\%]$ & $[100\%]$ & $[100\%]$ & $[100\%]$ & $[100\%]$\\
    \hline
C-II & $0.141$ & $0.674$ & $0.0384$ & $0.034$ & $0.065$ & $0.1$&1.94\\ 
     & $[17.38\%]$ & $[30.35\%]$ & $[31.47\%]$ & $[16.83\%]$ & $[13.65\%]$ & $[98.04\%]$ & $[97\%]$\\
\hline
C-III & $0.138$ & $0.664$ & $0.0224$ & $0.033$ & $0.064$ & $0.097$&1.92\\ 
    & $[17.02\%]$ & $[29.90\%]$ & $[18.36\%]$ & $[16.34\%]$ & $[13.44\%]$ & $[95.10\%]$ & $[96\%]$\\
 \hline
C-IV & $0.011$ & $0.0484$ & $0.0065$ & $0.0028$ &  $0.0085$ & $0.0542$&1.1\\ 
     & $[1.36\%]$ & $[2.18\%]$ & $[5.33\%]$ & $[1.39\%]$ & $[1.78\%]$ & $[53.14\%]$ & $[55\%]$\\
 \hline 
 C-V & $0.009$ & $0.042$ & $0.0059$ & $0.002$ & $0.0038$ & $0.047$&0.91\\ 
     & $[1.11\%]$ & $[1.89\%]$ & $[4.84\%]$ & $[0.99\%]$ & $[0.80\%]$ & $[46.08\%]$ & $[45.50\%]$\\
  \hline
C-VI & $0.0038$ & $0.0224$ & $0.0031$ & $0.0006$ &  $0.0018$ & $0.033$ &0.61\\ 
    & $[0.47\%]$ & $[1\%]$ & $[2.54\%]$ & $[0.30\%]$ & $[0.38\%]$ & $[32.35\%]$ & $[30.50\%]$\\
\hline
 \hline
  $e^\pm e^\pm+2J$  & $W^{\pm}W^{\pm}jj$ & $W^{\pm}Zjj$ & $W^{\pm}W^{\mp}Zj$ & $t\bar{t}W^{\pm}$ & $t\bar{t}Z$ & sig($1$ TeV)&sig($1$ TeV)\\
    &(fb)&(fb)&(fb)&(fb)&(fb)&(fb)&(fb)\\
    &&&&&&Case-I&Case-II\\
 \hline
 \hline
C-I & $0.42$ & $1.94$ & $0.0933$ & $0.0944$ & $0.34$  & $0.065$&1.3  \\ 
    & $[100\%]$ & $[100\%]$ & $[100\%]$ & $[100\%]$ & $[100\%]$ & $[100\%]$ & $[100\%]$\\
    \hline
C-II  & $0.087$ & $0.65$ & $0.0161$ & $0.0184$ & $0.060$ & $0.063$&1.25  \\ 
      & $[20.71\%]$ & $[33.50\%]$ & $[17.26\%]$ & $[19.49\%]$ & $[17.65\%]$ & $[96.92\%]$ & $[96.15\%]$\\
   \hline
C-III & $0.085$ & $0.64$ & $0.008$ & $0.0181$ & $0.056$ & $0.063$ & 1.24  \\ 
      & $[20.24\%]$ & $[32.99\%]$ & $[8.57\%]$ & $[19.17\%]$ & $[16.47\%]$ & $[96.92\%]$ & $[95.38\%]$\\
    \hline
 C-IV & $0.031$ & $0.054$ & $0.0055$ & $0.0021$ &  $0.0067$ & $0.038$&0.74  \\ 
      & $[7.38\%]$ & $[2.78\%]$ & $[5.90\%]$ & $[2.22\%]$ & $[1.97\%]$ & $[58.46\%]$ & $[56.92\%]$\\
      \hline
C-V & $0.0052$ & $0.046$ & $0.0045$ & $0.00143$ & $0.0047$ & $0.031$&0.614   \\ 
    & $[1.24\%]$ & $[2.37\%]$ & $[4.82\%]$ & $[1.51\%]$ & $[1.38\%]$ & $[47.69\%]$ & $[47.23\%]$\\
    \hline
 C-VI& $0.0024$ & $0.024$ & $0.0025$ & $0.00072$ &  $0.0023$ & $0.028$&0.541  \\ 
    & $[0.57\%]$ & $[1.24\%]$ & $[2.68\%]$ & $[0.76\%]$ & $[0.68\%]$ & $[43.10\%]$ & $[41.61\%]$\\
\hline
 \end{tabular}
 \end{center}
\caption{Cut flow for SSDL plus two fat-jet signal. Upper block represents the muon and lower block represents the electron signals. We consider $M_{Z^\prime}=3$ TeV at $\sqrt{s}=100$ TeV.}
\label{tab4}
\end{table}

\subsection{Trilepton plus fat-jet in association with missing energy}
\label{ssdl-3}
After the pair production of the RHNs from $Z^\prime$, each of the RHNs will dominantly decay into $\ell W$ mode. One of the $W$ can decay hadronically producing a fat-jet and the other $W$ can decay leptonically. Hence a trilepton signal will be generated with a fat-jet in association with missing energy. In the leptonic decay of the $W$ boson we do not consider tau lepton in this analysis. In this case the selection and advanced cuts will remain the same as the SSDL case. Hence we do not show any histograms. The fat-jet will be created from the boosted $W$ produced from the decay of one of the RHNs. We consider $M_{Z^\prime}=$3 TeV and RHNs are degenerate with two benchmark masses at $M_{N_{1,2}}=500$ GeV and $M_{N_{1,2}}=1$ TeV. The trilepton signal consists of two same sign leptons and the other one must have the opposite sign. Therefore the combination of three lepton system has either $`+1$' or $`-1$' charge taking all possible combinations of the leptons. 

We generate SM backgrounds with pre-selection cuts including the transverse momentum of the jets $p_T^j > 10$ GeV and pseudo-rapidity of the jets $|\eta^j| < 2.5$. We consider the separation cuts in the $\eta-\phi$ plane between jets and leptons as $\Delta R_{j\ell} > 0.4$ and lepton-lepton as $\Delta R_{\ell \ell} > 0.4$. At the 27 TeV and 100 TeV we consider the transverse momentum of the jets as $p_T^j > 20$ GeV with other pre-selection cuts. 

We generate the backgrounds $W^\pm h+$ jets, $W^\pm Z+$jets, $ZZ+$jets, $W^\pm W^\pm W^\mp+$jets, $ZZW^\pm$, $W^\pm W^\mp Z$, $t\overline{t} W$ and $t\overline{t}Z$, to study trilepton final state plus one fat-jet in association with missing energy. For these backgrounds we consider leptonic decay modes of the gauge bosons. From these backgrounds we consider  $\mu^\pm \mu^\pm \mu^\pm (e^\pm)$ and $e^\pm e^\pm e^\pm (\mu^\pm)$ modes respectively in association with a fat-jet and missing energy which will mimic the trilepton plus missing energy and fat-jet signal from the RHN pair. The SM background cross sections are given in Tab.~\ref{XsecBG} at proton colliders for different $\sqrt{s}$ after the application of the basic cuts.

\begin{table}[h]
\begin{center}
\tiny
  \begin{tabular}{|c|c|c|c|c| } 
 \hline 
    Backgrounds processes& Signals&14 TeV& 27 TeV& 100 TeV\\
\hline
     &&(fb)&(fb)&(fb)\\
\hline
$t\overline{t}W^{\pm}$&$\mu^\pm \mu^\pm \mu^\pm (e^\pm) $&7.3&21.33&127.5\\
&$e^\pm e^\pm e^\pm (\mu^\pm)$&5.36&15.67&93.7\\
\hline
$t\overline{t}Z$&$\mu^\pm \mu^\pm \mu^\pm (e^\pm) $&23.84&112.9&1510.0\\
&$e^\pm e^\pm e^\pm (\mu^\pm)$&17.51&82.9&1110.0\\
\hline
$W^\pm h j$&$\mu^\pm \mu^\pm \mu^\pm (e^\pm) $&6.8&15.1&87.9\\
&$e^\pm e^\pm e^\pm (\mu^\pm)$&5.0&11.1&64.6\\
\hline
$W^\pm Z j$&$\mu^\pm \mu^\pm \mu^\pm (e^\pm) $&634.37&1248.2&11000.0\\
&$e^\pm e^\pm e^\pm (\mu^\pm)$&466.07&916.9&8050.0\\
\hline
$W^\pm W^\pm W^\mp j$&$\mu^\pm \mu^\pm \mu^\pm (e^\pm) $&2.55&8.5&64.7\\
&$e^\pm e^\pm e^\pm (\mu^\pm)$&1.88&6.22&47.5\\
\hline
$ZZj$&$\mu^\pm \mu^\pm \mu^\pm (e^\pm) $&94.75&256.9&1500.0\\
&$e^\pm e^\pm e^\pm (\mu^\pm)$&69.61&188.8&1300.0\\
\hline
$ZZW^\pm$&$\mu^\pm \mu^\pm \mu^\pm (e^\pm) $&1.043&11.9&7.582\\
&$e^\pm e^\pm e^\pm (\mu^\pm)$&0.78&6.21&10.35\\
\hline
$W^\pm W^\mp Z$&$\mu^\pm \mu^\pm \mu^\pm (e^\pm) $&3.39&9.1&48.04\\
&$e^\pm e^\pm e^\pm (\mu^\pm)$&2.5&6.7&35.3\\
\hline
 \end{tabular}
 \end{center}
\caption{SM backgrounds after imposing the basic cuts at different pp colliders with $\sqrt{s}=14$ TeV, $27$ TeV and $100$ TeV for trilepton plus jets in association with missing energy process.}
\label{XsecBG}
\end{table}

In this $\rm trilepton+jet+MET$ final state, the leading two leptons in $p_{T}$ order are expected to come directly from the decay of the RHN ($N \to W\ell$) and the third lepton is mostly coming from further decays of one of the $W$ boson (we call it a trailing lepton). As the RHN is very heavy, the leading two leptons have larger $p_{T}$ than the third lepton. Here we use a large $p_{T}$ cut for leading and sub-leading leptons and a low $p_{T}$ cut for the third lepton. Another $W$ boson comes from one of the RHN if decays hadronically form collimated jet or fat-jet as this $W$ boson is coming from heavy RHN decay, so it is boosted. This fat-jet can have a large $p_{T}$. Using softdrop algorithm, we estimate the jet mass of the fat-jet for the signal and background. The nature of the distributions regarding the leading and subleading leptons and fat-jet are almost the same as those of the SSDL case. Therefore we do not add the histograms for trilepton mode.


At the $14$ TeV and $27$ TeV hadron colliders we select the events with at least one fat-jet with $p_T^J > 180$ GeV (C-I). Three leptons are selected as leading, sub-leading and trailing respectively with $ p_{T}^{\ell_{1}}> 200$ GeV,  $p_{T}^{\ell_{2}} > 100$ GeV and  $p_{T}^{\ell_{3}} >  20$ GeV (C-II). Finally the softdrop jet mass is selected within a window of $\pm 15$ GeV around the $W$ boson mass (C-III). At the $100$ TeV collider along with C-I and C-III we add different lepton $p_T$ cuts. At the 100 TeV three leptons are selected with $ p_{T}^{\ell_{1}}> 200$ GeV,  $p_{T}^{\ell_{2}} > 200$ GeV and  $p_{T}^{\ell_{3}} >  20$ GeV respectively (C-II). The cut flow with the efficiencies for $\mu^\pm \mu^\pm \ell^\mp$ and $e^\pm e^\pm \ell^\mp$ signals and backgrounds are shown in the upper and lower panels of Tab.~\ref{tabI-3l-2} for $\sqrt{s}=14$ TeV. We study 500 GeV and 1 TeV RHNs as benchmark points for Cases-I and II. Like the SSDL case here also we consider the degenerate RHNs. The leading background is originated from $WZ+$jets process where as the sub-leading contribution from the background is coming from $WWZ$ process. The other backgrounds are 2-3 orders of magnitude smaller than the leading background. 
\begin{table}[h]
\begin{center}
\tiny
  \begin{tabular}{ |c|c|c|c|c|c|c|c|c|c|c|c|c| } 
 \hline
   $ \mu^{\pm} \mu^{\pm} \mu^{\mp}/e^{\mp}$&$t\bar{t}W^{\pm}$&$t\bar{t}Z$ & $W^{\pm}h j$ & $W^{\pm}Zj$ & $W^{\pm}W^{\pm}W^{\mp}j$ & $ZZj$&$ZZW^{\pm}$&$W^{\pm}W^{\mp}Z$&Sig($500$ GeV)&Sig($1$ TeV)&Sig($500$ GeV)&Sig($1$ TeV)\\
    &&&&&&&&&Case-I& Case-I&Case-II&Case-II\\
    &&&&&&&&&(fb)&(fb)&(fb)&(fb)\\
   \hline
  C-I & 0.0344&0.674&0.0084&6.461&0.0274&0.28&0.015&0.114&0.0016&0.00134&0.051&0.028\\
      & $[100\%]$ & $[100\%]$ & $[100\%]$ & $[100\%]$ & $[100\%]$ & $[100\%]$ & $[100\%]$ & $[100\%]$ & $[100\%]$ & $[100\%]$ & $[100\%]$ & $[100\%]$\\
  \hline
 C-II &0.0015&0.0464&0.000302&0.464&0.003&0.024&0.005&0.0064&0.00142&0.0013&0.047&0.027\\
     & $[4.36\%]$ & $[6.88\%]$ & $[3.60\%]$ & $[7.18\%]$ & $[10.95\%]$ & $[8.57\%]$ & $[33.30\%]$ & $[5.61\%]$ & $[88.75\%]$ & $[97.01\%]$ & $[92.16\%]$ & $[96.43\%]$\\
  \hline
  C-III&0.00025&0.01&0.00013&0.0323&0.00035&0.003&0.002&0.003&0.0008&0.0009&0.0254&0.019\\
      & $[0.73\%]$ & $[1.48\%]$ & $[1.55\%]$ & $[0.50\%]$ & $[1.277\%]$ & $[1.07\%]$ & $[13.30\%]$ & $[2.63\%]$ & $[50\%]$ & $[67.16\%]$ & $[49.80\%]$ & $[67.85\%]$\\
   \hline
   $e^{\pm} e^{\pm} e^{\mp}/\mu^{\mp}$&$t\bar{t}W^{\pm}$&$t\bar{t}Z$ & $W^{\pm}h j$ & $W^{\pm}Zj$ & $W^{\pm}W^{\pm}W^{\mp}j$ & $ZZj$&$ZZW^{\pm}$&$W^{\pm}W^{\mp}Z$&Sig($500$ GeV)&Sig($1$ TeV)&Sig($500$ GeV)&Sig($1$ TeV)\\
    &&&&&&&&&Case-I& Case-I&Case-II&Case-II\\
    &&&&&&&&&(fb)&(fb)&(fb)&(fb)\\
   \hline 
    \hline
C-I  &0.0253&0.5&0.0036&4.75&0.02&0.207&0.011&0.09&0.0012&0.001&0.0374&0.013\\
     & $[100\%]$ & $[100\%]$ & $[100\%]$ & $[100\%]$ & $[100\%]$ & $[100\%]$ & $[100\%]$ & $[100\%]$ & $[100\%]$ & $[100\%]$ & $[100\%]$ & $[100\%]$\\
  \hline
C-II  &0.0011&0.034&0.00022&0.333&0.00217&0.0174&0.004&0.005&0.001&0.00096&0.0345&0.012\\
      & $[4.35\%]$ & $[6.80\%]$ & $[6.10\%]$ & $[7.01\%]$ & $[10.85\%]$ & $[8.40\%]$ & $[36.36\%]$ & $[5.50\%]$ & $[83.30\%]$ & $[96\%]$ & $[92.24\%]$ & $[92.31\%]$\\
  \hline
 C-III &0.0002&0.007&0.0001&0.024&0.00027&0.0021&0.00161&0.0022&0.0006&0.00063&0.0183&0.011\\
       & $[0.79\%]$ & $[1.40\%]$ & $[2.70\%]$ & $[0.50\%]$ & $[1.35\%]$ & $[1.01\%]$ & $[14.63\%]$ & $[2.40\%]$ & $[50\%]$ & $[63\%]$ & $[48.93\%]$ & $[84.61\%]$\\
   \hline
      \end{tabular}
 \end{center}
\caption{Cut flow for trilepton plus fat-jet signal with missing energy. Upper block represents the muon and lower block represents the electron signals. We consider $M_{Z^\prime}=3$ TeV at $\sqrt{s}=14$ TeV.}
\label{tabI-3l-2}
\end{table}
Solving the signal and background relation for a particular significance applied in Sec.~\ref{ssdl-1}, we estimate the luminosity to achieve that significance. 
We find that the significance of the two benchmarks in Case-I are very close and roughly below $1-\sigma$. In Case-II, $5-\sigma$ significance can be attained at $3$ ab$^{-1}$ luminosity for $M_N=500$ GeV and for $M_N=1$ TeV $5-\sigma$ significance could be attained at $1.5$ ab$^{-1}$. 
\begin{table}[h]
\begin{center}
\tiny
  \begin{tabular}{ |c|c|c|c|c|c|c|c|c|c|c| } 
 \hline
   $\mu^\pm\mu^\pm\mu^\mp/e^\mp$ &$t\bar{t}W^{\pm}$&$t\bar{t}Z$ & $W^{\pm}h j$ & $W^{\pm}Zj$ & $W^{\pm}W^{\pm}W^{\mp}j$ & $ZZj$&$ZZW^{\pm}$&$W^{\pm}W^{\mp}Z$&Sig($1$ TeV)&Sig($1$ TeV)\\
    &&&&&&&&&Case-I&Case-II\\
    &&&&&&&&&(fb)&(fb)\\
   \hline
   \hline
  C-I&0.107&3.82&0.0171&22.94&0.112&0.931&0.042&0.207&0.012&0.252\\
     & $[100\%]$ & $[100\%]$ & $[100\%]$ & $[100\%]$ & $[100\%]$ & $[100\%]$ & $[100\%]$ & $[100\%]$ & $[100\%]$ & $[100\%]$\\
  \hline
  C-II&0.007&0.634&0.0008&3.35&0.0203&0.11&0.006&0.044&0.011&0.24\\
      & $[6.54\%]$ & $[16.60\%]$ & $[4.68\%]$ & $[14.60\%]$ & $[18.12\%]$ & $[11.81\%]$ & $[14.28\%]$ & $[21.26\%]$ & $[91.60\%]$ & $[95.24\%]$ \\
  \hline
  C-III&0.005&0.133&0.0004&0.455&0.0014&0.01&0.0023&0.02&0.008&0.17\\
       & $[4.67\%]$ & $[3.48\%]$ & $[2.34\%]$ & $[1.98\%]$ & $[1.25\%]$ & $[1.07\%]$ & $[5.48\%]$ & $[9.66\%]$ & $[66.60\%]$ & $[67.46\%]$\\
   \hline
   $e^\pm e^\pm e^\mp/\mu^\mp$ &$t\bar{t}W^{\pm}$&$t\bar{t}Z$ & $W^{\pm}h j$ & $W^{\pm}Zj$ & $W^{\pm}W^{\pm}W^{\mp}j$ & $ZZj$&$ZZW^{\pm}$&$W^{\pm}W^{\mp}Z$&Sig($1$ TeV)&Sig($1$ TeV)\\
    &&&&&&&&&Case-I&Case-II\\
    &&&&&&&&&(fb)&(fb)\\
     \hline
     \hline
  C-I&0.08&2.81&0.0125&16.87&0.083&0.685&0.024&0.153&0.008&0.18\\
     & $[100\%]$ & $[100\%]$ & $[100\%]$ & $[100\%]$ & $[100\%]$ & $[100\%]$ & $[100\%]$ & $[100\%]$ & $[100\%]$ & $[100\%]$\\
  \hline
  C-II&0.0054&0.466&0.00057&2.5&0.016&0.0984&0.0042&0.025&0.007&0.15\\
      & $[6.75\%]$ & $[16.58\%]$ & $[4.56\%]$ & $[14.82\%]$ & $[19.28\%]$ & $[14.36\%]$ & $[17.50\%]$ & $[16.34\%]$ & $[87.50\%]$ & $[83.30\%]$\\
  \hline
  C-III&0.00035&0.103&0.0003&0.34&0.001&0.008&0.0017&0.0072&0.0062&0.12\\
       & $[0.44\%]$ & $[3.66\%]$ & $[2.40\%]$ & $[2.01\%]$ & $[1.20\%]$ & $[1.17\%]$ & $[7.08\%]$ & $[4.70\%]$ & $[77.50\%]$ & $[66.60\%]$\\
 \hline
 \end{tabular}
 \end{center}
\caption{Cut flow for trilepton plus fat-jet signal with missing energy. Upper block represents the muon and lower block represents the electron signals. We consider $M_{Z^\prime}=3$ TeV at $\sqrt{s}= 27$ TeV.}
\label{tabI-3l-1}
\end{table}
The cut flow and the efficiencies for $\mu^\pm \mu^\pm \ell^\mp$ signal and backgrounds at $\sqrt{s}=27$ TeV are shown in the upper panel and those for the $e^\pm e^\pm \ell^\mp$ events are shown in the lower panel of Tab.~\ref{tabI-3l-1}. We study 1 TeV RHN as a benchmark for Cases-I and II. The leading background is originated from $WZ+$jets processes, whereas the sub-leading contribution to the background comes from $WWZ$ process, respectively. The other backgrounds are 2-3 orders of magnitude smaller than the leading background. We find that significance is below $2-\sigma$ up to a high luminosity for Case-I, however, that for Case-II can be $5-\sigma$ at $200$ fb$^{-1}$ luminosity.
\begin{table}[h]
\begin{center}
\tiny
  \begin{tabular}{ |c|c|c|c|c|c|c|c|c|c|c| } 
 \hline
   $ \mu^{\pm} \mu^{\pm} \mu^{\mp}/e^{\mp}$&$t\bar{t}W^{\pm}$&$t\bar{t}Z$ & $W^{\pm}h j$ & $W^{\pm}Zj$ & $W^{\pm}W^{\pm}W^{\mp}j$ & $ZZj$&$ZZW^{\pm}$&$W^{\pm}W^{\mp}Z$&Sig($1$ TeV)&Sig($1$ TeV)\\
    &&&&&&&&&Case-I& Case-II\\
    &&&&&&&&&(fb)&(fb)\\
   \hline
C-I&0.455&44.7&0.0811&133.9&0.771&6.854&0.14&0.746&0.124&2.54\\
   & $[100\%]$ & $[100\%]$ & $[100\%]$ & $[100\%]$ & $[100\%]$ & $[100\%]$ & $[100\%]$ & $[100\%]$ & $[100\%]$ & $[100\%]$\\
  \hline
  C-II&0.0083&1.12&0.00302&7.3&0.0622&0.338&0.11&0.0502&0.113&2.38\\
      & $[1.82\%]$ & $[2.50\%]$ & $[3.72\%]$ & $[5.45\%]$ & $[8.07\%]$ & $[4.93\%]$ & $[78.57\%]$ & $[6.73\%]$ & $[91.13\%]$ & $[93.70\%]$\\
  \hline
C-III&0.00151&0.272&0.0064&0.66&0.0057&0.09&0.034&0.0149&0.065&1.36\\
     & $[0.33\%]$ & $[0.61\%]$ & $[7.90\%]$ & $[0.49\%]$ & $[0.74\%]$ & $[1.31\%]$ & $[24.28\%]$ & $[1.99\%]$ & $[52.42\%]$ & $[53.54\%]$\\
  \hline
   $e^{\pm} e^{\pm} e^{\mp}/\mu^{\mp}$&$t\bar{t}W^{\pm}$&$t\bar{t}Z$ & $W^{\pm}h j$ & $W^{\pm}Zj$ & $W^{\pm}W^{\pm}W^{\mp}j$ & $ZZj$&$ZZW^{\pm}$&$W^{\pm}W^{\mp}Z$&Sig($1$ TeV)&Sig($1$ TeV)\\
    &&&&&&&&&Case-I& Case-II\\
    &&&&&&&&&(fb)&(fb)\\
   \hline 
    \hline
C-I&0.3344&32.84&0.06&98.42&0.57&5.035&0.101&0.525&0.106&1.83\\
   & $[100\%]$ & $[100\%]$ & $[100\%]$ & $[100\%]$ & $[100\%]$ & $[100\%]$ & $[100\%]$ & $[100\%]$ & $[100\%]$ & $[100\%]$\\
  \hline
  C-II&0.0034&0.82&0.0022&5.343&0.046&0.772&0.01&0.084&0.098&1.74\\
      & $[1.02\%]$ & $[2.50\%]$ & $[3.60\%]$ & $[5.43\%]$ & $[8.07\%]$ & $[15.33\%]$ & $[9.90\%]$ & $[16\%]$ & $[92.45\%]$ & $[95.08\%]$\\
  \hline
  C-III&0.002&0.12&0.0017&0.483&0.005&0.076&0.0021&0.025&0.060&0.93\\
       & $[0.60\%]$ & $[0.36\%]$ & $[2.83\%]$ & $[0.49\%]$ & $[0.88\%]$ & $[1.51\%]$ & $[2.08\%]$ & $[4.76\%]$ & $[56.60\%]$ & $[50.82\%]$\\
   \hline
   \end{tabular}
 \end{center}
\caption{Cut flow for trilepton plus fat-jet signal with missing energy. Upper block represents the muon and lower block represents the electron signals. We consider $M_{Z^\prime}=3$ TeV at $\sqrt{s}=100$ TeV.}
\label{tabI-3l-2-100}
\end{table}
The cut flow and the efficiencies for $\mu^\pm \mu^\pm \ell^\mp$ and $e^\pm e^\pm \ell^\mp \ell^\mp$ events at $\sqrt{s}= 100$ TeV are shown in the upper and lower panels of Tab.~\ref{tabI-3l-2-100}. We notice that a $5-\sigma$ significance can be attained at $2$ ab$^{-1}$ in Case-I whereas the same can be achieved at the $15$ fb$^{-1}$ luminosity in Case-II. 
\subsection{Bounds on $M_N-M_{Z^\prime}$ plane}
\label{Sec4}
\begin{figure}[h]
\centering
\includegraphics[width=0.495\textwidth,angle=0]{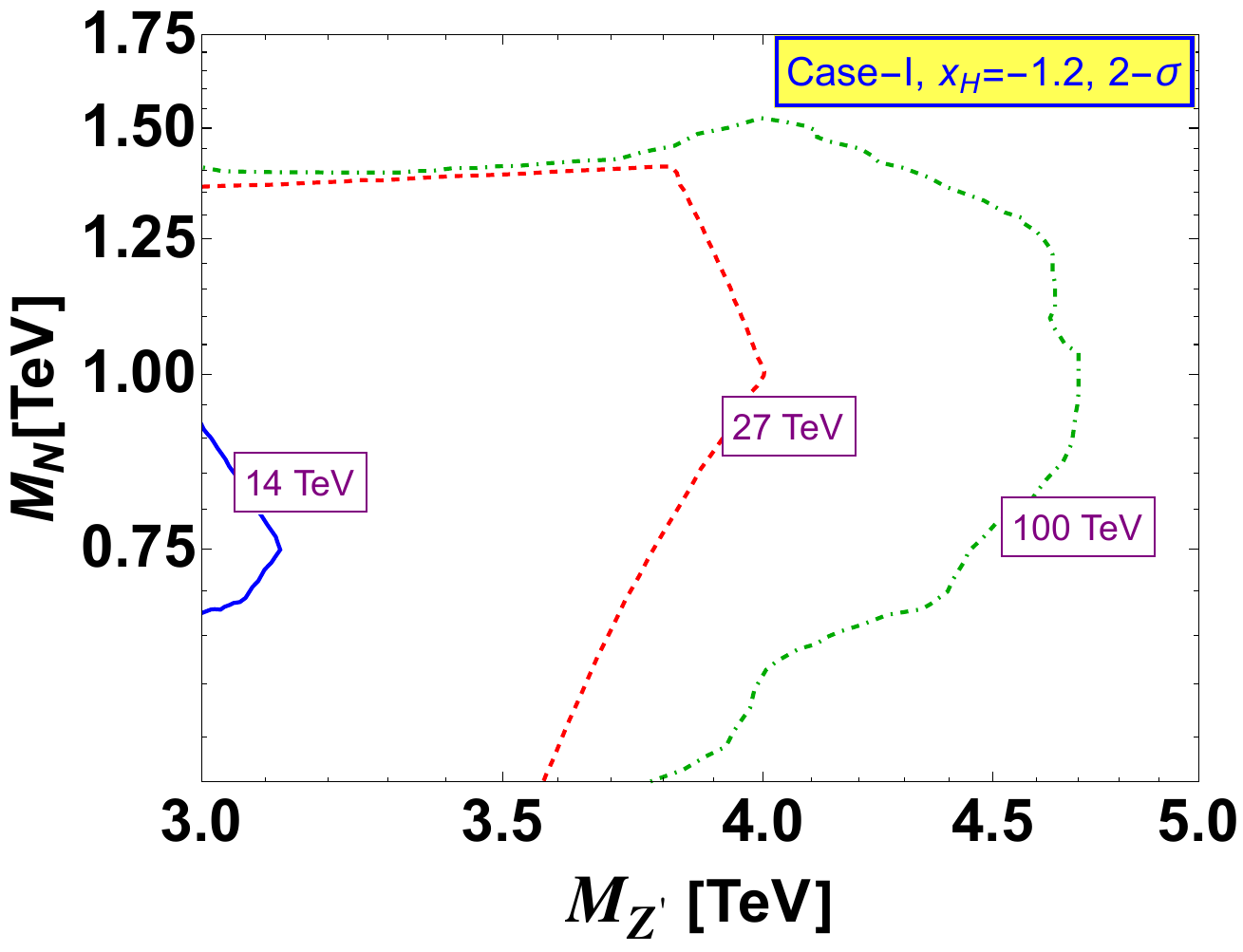}
\includegraphics[width=0.47\textwidth,angle=0]{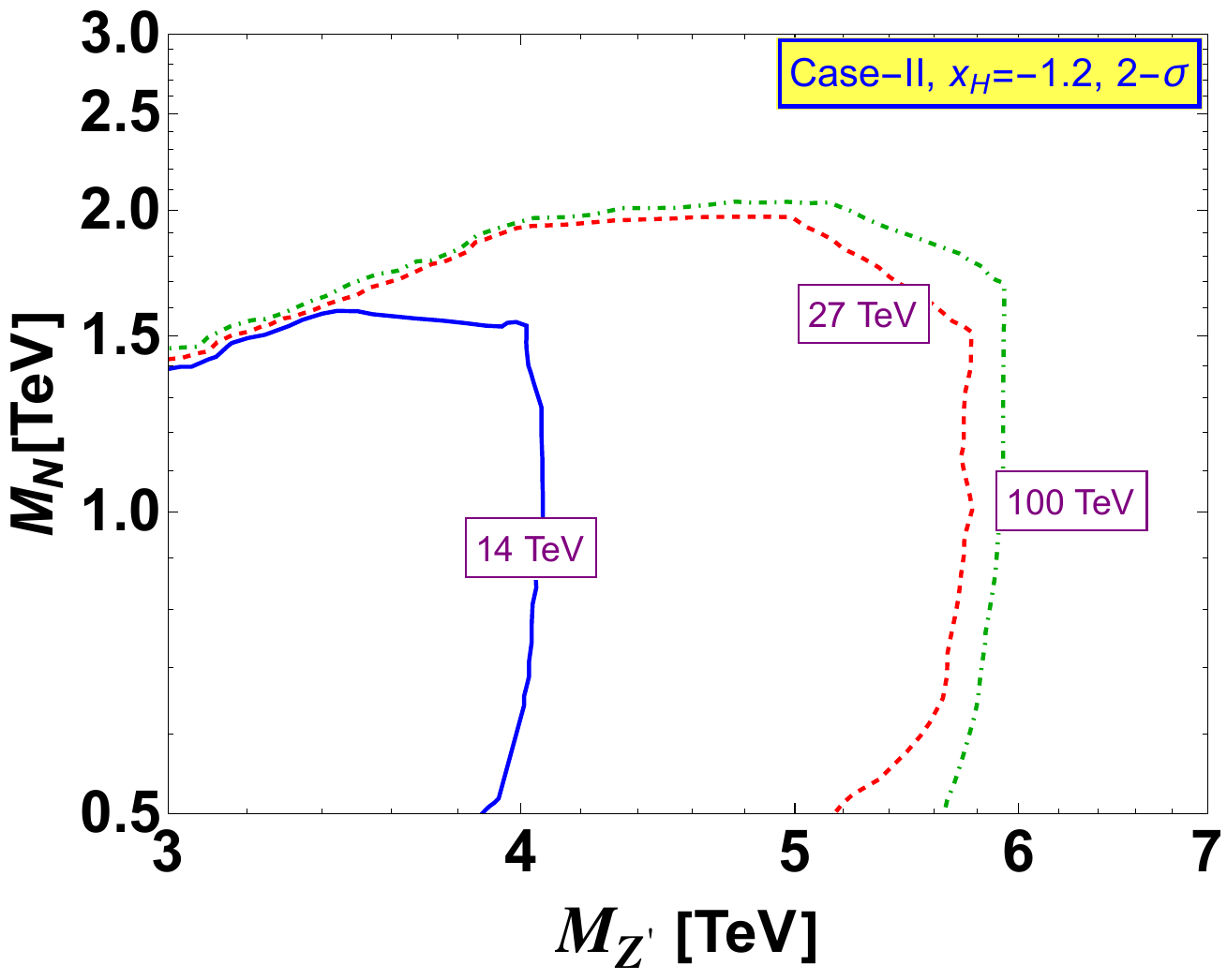}
\caption{$2-\sigma$ exclusion at different hadron colliders on $M_N-M_{Z^\prime}$ plane from SSDL+2$J$ signature. We consider $3$ ab$^{-1}$ luminosity for $14$ TeV and $27$ TeV colliders and $30$ fb$^{-1}$ luminosity for $100$ TeV respectively.}
\label{MN-MZp-x}
\end{figure}
Applying the constraints on the $g^\prime-M_{Z^\prime}$ plane we produce the RHN pair from $Z^\prime$ using Cases-I and II.
We study the boosted objects from the RHNs. Considering the SSDL final state manifesting the Majorana nature of the 
RHNs we perform a scan over $3$ TeV $\leq M_{Z^\prime}< 7$ TeV considering $500$ GeV $\leq M_N < \frac{M_{Z^\prime}}{2}$.
We consider the leading decay mode $\ell W$ from each RHN and the heavy mass will boost each $W$ boson from the RHN to produce fat-jet. 
Hence we study the SSDL+ 2 J signature at hadron colliders with $\sqrt{s}=14$ TeV, $27$ TeV and $100$ TeV respectively. 
Following the selection and advanced cuts described in Sec.~\ref{ssdl-1}, we produce the signal and study the backgrounds to produce a $2-\sigma$ exclusion limit in the $M_N-M_{Z^\prime}$ plane. The contours are produced using $3$ ab$^{-1}$ luminosity for $14$ TeV and $27$ TeV colliders and $30$ fb$^{-1}$ luminosity for $100$ TeV collider respectively. These contours for Case-I (II) are shown in the left (right) panel of Fig.~\ref{MN-MZp-x}. Hence we infer SSDL plus two fat-jets provides an interesting handle to study the RHN pair production from $Z^\prime$. In this context, we mention that trilepton plus single fat-jet signature is another interesting aspect that could be tested from the RHN pair production, however, to test this channel we require a high luminosity at $27$ TeV and $100$ TeV colliders. In that case, we could probe a relatively smaller parameter space compared to the SSDL scenario in the proton colliders. 
\section{Heavy neutrino pair production at the $e^+e^-$ collider}
\label{RHNemep}
\begin{figure}
\includegraphics[scale=0.18]{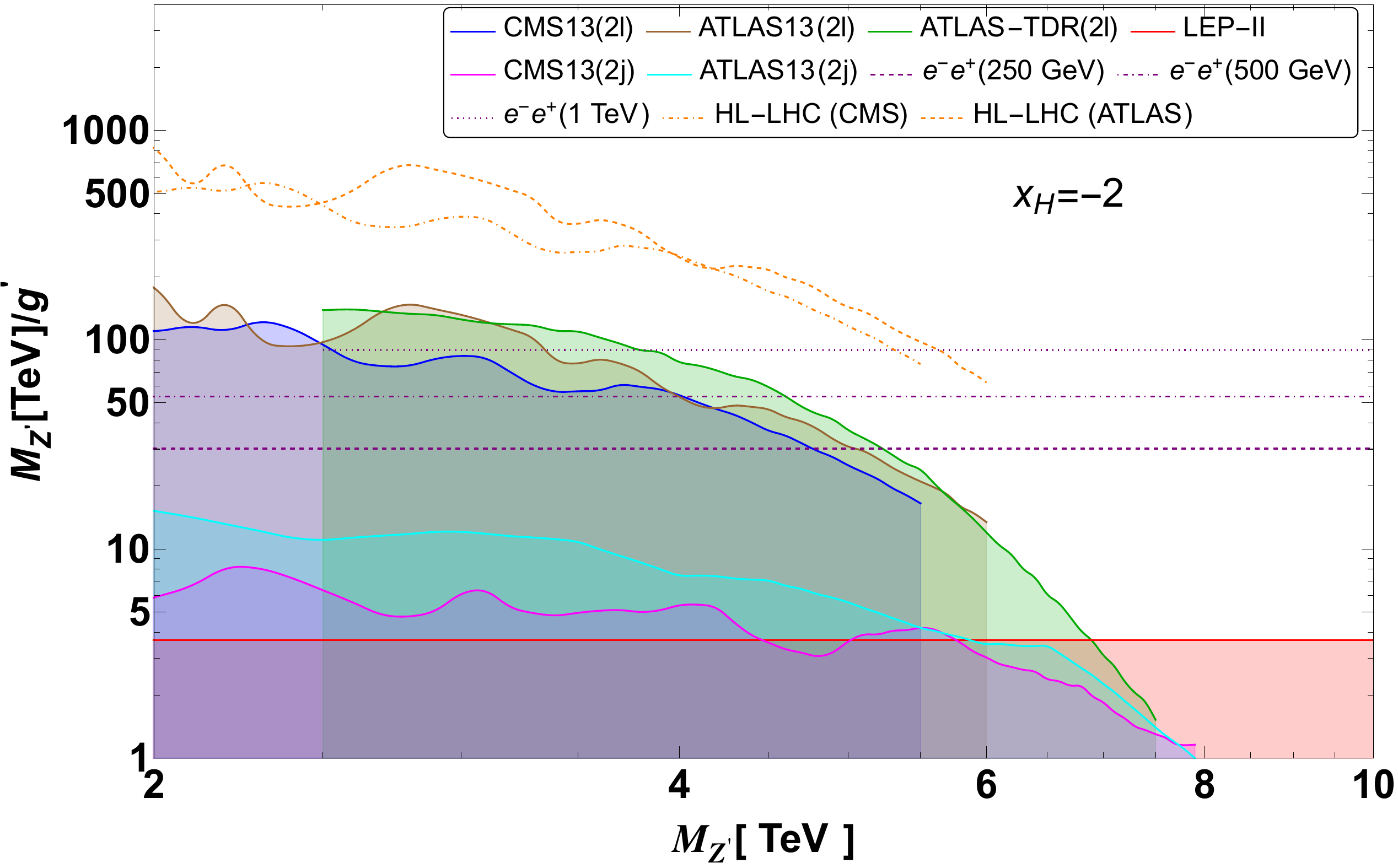} 
\includegraphics[scale=0.18]{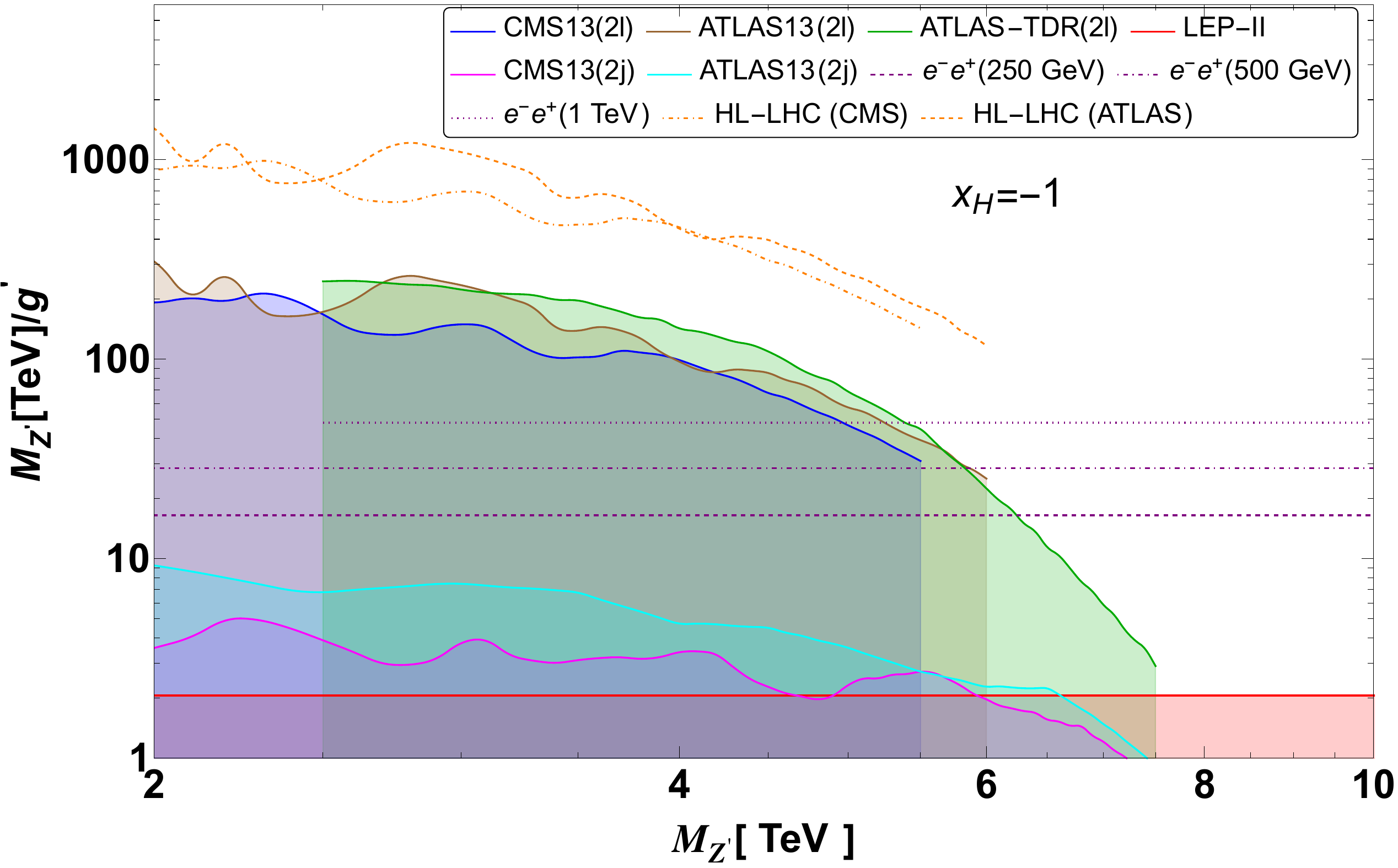}
\includegraphics[scale=0.18]{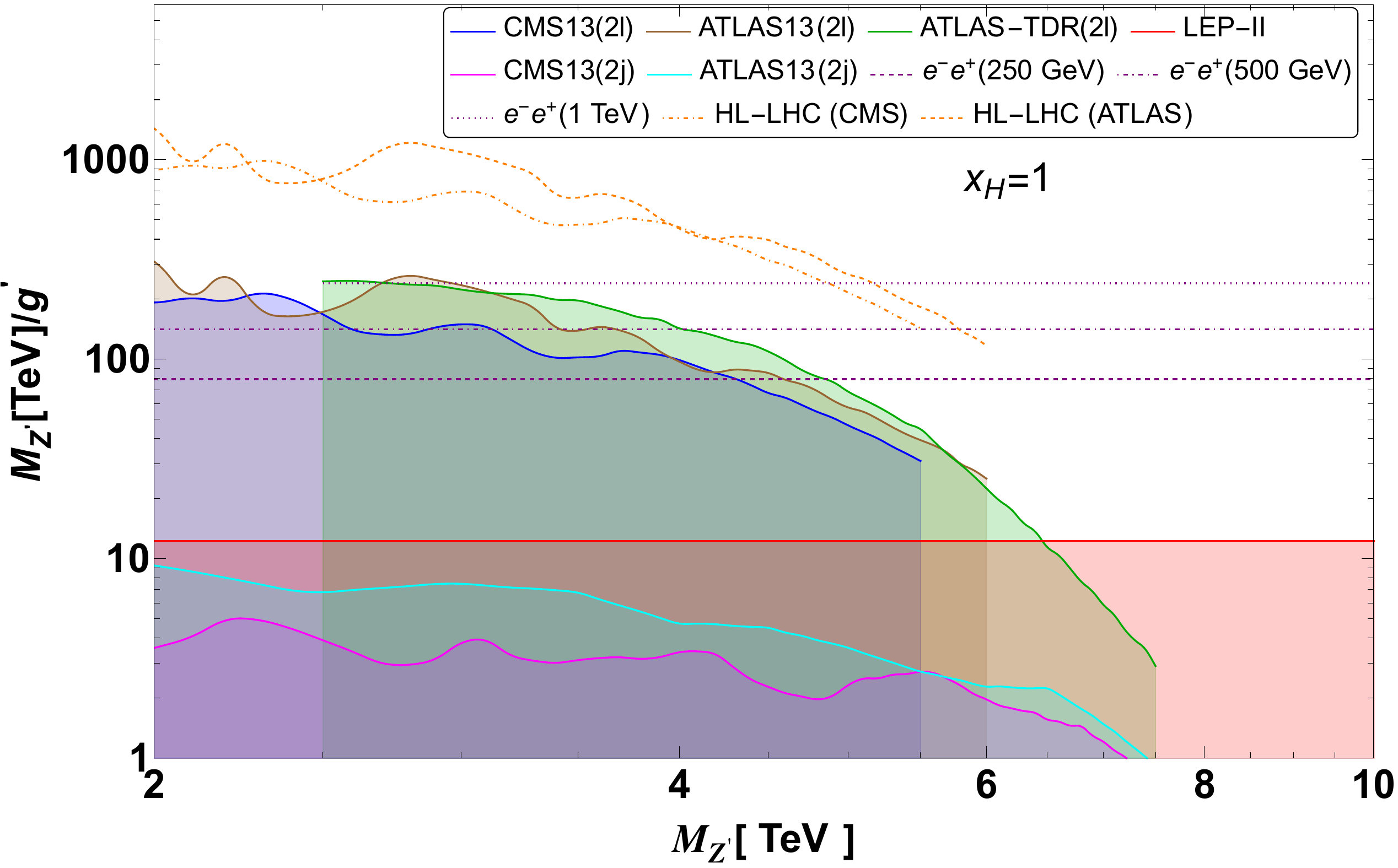}
\caption{Lower limits on $M_{Z^\prime}/g^\prime$ as a function of $M_{Z^\prime}$ for $x_H=-2$ (top, left), $-1$ (top, right) and $1$ (bottom) respectively from dilepton searches at ATLAS \cite{Aad:2019fac}, CMS \cite{CMS:2021ctt} and ATLAS-TDR \cite{CERN-LHCC-2017-018}, dijet searches from ATLAS \cite{ATLAS:2019bov} and CMS \cite{Sirunyan:2018xlo} respectively. Considering $M_{Z^\prime} >> \sqrt{s}$ we estimate the limits on $M_{Z^\prime}/g^\prime$ as a function of $M_{Z^\prime}$ from LEP-II \cite{LEP:2003aa} and prospective ILC \cite{LCCPhysicsWorkingGroup:2019fvj}. The shaded regions are excluded by respective experiments.}
\label{XXX}
\end{figure}

Pair production of heavy neutrinos at the electron positron collider is another important aspect of this model which can be studied at $\sqrt{s}=250$ GeV, $500$ GeV and $1$ TeV for Cases-I and II. We calculate the heavy neutrino pair production at $e^-e^+$ collider in terms of $M_{Z^\prime}$, $N_N$ and $x_H$ as
\bea
\sigma(e^-e^+ \to {Z^\prime}^\ast \to N_i N_i) = \frac{{g^\prime}^4 Q_N^2 s}{192 \pi}\frac{(8+ 12 x_H + 5 x_H^2)}{(s-M_{Z^\prime}^2)^2+ M_{Z^\prime}^2 \Gamma_{Z^\prime}^2} \Big(1-4\frac{M_N^2}{s}\Big)^{\frac{3}{2}}
\label{RHN-ee}
\eea
where $Q_N$ is the $U(1)_X$ charge of the heavy neutrino under the $U(1)_X$ gauge group and $\Gamma_{Z^\prime}$ is the total decay width of $Z^\prime$ in Cases-I and II respectively. In the limit of $M_{Z^\prime} > \sqrt{s}$ the pair production of the heavy neutrinos from Eq.~\ref{RHN-ee} reduces to 
\bea
\sigma(e^-e^+ \to {Z^\prime}^\ast \to N_i N_i) \simeq \Big(\frac{{g^\prime}^4}{{M_{Z^\prime}}^4}\Big) \frac{Q_N^2 s(8+ 12 x_H + 5 x_H^2)}{192 \pi} \Big(1-4\frac{M_N^2}{s}\Big)^{\frac{3}{2}}
\label{RHN-ee2}
\eea
We note that the cross section is directly proportional to $\frac{g^{\prime}}{M_{Z^\prime}}$. Before estimating the cross sections at different $\sqrt{s}$ we calculate bounds on $g^\prime$ depending on $M_{Z^\prime}$ using the dilepton cross sections from ATLAS \cite{Aad:2019fac}, CMS \cite{CMS:2021ctt} and ATLAS technical design report \cite{CERN-LHCC-2017-018} and dijet cross sections from ATLAS \cite{ATLAS:2019bov} and CMS \cite{Sirunyan:2018xlo} using Eq.~\ref{gp} considering $M_N > \frac{M_{Z^\prime}}{2}$ which exerts strongest limit on the $U(1)_X$ coupling. The constraints from LEP \cite{LEP:2003aa,Carena:2004xs,Schael:2013ita,Das:2021esm} are obtained using $M_{Z^\prime} > \sqrt{s}$. Hence we obtain the limits on the quantity $\frac{M_{Z^\prime}}{g^\prime}$ in terms of $M_{Z^\prime}$ for different $x_H$ which measures the VEV of the $U(1)_X$ theory. The $Z^\prime$ phenomenology at the $e^-e^+$ collider been studied in  \cite{Das:2021esm} where it has been shown that for $x_H=-2$ there is no interaction between the left handed fermion doublet and $Z^\prime$ which is $U(1)_R$ scenario. It can also be noted that for $x_H=-1$ the interaction between the right handed electron and $Z^\prime$ vanishes. Therefore we consider these two charges as they directly affect the interaction between electron and $Z^\prime$ for the heavy neutrino pair production. In addition to that we consider $x_H=1$ where left handed lepton doublet and right handed electron interact with $Z^\prime$, however, right handed down type quark does not have coupling with $Z^\prime$ and it has no direct impact in the heavy neutrino pair production process at the $e^-e^+$ collider. The bounds on $\frac{M_{Z^\prime}}{g^\prime}$ for $x_H=-2$, $-1$ and $1$ depending on $M_{Z^\prime}$ from LEP and the $e^-e^+$ colliders considering $M_{Z^\prime} > \sqrt{s}$ are shown as the parallel lines with respect to the horizontal axis in Fig.~\ref{XXX}. The shaded regions in this figure are excluded by dilepton and dijet searches from LHC and limits on the effective scale form LEP-II experiment respectively. These lower bounds show the scale of the VEV could be probed in these experiments. We estimate the bounds on $\frac{M_{Z^\prime}}{g^\prime}$ at the HL-LHC scaling the ATLAS $(\mathcal{L}_{\rm current}=139$ fb$^{-1})$ and CMS $(\mathcal{L}_{\rm current}=140$ fb$^{-1})$ results for $M_N > \frac{M_{Z^\prime}}{2}$ using $g^\prime \simeq g^\prime_{\rm current} \sqrt{\frac{\mathcal{L}_{\rm current}}{\mathcal{L}_{\rm future}}}$ where $\mathcal{L}_{\rm future}=3000$ fb$^{-1}$. These prospective bounds estimated by scaling the ATALS (CMS) bounds are shown by the orange dashed (dot-dashed) line in Fig.~\ref{XXX}.

\begin{figure}
\includegraphics[scale=0.35]{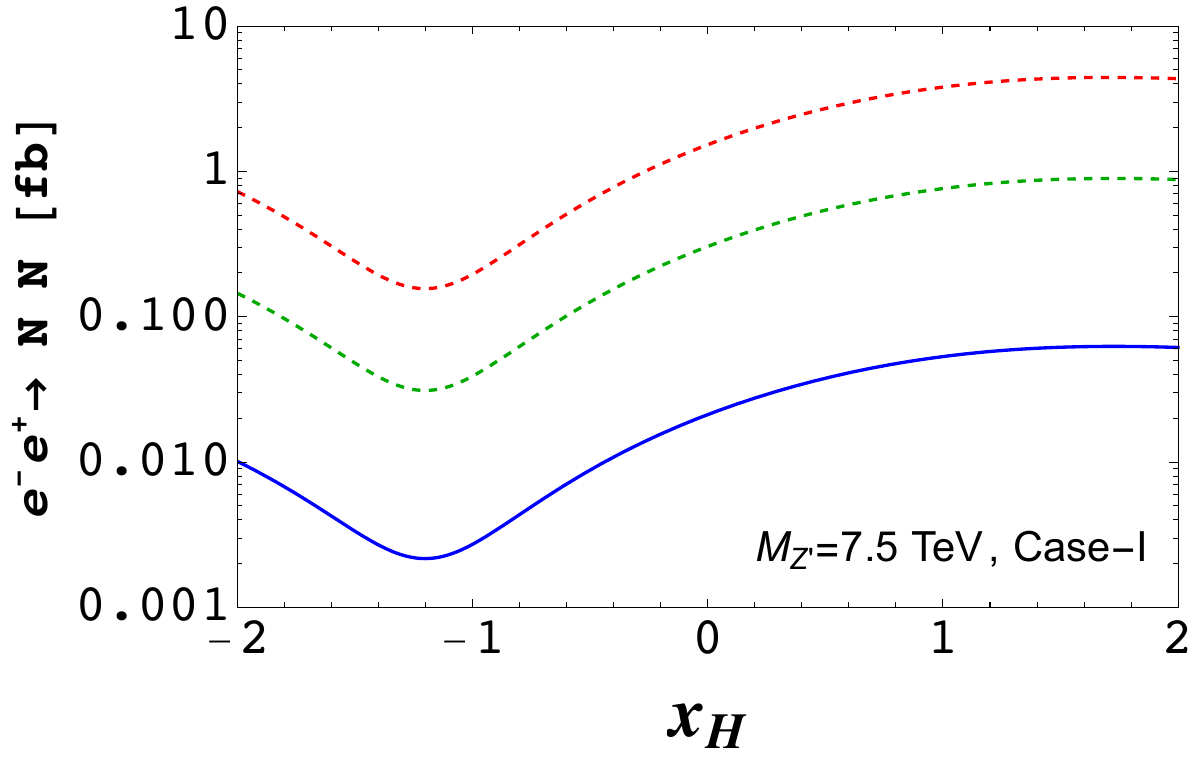} 
\includegraphics[scale=0.35]{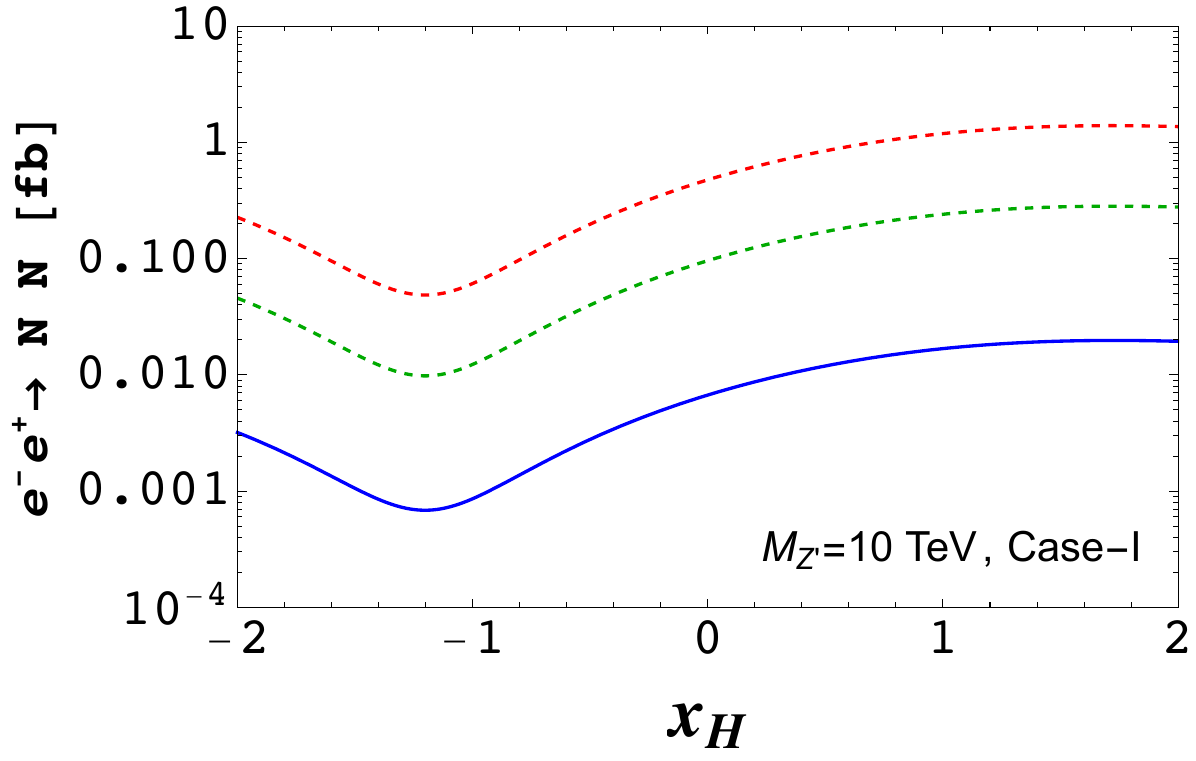}
\includegraphics[scale=0.35]{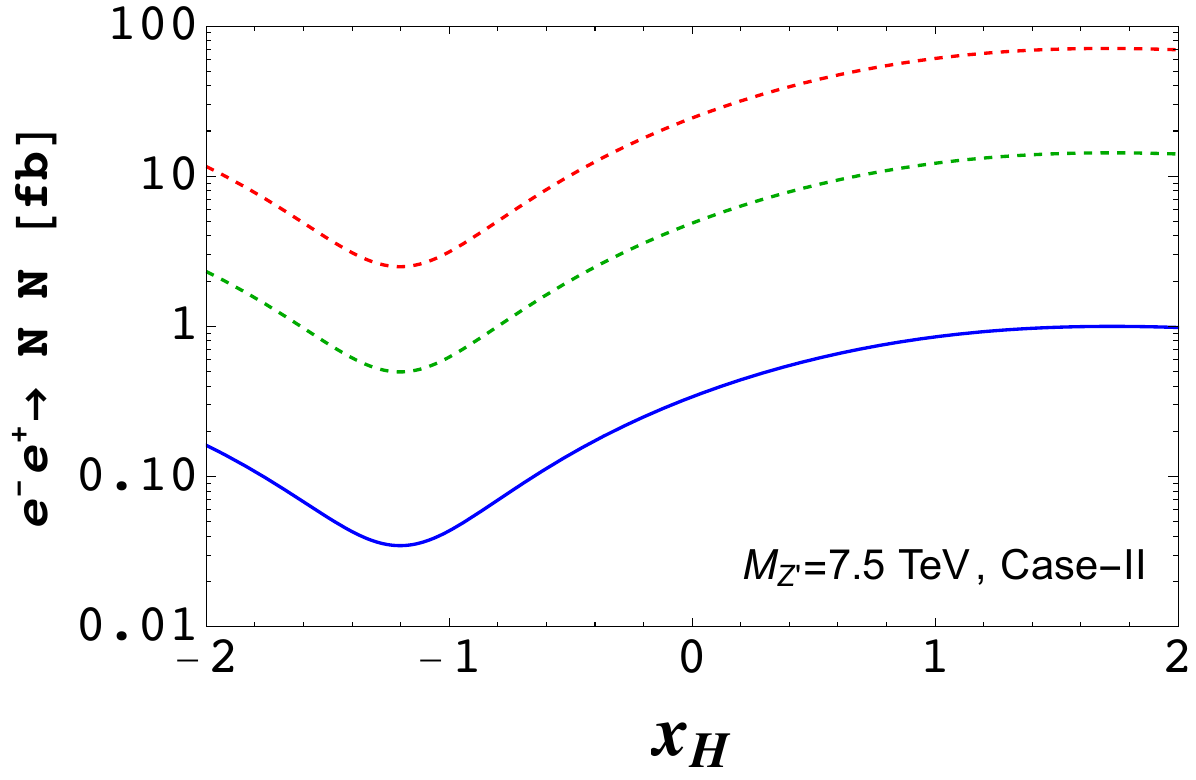} 
\includegraphics[scale=0.35]{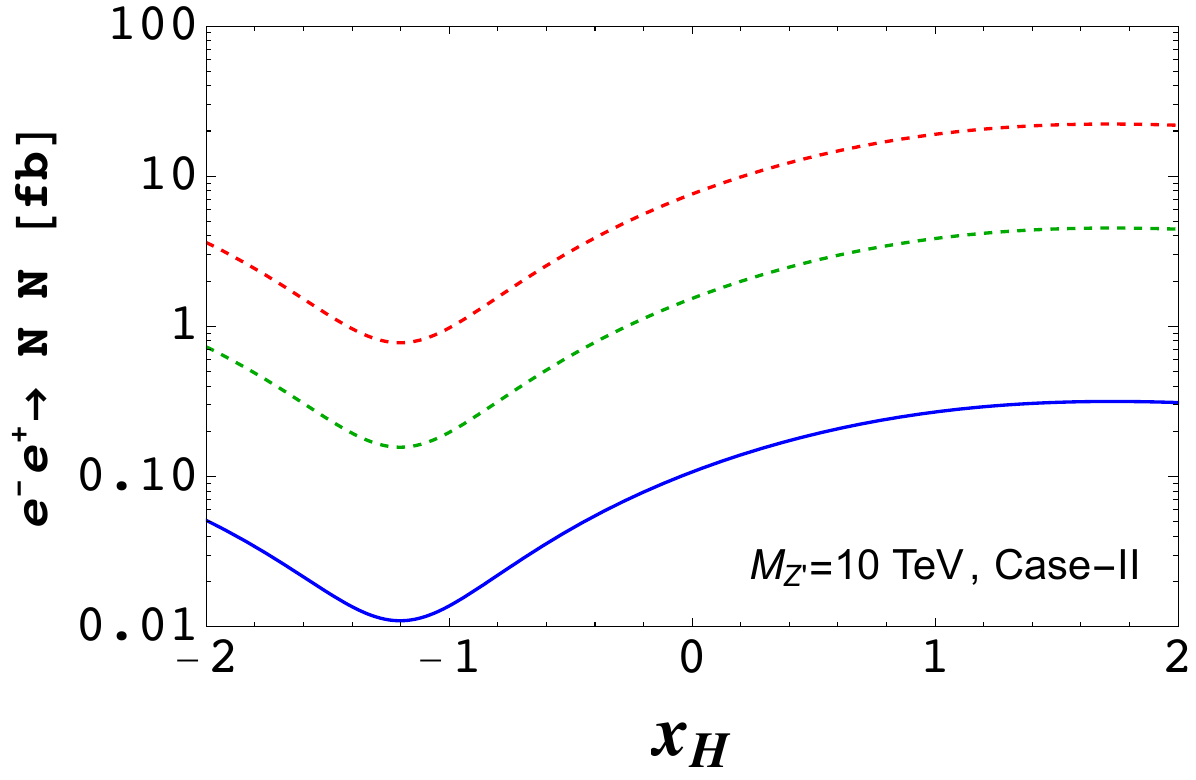}
\caption{Heavy neutrino pair production cross section normalized by the U$(1)_X$ gauge coupling at electron positron colliders for $M_{N}=100$ GeV fixing $M_{Z^\prime}=7.5$ TeV (left panel) and 10 TeV (right panel) depending on $x_H$ for Cases-I (upper panel) and II (lower panel). We consider three different center of mass energies $\sqrt{s}=250$ GeV, 500 GeV and 1 TeV from bottom to top respectively in each panel.}
\label{XXX1}
\end{figure}

We calculate the cross section for the heavy neutrino pair production being normalized by the $U(1)_X$ gauge coupling for $M_{Z^\prime}=7.5$ TeV (left panel) and 10 TeV (right panel) depending on $x_H$ for different $\sqrt{s}$ in Fig.~\ref{XXX1} fixing $M_N=100$ GeV. We find that the cross sections attain certain values at $x_H=-2$ and decreases with the increase in $x_H$ attaining a minimum at $x_H=-1.2$. Finally the cross section increases with the increase in $x_H$ and attains a maximum at $x_H =1$ and maintains a constant value with $x_H \geq 1$. We show the Case-I (II) in the upper (lower) panel of Fig.~\ref{XXX1} for $\sqrt{s}=250$ GeV, 500 GeV and 1 TeV in each panel from bottom to top respectively. In this calculation we consider only one generation of the heavy neutrino for simplicity. Production of more than one generation of heavy neutrino pair can be possible, however, due to the universal $U(1)_X$ gauge coupling of our model, the corresponding cross sections can be multiplied by the number of generations. 
\subsection{SSDL+2j signal}
\begin{figure}[h]
\centering
\includegraphics[width=0.32\textwidth,angle=0]{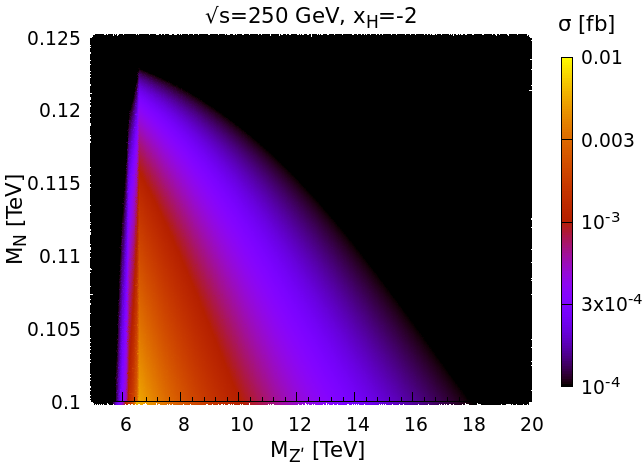}
\includegraphics[width=0.32\textwidth,angle=0]{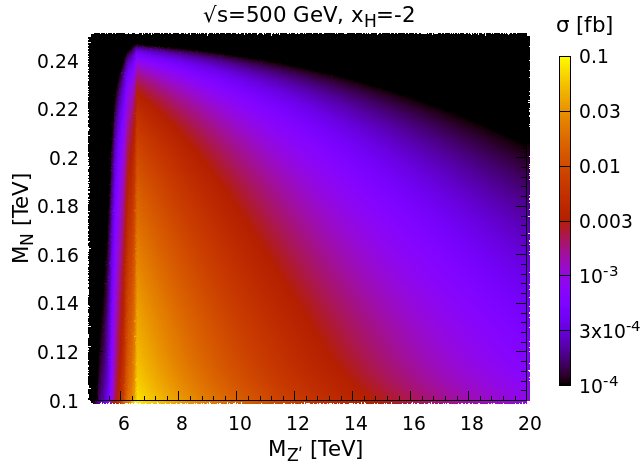}
\includegraphics[width=0.32\textwidth,angle=0]{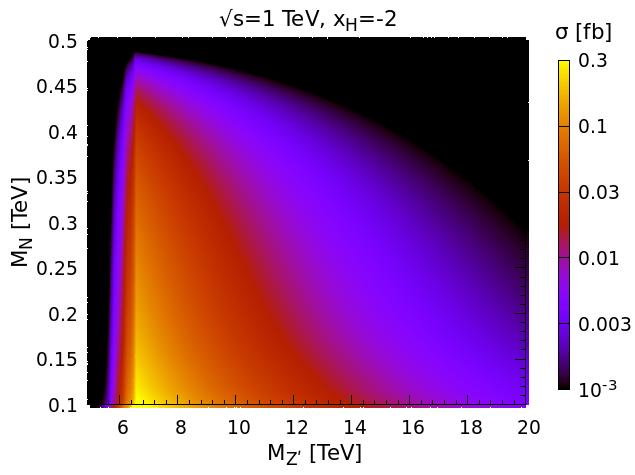}
\includegraphics[width=0.32\textwidth,angle=0]{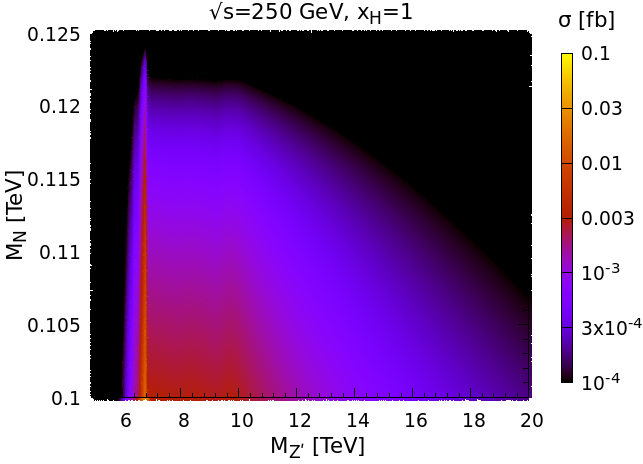}
\includegraphics[width=0.32\textwidth,angle=0]{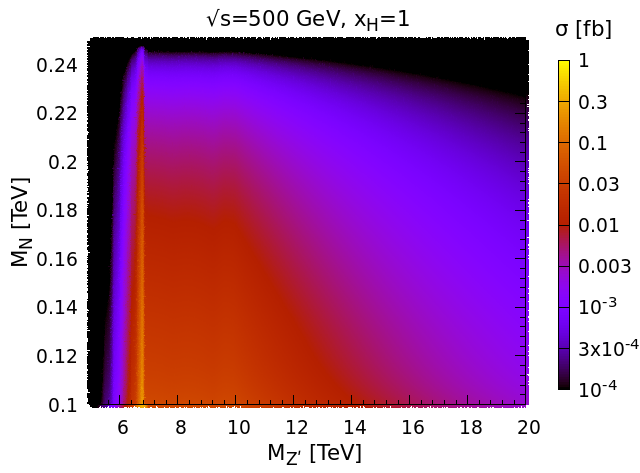}
\includegraphics[width=0.32\textwidth,angle=0]{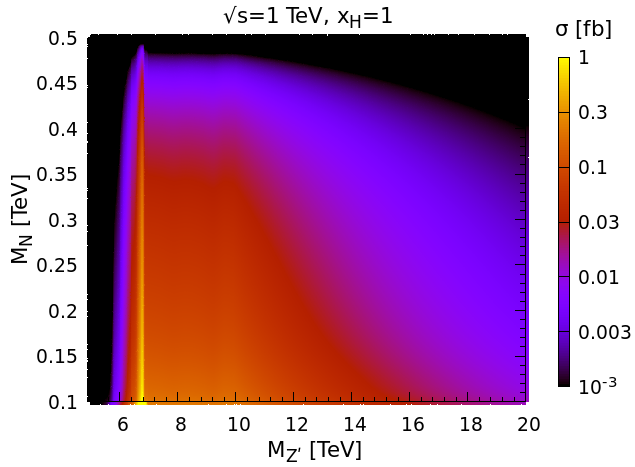}
\caption{The density plot representing dilepton plus four jet cross section on $M_N-M_{Z^\prime}$ plane for Case-I considering $x_H=-2$ (upper panel) and 1 (lower panel) at different center of mass energies such as 250 GeV, 500 GeV and 1TeV from left to right. The bar chart represents the cross sections in fb.}
\label{MN-MZp-ee-case1-dilepton}
\end{figure}

\begin{figure}[h]
\centering
\includegraphics[width=0.32\textwidth,angle=0]{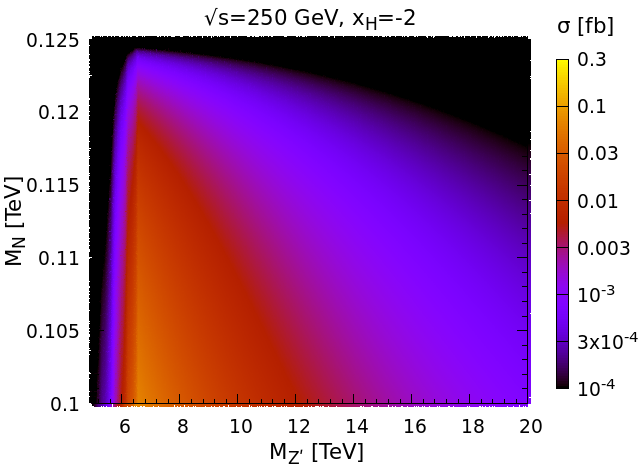}
\includegraphics[width=0.32\textwidth,angle=0]{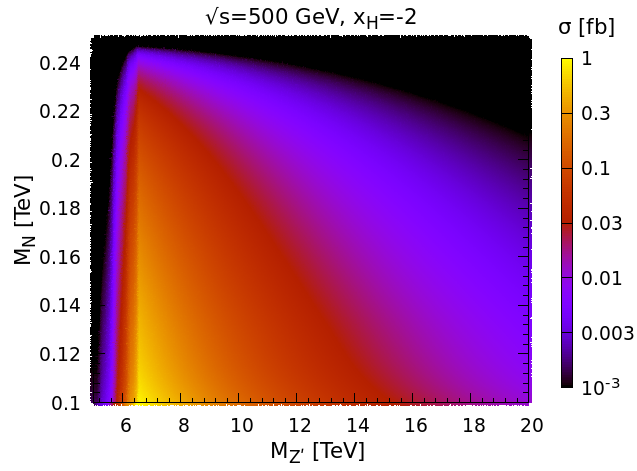}
\includegraphics[width=0.32\textwidth,angle=0]{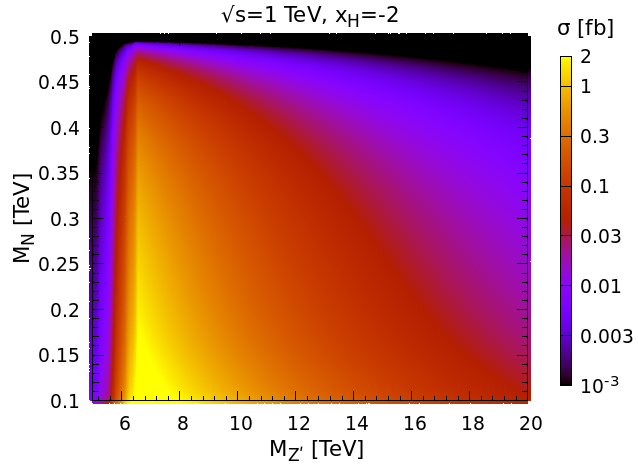}
\includegraphics[width=0.32\textwidth,angle=0]{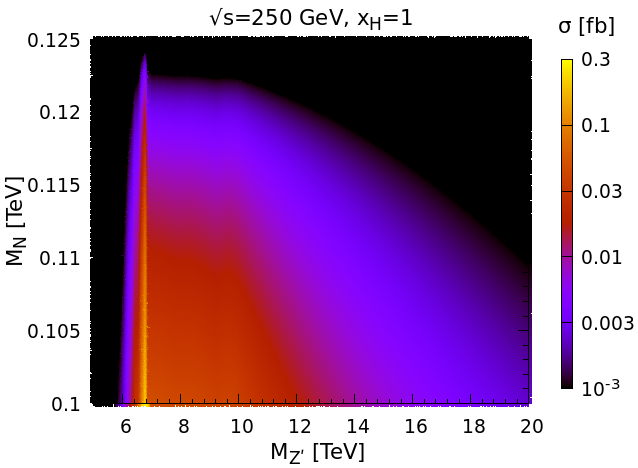}
\includegraphics[width=0.32\textwidth,angle=0]{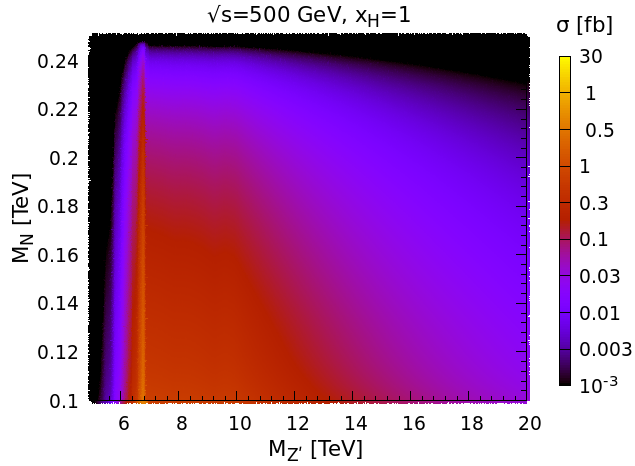}
\includegraphics[width=0.32\textwidth,angle=0]{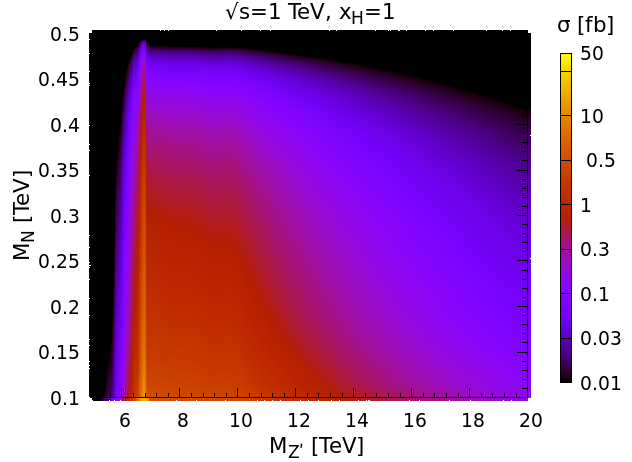}
\caption{The density plot representing dilepton plus four jet cross section on $M_N-M_{Z^\prime}$ plane for Case-II considering $x_H=-2$ (upper panel) and 1 (lower panel) at different center of mass energies such as 250 GeV, 500 GeV and 1TeV from left to right. The bar chart represents the cross sections in fb.}
\label{MN-MZp-ee-case2-dilepton}
\end{figure}

In electron positron colliders we can produce the heavy neutrinos in pair and each heavy neutrino decay into the leading mode following $N \to \ell W$.
In this analysis we consider SSDL plus four jet signal. 
The jets are coming from the hadronic decay of one of the $W$ bosons. 
To study this signal we consider $x_H=-2$ and 1 for $\sqrt{s}=250$ GeV, 500 GeV and 1 TeV respectively for $5$ TeV $\leq M_{Z^\prime} \leq 20$ TeV. 
From Fig.~\ref{XXX1} we find that heavy neutrino pair production cross section is small at $x_H=-1$ compared to $x_H=-2$ and 1. 
Therefore in further analyses we proceed with $x_H=-2$ and $1$.  
We considered $100$ GeV $\leq M_N \leq 125$ GeV for $\sqrt{s}=250$ GeV, $100$ GeV $\leq M_N \leq 250$ GeV for $\sqrt{s}=500$ GeV, and $100$ GeV $\leq M_N \leq 500$ GeV for $\sqrt{s}=1$ TeV respectively to produce the density plots for Cases-I and II in Figs.~\ref{MN-MZp-ee-case1-dilepton} and \ref{MN-MZp-ee-case2-dilepton} respectively. The bar chart in the right of each panel represents corresponding cross sections at different $\sqrt{s}$ for different $x_H$. The results in Case-II is higher than the Case-I due to different $U(1)_X$ charges of the heavy neutrinos. 
Due to the heavy neutrino pair production process, the cross section becomes negligibly small at the threshold $M_N \sim \frac{\sqrt{s}}{2}$. 
The cross section almost remains the same before the vicinity of this threshold beyond which the cross section sharply drops when $M_N \sim \frac{\sqrt{s}}{2}$. 
The density plots also resemble the nature of the production cross section depending on $x_H$ as shown in Fig.~\ref{XXX1}.
According to the choice of $x_H$ we obtain the maximum cross section at $x_H=1$. 
The cross section increases with the increase in $\sqrt{s}$. In this analysis we consider the strongest bounds on $g^\prime$ for different $M_{Z^\prime}$ at $x_H=-2$ and 1 respectively which can be easily estimated form Fig.~\ref{XXX}. The results for $x_H=-2$ (1) is shown in the upper (lower) panel of each figure for $\sqrt{s}=250$ GeV, $500$ GeV and 1 TeV from left to right. 
\subsection{3$\ell$+2j+missing energy signal}
\begin{figure}[h]
\centering
\includegraphics[width=0.32\textwidth,angle=0]{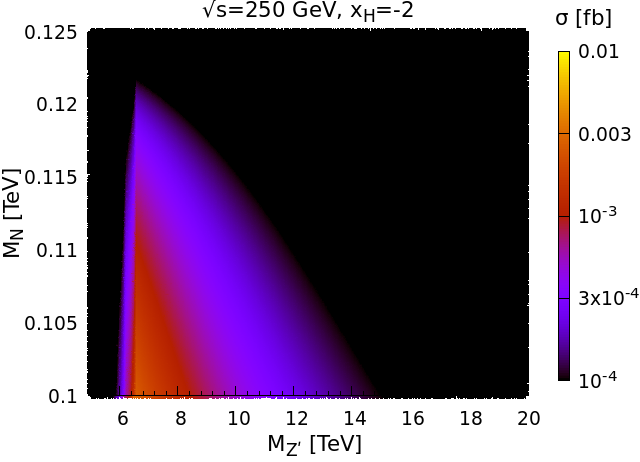}
\includegraphics[width=0.32\textwidth,angle=0]{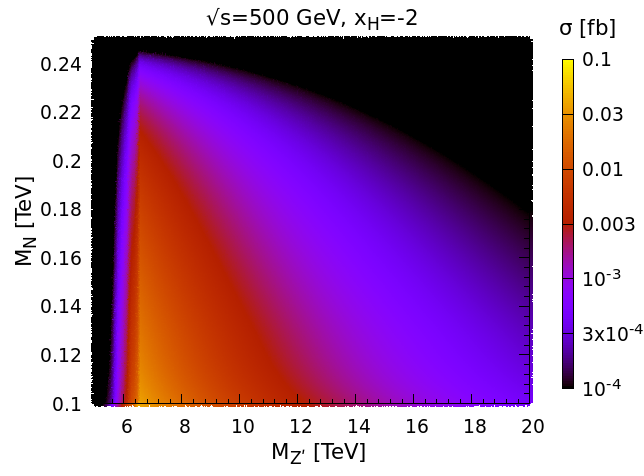}
\includegraphics[width=0.32\textwidth,angle=0]{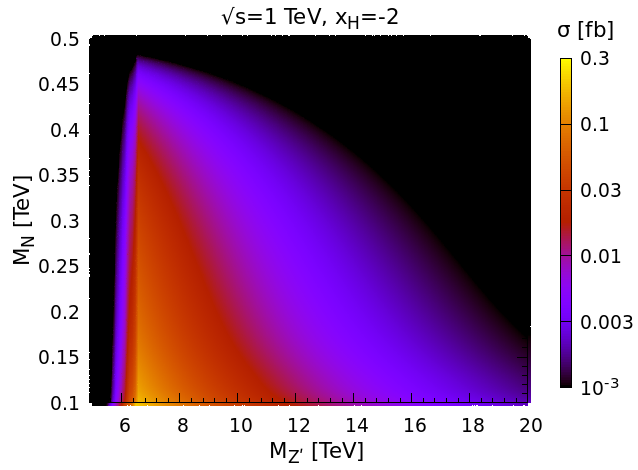}\\
\includegraphics[width=0.32\textwidth,angle=0]{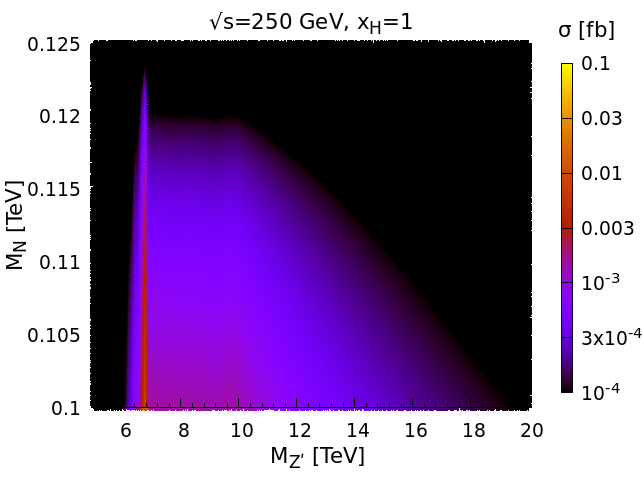}
\includegraphics[width=0.32\textwidth,angle=0]{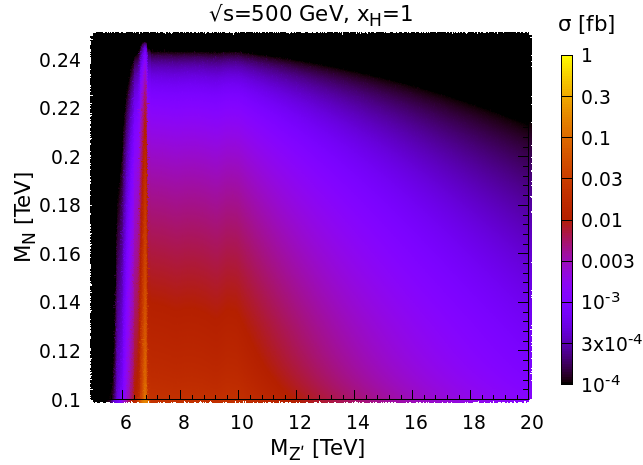}
\includegraphics[width=0.32\textwidth,angle=0]{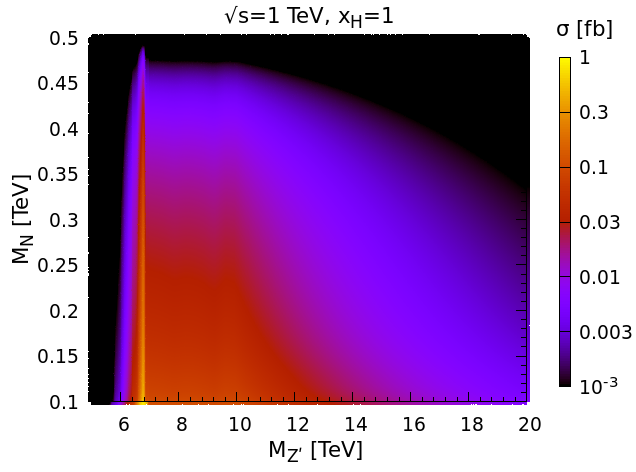}
\caption{The density plot representing trilepton plus two jet cross section in association with missing energy on $M_N-M_{Z^\prime}$ plane for Case-I considering $x_H=-2$ (upper panel) and 1 (lower panel) at different center of mass energies such as 250 GeV, 500 GeV and 1TeV from left to right. The bar chart represents the cross sections in fb.}
\label{MN-MZp-ee-case1-1}
\end{figure}

\begin{figure}[h]
\centering
\includegraphics[width=0.32\textwidth,angle=0]{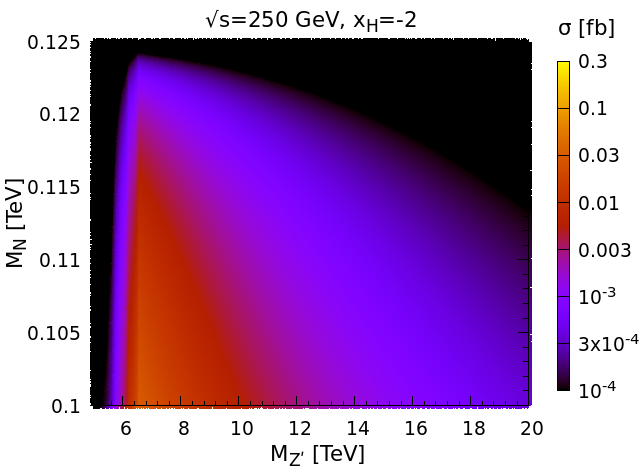}
\includegraphics[width=0.32\textwidth,angle=0]{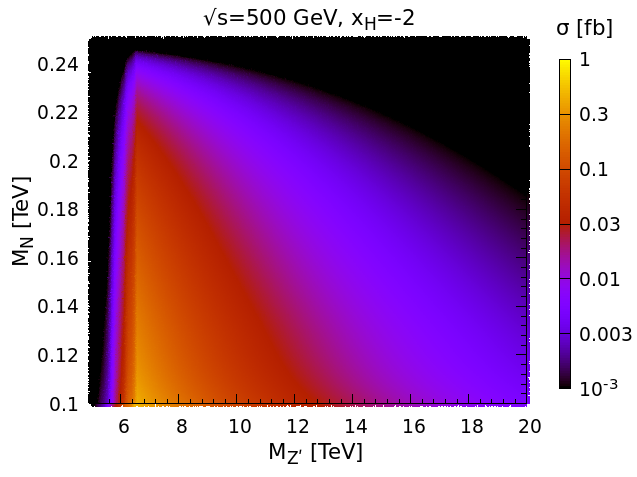}
\includegraphics[width=0.32\textwidth,angle=0]{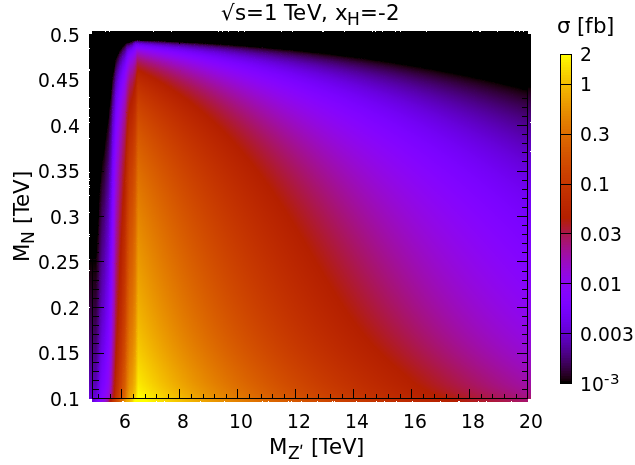}
\includegraphics[width=0.32\textwidth,angle=0]{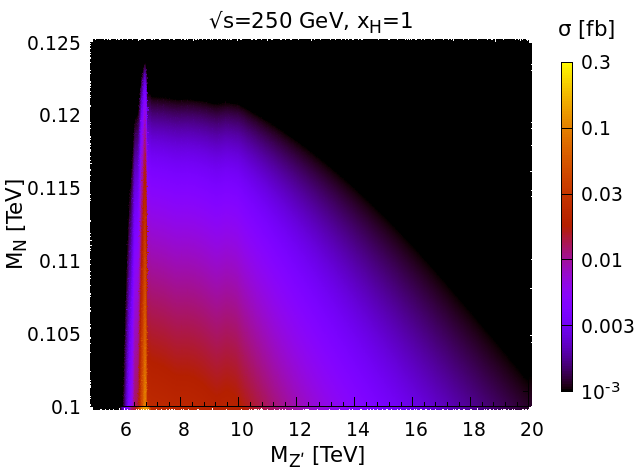}
\includegraphics[width=0.32\textwidth,angle=0]{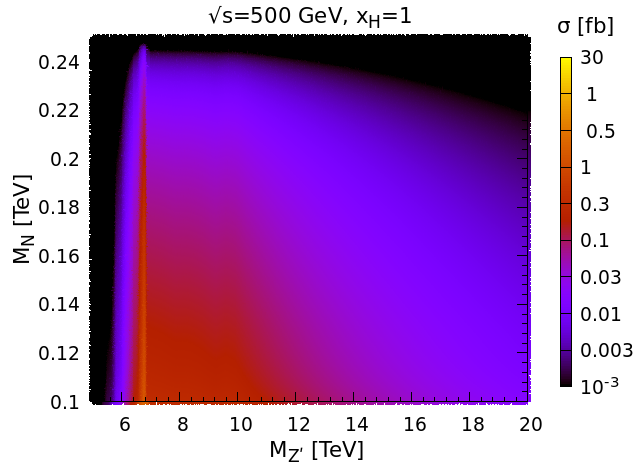}
\includegraphics[width=0.32\textwidth,angle=0]{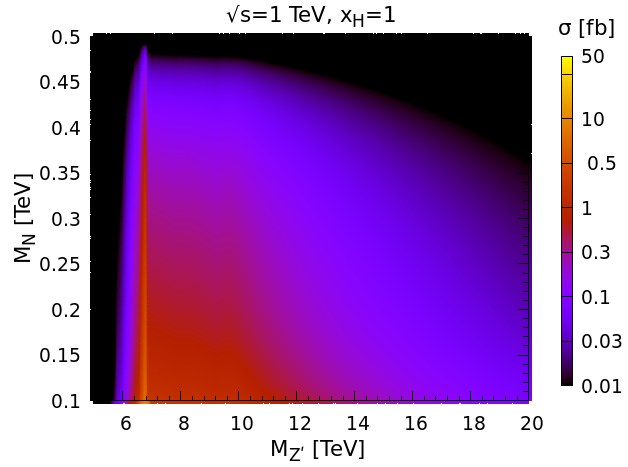}
\caption{The density plot representing trilepton plus two jet cross section in association with missing energy on $M_N-M_{Z^\prime}$ plane for Case-II considering $x_H=-2$ (upper panel) and 1 (lower panel) at different center of mass energies such as 250 GeV, 500 GeV and 1TeV from left to right. The bar chart represents the cross sections in fb.}
\label{MN-MZp-ee-case2-1}
\end{figure}

In electron positron colliders we can produce the heavy neutrinos in pair and each heavy neutrino decay into the leading mode following $N \to \ell W$.
In this analysis we consider a trilepton signal in association with two jets and missing energy. 
The jets are coming from the hadronic decay of one of the $W$ bosons whereas the third lepton and missing momentum is coming the leptonic decay of the other $W$ boson. In this case we consider two generations of the heavy neutrinos comprising the trilepton signal where the leptons include all possible combinations with electron and muon with trilepton charge combination as $+1$ and $-1$. To study this signal we consider $x_H=-2$ and 1 for $\sqrt{s}=250$ GeV, 500 GeV and 1 TeV respectively for $5$ TeV $\leq M_{Z^\prime} \leq 20$ TeV. We considered $100$ GeV $\leq M_N \leq 125$ GeV for $\sqrt{s}=250$ GeV, $100$ GeV $\leq M_N \leq 250$ GeV for $\sqrt{s}=500$ GeV, and $100$ GeV $\leq M_N \leq 500$ GeV for $\sqrt{s}=1$ TeV respectively to produce the density plots for Cases-I and II in Figs.~\ref{MN-MZp-ee-case1-1} and \ref{MN-MZp-ee-case2-1} respectively. The results in Case-II is roughly more than one order of magnitude higher than the results of Case-I due to different $U(1)_X$ charges of the heavy neutrinos. Due to the heavy neutrino pair production process, the cross section becomes negligibly small at the threshold $M_N \sim \frac{\sqrt{s}}{2}$. 
The cross section almost remains the same before the vicinity of $M_N \sim \frac{\sqrt{s}}{2}$ after that it drops sharply to zero. 
According to the choice of $x_H$ we obtain the maximum cross section at $x_H=1$.
The cross section increases with the increase in $\sqrt{s}$.
\subsection{Bounds on $M_N-M_{Z^\prime}$ plane}
\begin{figure}[h]
\centering
\includegraphics[width=0.43\textwidth,angle=0]{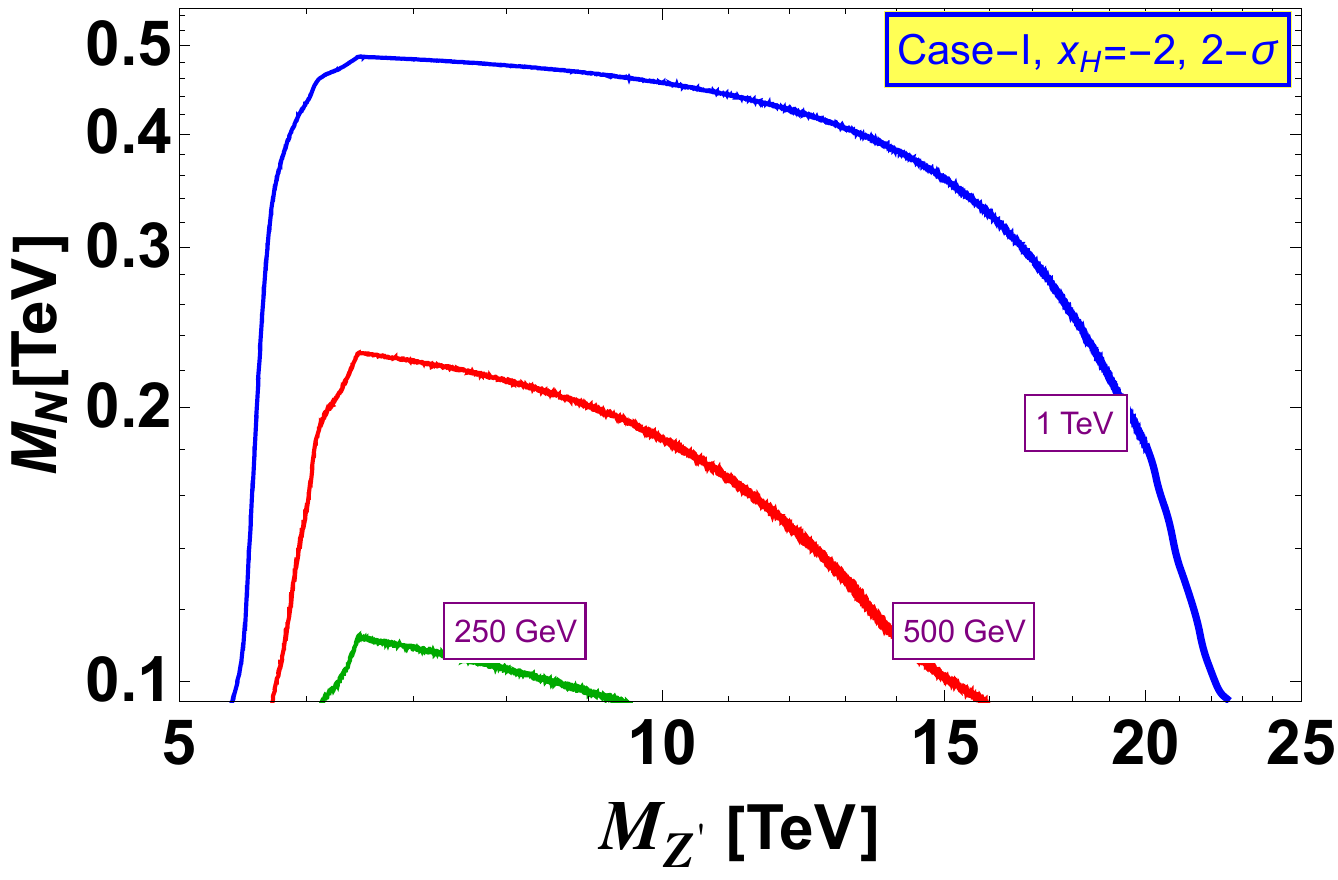}
\includegraphics[width=0.43\textwidth,angle=0]{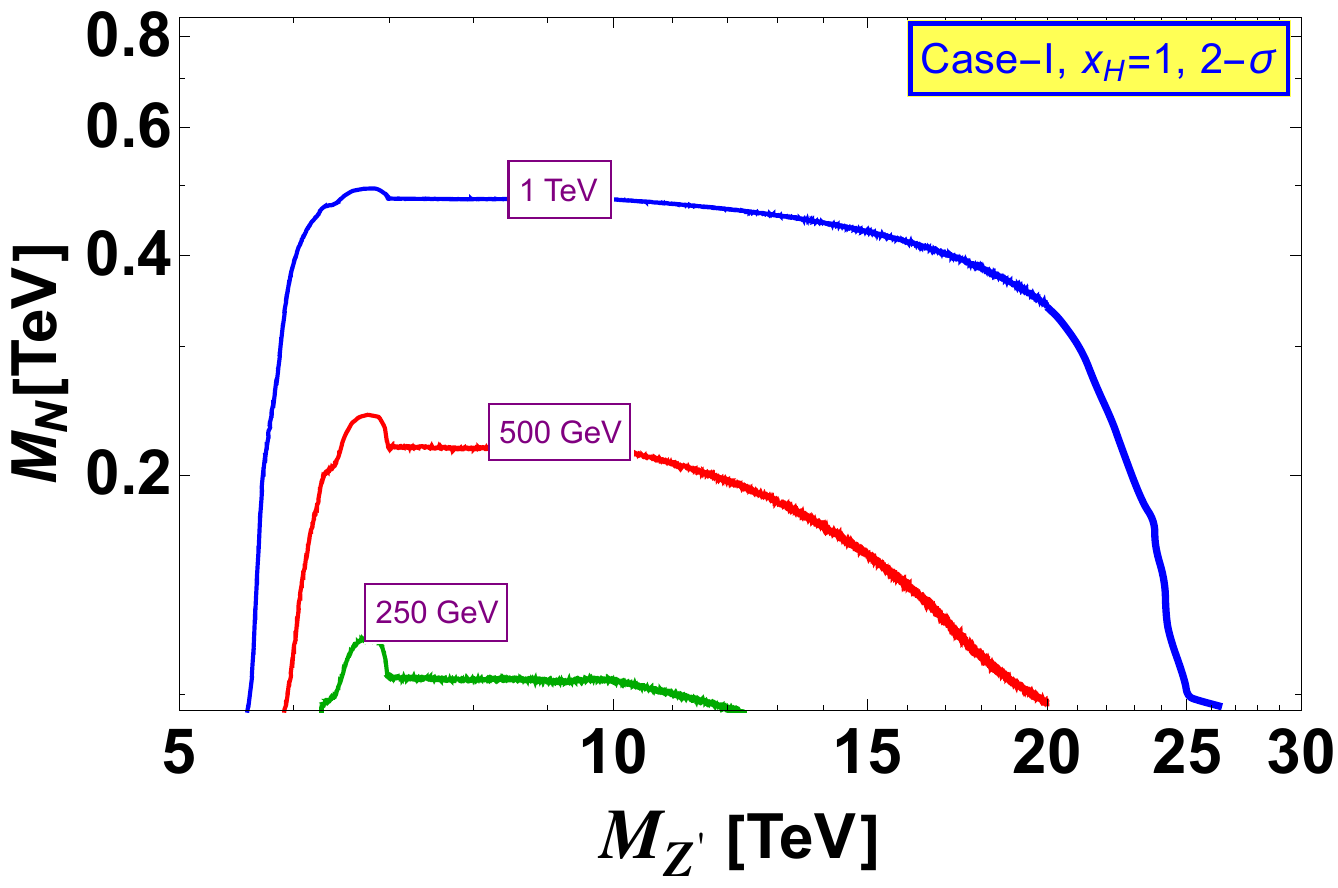}
\includegraphics[width=0.43\textwidth,angle=0]{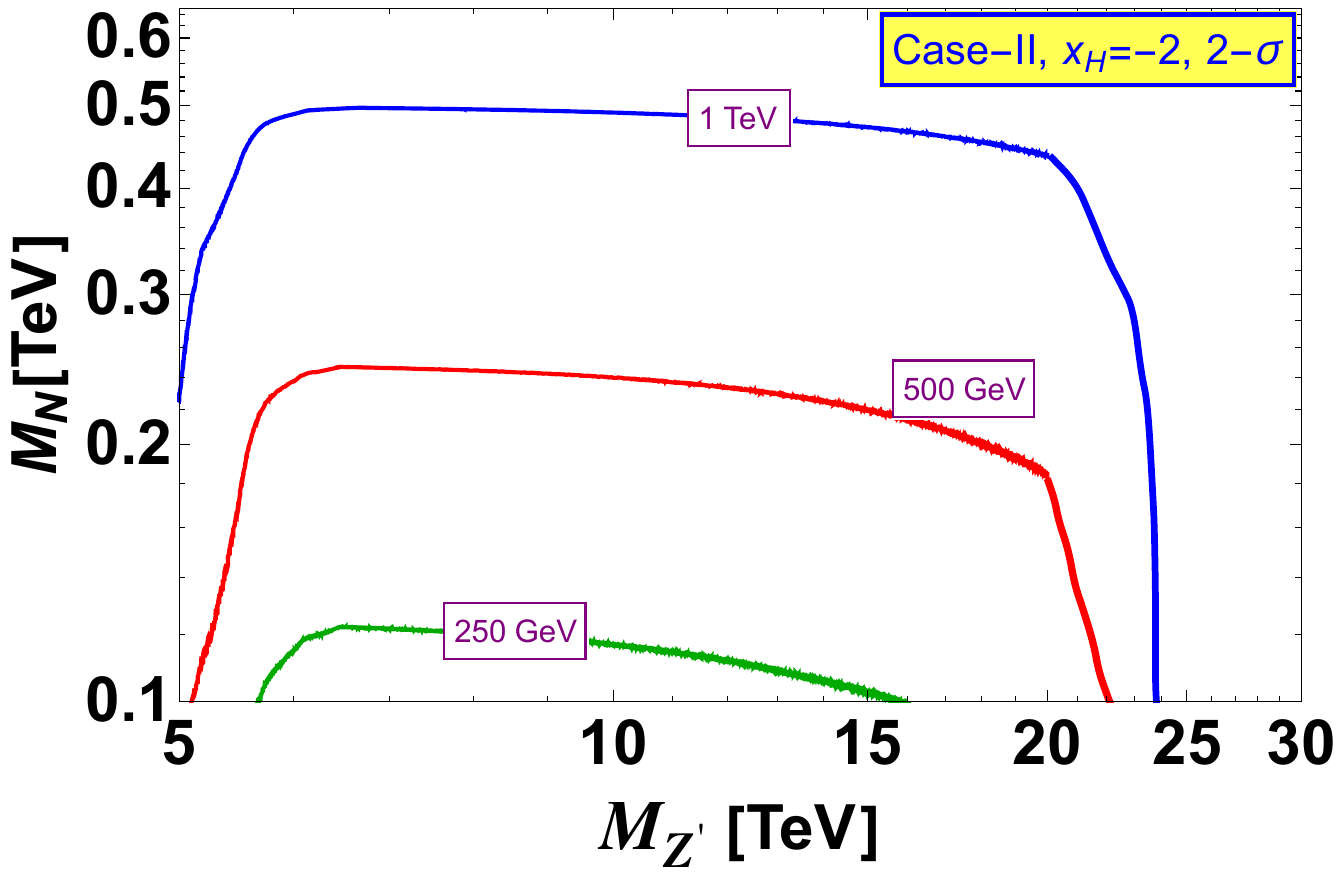}
\includegraphics[width=0.43\textwidth,angle=0]{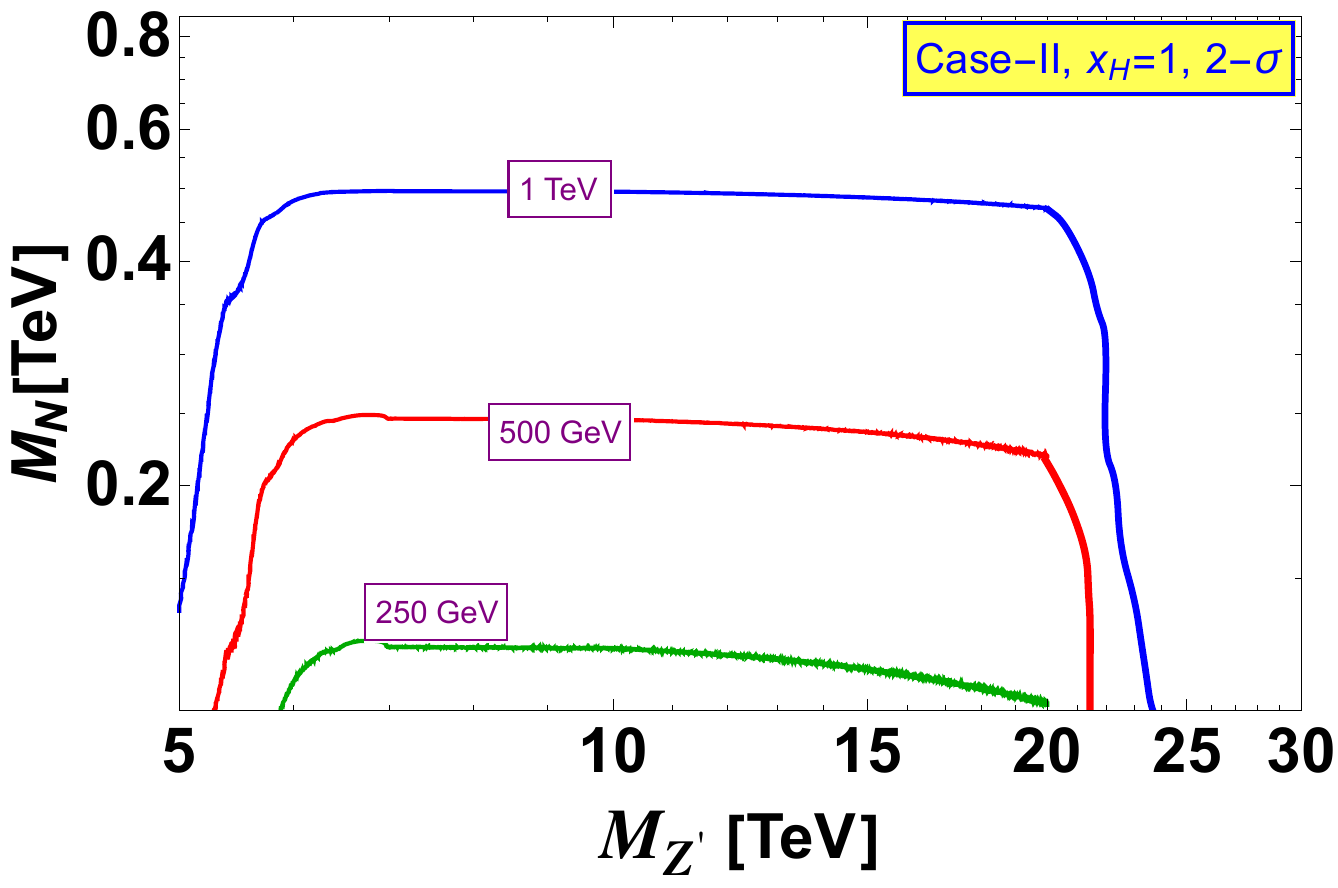}
\caption{2$-\sigma$ contour plot on $M_N-M_{Z^\prime}$ plane in the upper (lower) panel for Case-I(II) considering $x_H=-2$ (left) and $1$ (right) at 2 ab$^{-1}$, 4 ab$^{-1}$ and 8 ab$^{-1}$ luminosities in 250 GeV, 500 GeV and 1 TeV $e^-e^+$ colliders respectively for the combined SSDL plus four jet signal.}
\label{MN-MZp-2-sigma-2l}
\end{figure}

We simulate the SSDL plus four jet signal the generic backgrounds using MadGraph \cite{Alwall:2011uj, Alwall:2014hca}, hadronizing the events by PYTHIA8 \cite{Sjostrand:2007gs} followed by the detector simulation using the ILD card in Delphes \cite{deFavereau:2013fsa} for different $x_H$ and $\sqrt{s}$ varying $M_{N}$ and $M_{Z^\prime}$ according to Figs.~\ref{MN-MZp-ee-case1-dilepton} and \ref{MN-MZp-ee-case2-dilepton} for Cases-I and II respectively to prepare a 2$-\sigma$ limit plot in the $M_N-M_{Z^\prime}$ plane. We apply $p_T^j > 20$ GeV, $p_T^\ell> 10$ GeV, $|\eta^{j,l}| < 2.5$ to estimate the SSDL background from $W^\pm W^\pm$ and $4Z$ channels. The SSDL plus four jet process can be generated from $W^\pm W^\pm$ process in association with missing energy where two same sign $W$ will decay leptonically into same flavor and the remaining ones will decay hadronically. The $e^\pm e^\pm+4j$ background has extremely small cross section at 250 GeV, however, the cross section becomes 0.0015 fb and 0.0082 fb at 500 GeV and 1 TeV respectively. The $\mu^\pm \mu^\pm+4j$ background also has extremely small cross section at 250 GeV, however, the cross section becomes 0.0015 fb and 0.0082 fb at 500 GeV and 1 TeV respectively. From $4Z$ we consider two of the $Z$ bosons decay leptonically and rest of the two decay hadronically. The cross section for this process at 500 GeV is $3\times 10^{-6}$ fb and  $1.2\times 10^{-5}$ fb respectively with the electrons. Similar cross sections can be obtained for muons. We also produce the $ZZ$+jets background where each $Z$ decays into $e^- e^+$ $(\mu^- \mu^+)$ modes giving rise to SSDL pairs. The cross sections for the $e^-e^+$ modes at 250 GeV, 500 GeV and 1 TeV are $1.5\times10^{-5}$ fb, 0.00126 fb and 0.00132 fb respectively. The cross sections for the $\mu^- \mu^+$ modes at 250 GeV, 500 GeV and 1 TeV are $1.5 \times 10^{-5}$ fb, 0.00127 fb and 0.00133 fb respectively. We estimate the $t\overline{t}$ backgrounds at $\sqrt{s}=$ 500 GeV and 1 TeV applying $p_T^j > 20$ GeV, $p_T^\ell> 10$ GeV, $|\eta^{j,l}| < 2.5$. We can not generate this background at $\sqrt{s}=250$ GeV which is energetically disallowed. We consider the final state $4j2b$ which has the cross section 222.4 fb (68.0 fb) at $\sqrt{s}=$ 500 GeV (1 TeV). The cross section of $2e2\nu2b$ final state at $\sqrt{s}=$ 500 GeV (1 TeV) is 6.23 fb (1.88 fb). We find that a final state of $2\mu 2\nu2b$ has a cross section of 6.23 fb (1.9 fb) at $\sqrt{s}=500$ GeV (1 TeV). Finally we consider another combination of final state $e\mu2\nu2b$ form $t\overline{t}$ process having cross section of 12.4 fb (3.76 fb) at $\sqrt{s}=$ 500 GeV (1 TeV). We ensure that the final signal and SM backgrounds have SSDL pair only and impose the azimuthal angular cut on the leptons as $|\cos\theta_\ell| < 0.95$ defining $\theta_\ell = \tan^{-1}\Big(\frac{p_T^\ell}{p_Z^\ell}\Big)$ where $p_T^\ell$ as the transverse momentum and $p_z$ is the $z-$ component of the three momentum of the lepton respectively. We impose a missing energy cut for the events such that $E_T^{\rm miss} < 80$ GeV. The backgrounds coming from the top quark pair production do not survive after the application of these cuts. Considering the signals and backgrounds we estimate the $2-\sigma$ significance limit using $\frac{S}{\sqrt{S+B}}$ where $S$ stands for signal events and $B$ stands for backgrounds using luminosities as 2 ab$^{-1}$, 4 ab$^{-1}$ and 8 ab$^{-1}$ at 250 GeV, 500 GeV and 1 TeV $e^-e^+$ colliders following Ref.~\cite{Barklow:2015tja}. Corresponding limit plots are shown in Fig.~\ref{MN-MZp-2-sigma-2l} where the Case-I (II) is given in the upper (lower) panel for $x_H=-2$ (1) in the left (right) panel combining the electron and muon events. At $\sqrt{s}=250$ GeV we predict that $6.2$ TeV $\leq M_{Z^\prime} < 9.5$ TeV with $0.095$ TeV $\leq M_N \leq 0.115$ TeV could be probed at 2$-\sigma$ level for $x_H=-2$ in Case-I. Similarly we predict that $6.4$ TeV $\leq M_{Z^\prime} < 12.25$ TeV with $0.095$ TeV $\leq M_N \leq 0.12$ TeV could be probed at 2$-\sigma$ level for $x_H=1$ in Case-I. Due to the improved charges in case of Case-II we predict that at 2$-\sigma$ level $5.7$ TeV $\leq M_{Z^\prime} \leq 16$ TeV with $0.1$ TeV $\leq M_N \leq 0.12$ TeV could be probed at 250 GeV for $x_H=-2$. This range for the heavy $Z^\prime$ could be extended up to 20 TeV at 250 GeV keeping the range of $M_N$ almost same for Case-II considering $x_H=1$. These ranges widen in case of $M_{Z^\prime}$ and as well as $M_N$ for 500 GeV and 1 TeV colliders respectively. 

\begin{figure}[h]
\centering
\includegraphics[width=0.43\textwidth,angle=0]{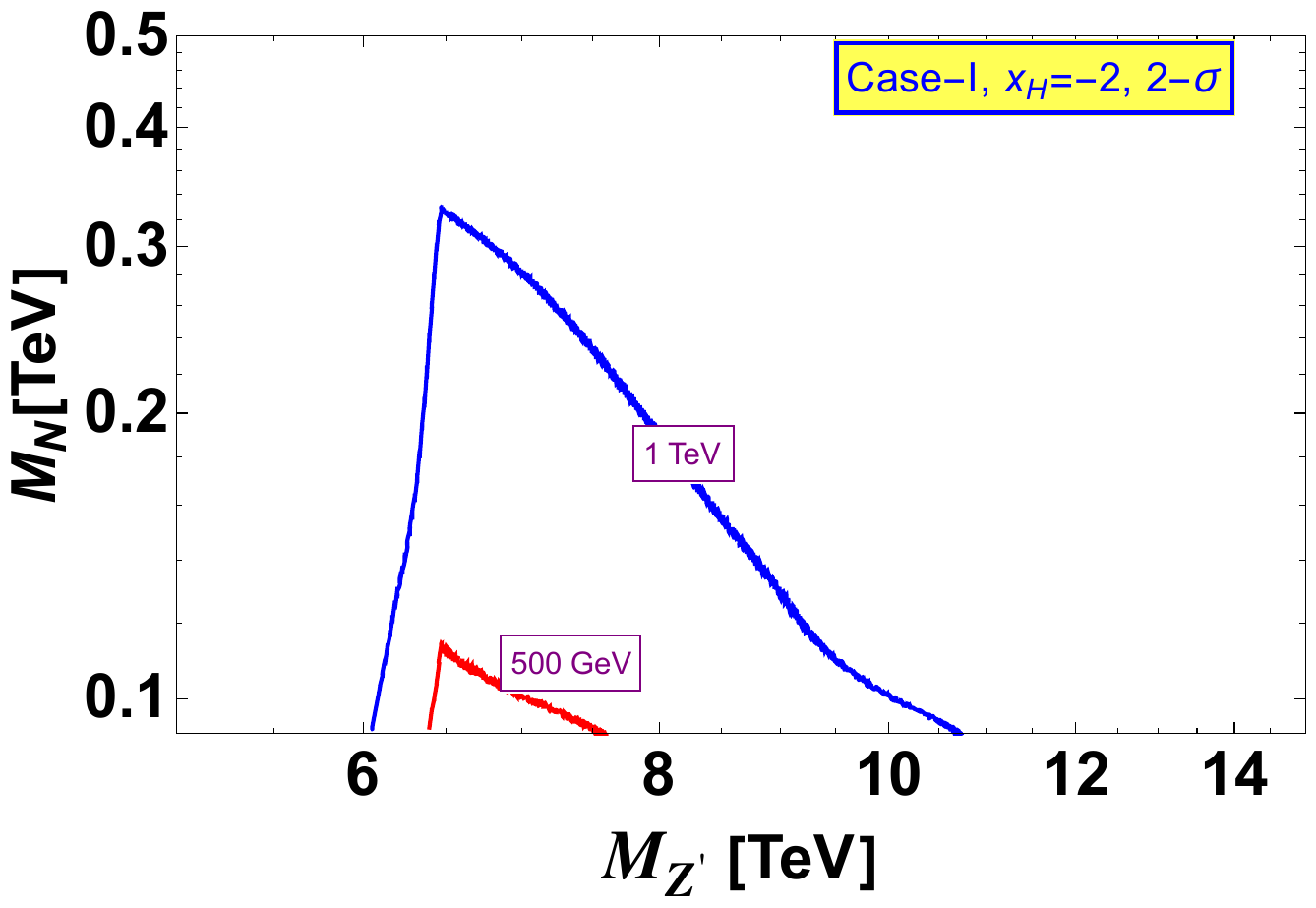}
\includegraphics[width=0.43\textwidth,angle=0]{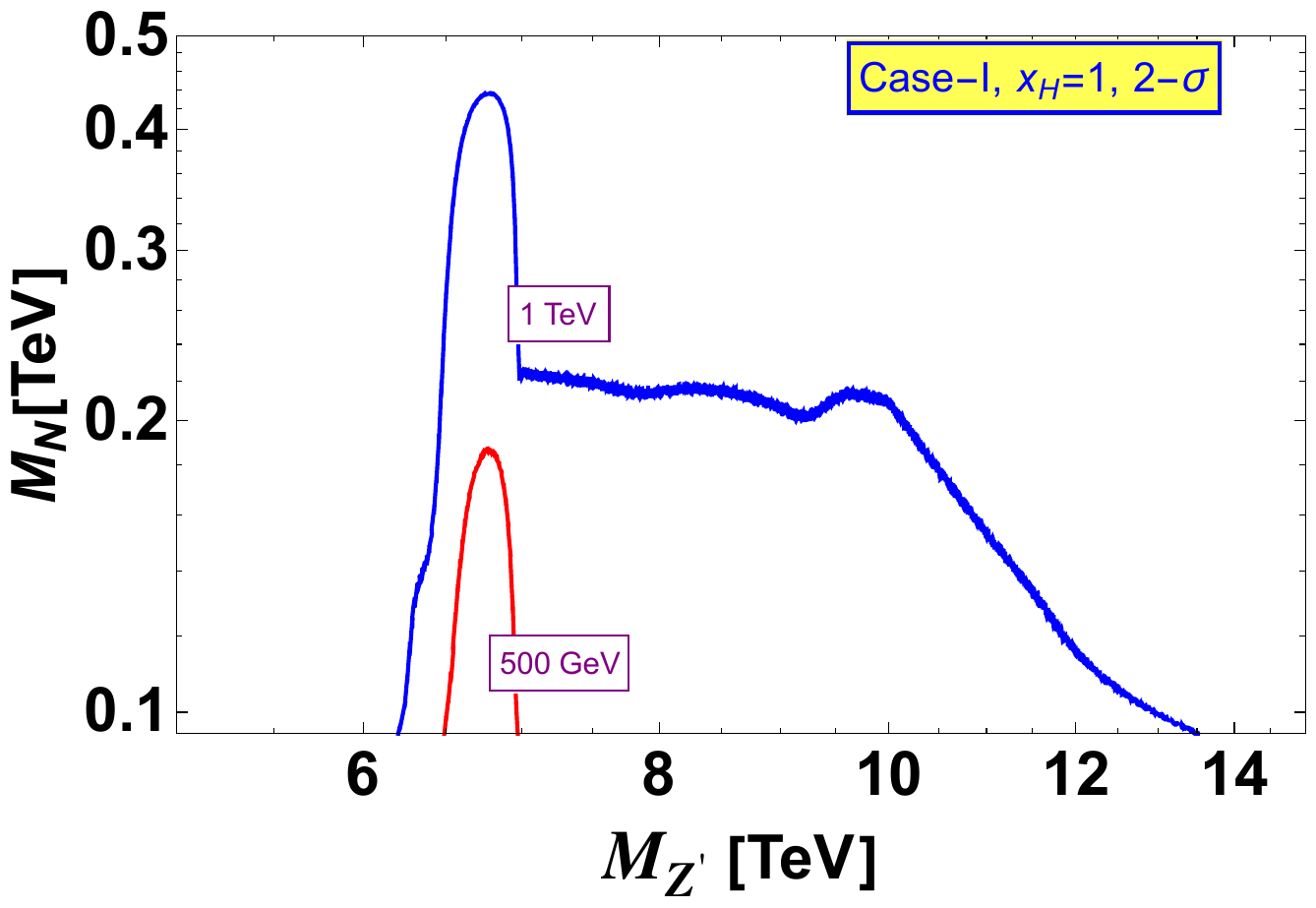}
\includegraphics[width=0.43\textwidth,angle=0]{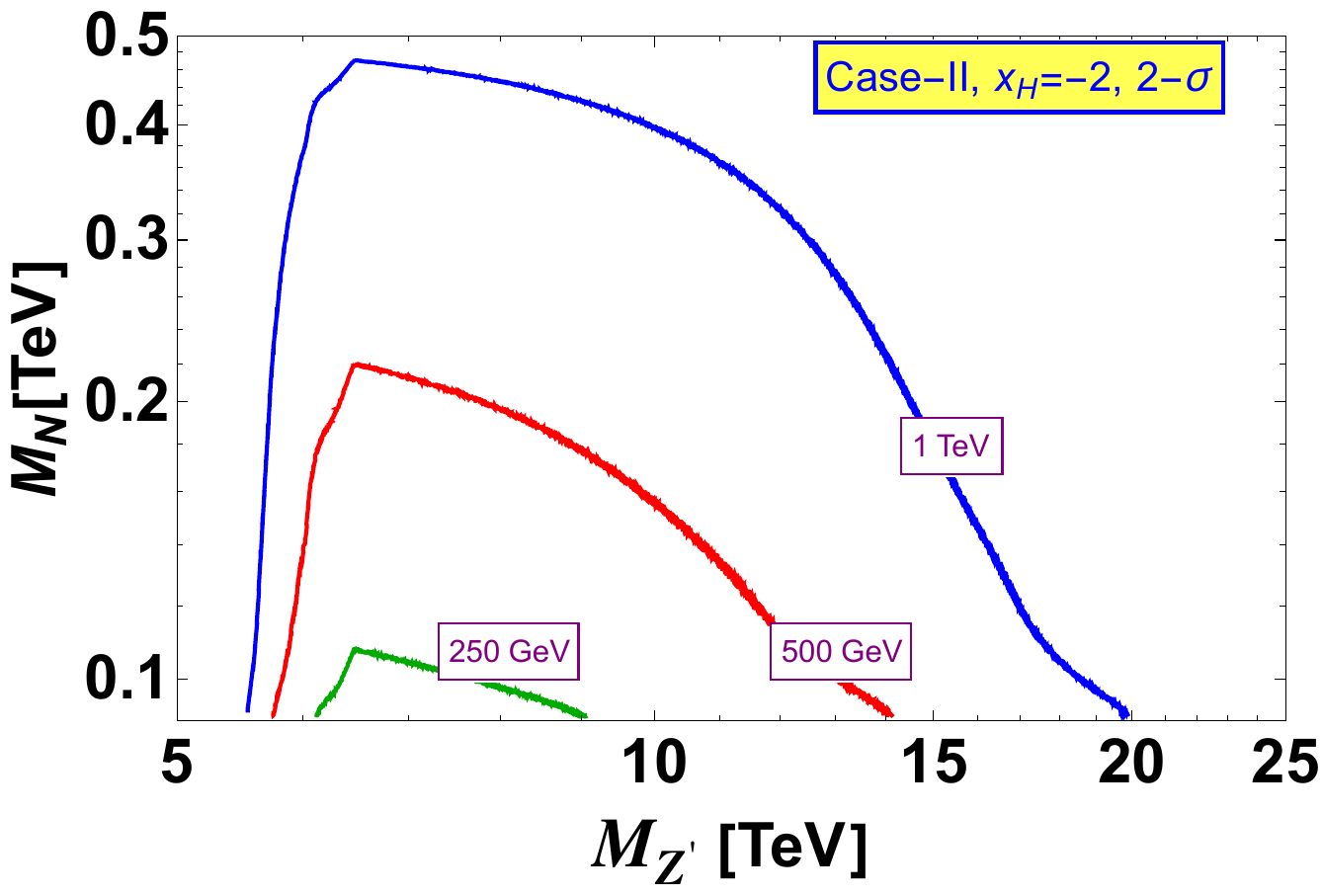}
\includegraphics[width=0.43\textwidth,angle=0]{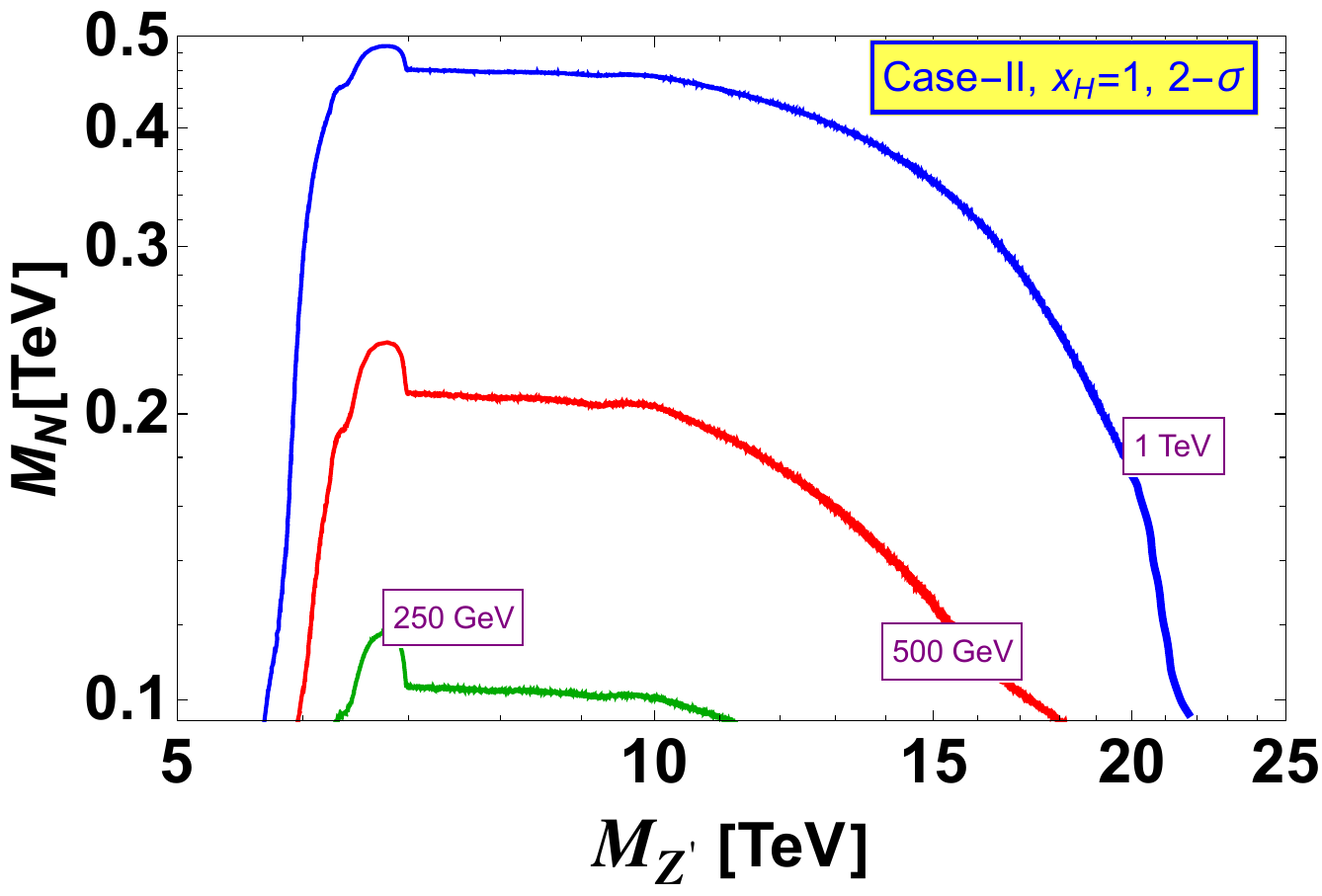}
\caption{2$-\sigma$ contour plot on $M_N-M_{Z^\prime}$ plane in the upper (lower) panel for Case-I(II) considering $x_H=-2$ (left) and $1$ (right) at 2 ab$^{-1}$, 4 ab$^{-1}$ and 8 ab$^{-1}$ luminosities in 250 GeV, 500 GeV and 1 TeV $e^-e^+$ colliders respectively for combined trilepton plus two jet in association with missing momentum.}
\label{MN-MZp-2-sigma}
\end{figure}

We simulate the trilepton plus two jet signal in association with missing energy and the generic backgrounds using MadGraph \cite{Alwall:2011uj, Alwall:2014hca}, hadronizing the events by PYTHIA8 \cite{Sjostrand:2007gs} followed by the detector simulation using the ILD card in Delphes \cite{deFavereau:2013fsa} for different $x_H$ and $\sqrt{s}$ varying $M_{N}$ and $M_{Z^\prime}$ according to Figs.~\ref{MN-MZp-ee-case1-1} and \ref{MN-MZp-ee-case2-1} for Cases-I and II respectively to prepare a 2$-\sigma$ limit plot in the $M_N-M_{Z^\prime}$ plane. We apply $p_T^j > 20$ GeV, $p_T^\ell> 10$ GeV, $|\eta^{j,l}| < 2.5$ to estimate the three electron plus two jets in association with missing energy generic SM background and the cross sections are 0.111 fb at $\sqrt{s}=250$ GeV, $1.05$ fb at $\sqrt{s}=500$ GeV and $3.53$ fb at $\sqrt{s}=1$ TeV respectively. Using the same cuts we simulate the three muon plus two jet background in association with missing energy and obtain the corresponding generic SM backgrounds cross sections as 0.1 fb at $\sqrt{s}=250$ GeV, 0.2 fb at $\sqrt{s}=500$ GeV and 0.2 fb at $\sqrt{s}=1$ TeV respectively. Using the mentioned cuts we produce the two electron and one muon generic SM background in association with two jets and missing momentum. The cross sections at $\sqrt{s}=250$ GeV, 500 GeV and 1 TeV are obtained as 0.061 fb, 0.6 fb and 1.63 fb respectively. Similarly we generate the two muon and one electron generic SM background events in association with two jets and missing energy at $\sqrt{s}=250$ GeV, 500 GeV and 1 TeV. We checked that $t\overline{t} Z$ channel gives extremely low cross sections $\mathcal{O}(10^{-4} \rm fb)$ after the application of the kinematic cuts. Hence we do not consider this process in further analysis. After applying the cuts we obtain the cross sections as 0.05 fb, 0.18 fb and 0.352 fb respectively. To estimate the SM backgrounds we considered charge combinations of the three charged leptons as $+1$ and $-1$.  Using the azimuthal angular cut on the leptons as $|\cos\theta_\ell| < 0.95$. To estimate the signal and generic SM background events we consider the luminosities as 2 ab$^{-1}$, 4 ab$^{-1}$ and 8 ab$^{-1}$ at 250 GeV, 500 GeV and 1 TeV $e^-e^+$ colliders following \cite{Barklow:2015tja}. We estimate the 2$-\sigma$ contours on the $M_N-M_{Z^\prime}$ plane in the upper (lower) panel of Fig.~\ref{MN-MZp-2-sigma} using $\frac{S}{\sqrt{S+ B}}$ using the maximum reach of the luminosities at different $e^- e^+$ colliders for Case-I (II) where result for $x_H=-2$ (1) is shown in the left (right) panel. We find that combined trilepton final states with electron and muon flavors can provide 2-$\sigma$ exclusion limits on $M_N-M_{Z^\prime}$ plane. The 2$-\sigma$ exclusion limit for Case-I at $x_H=-2$ is shown in the top left panel of Fig.~\ref{MN-MZp-2-sigma} for 6.4 TeV $\leq M_{Z^\prime} \leq 7.6$ TeV for 0.095 TeV $\leq M_N \leq 0.117$ TeV at $\sqrt{s}=500$ GeV whereas the range increases up to 6.1 TeV $\leq M_{Z^\prime} \leq 10.5$ TeV for 0.095 TeV $\leq M_N \leq 0.34$ TeV at $\sqrt{s}=1$ TeV. On the other hand for $x_H=1$, the cross section is higher and we can probe  6.5 TeV $\leq M_{Z^\prime} \leq 6.95$ TeV for 0.095 TeV $\leq M_N \leq 0.195$ TeV at $\sqrt{s}=500$ GeV and 6.25 TeV $\leq M_{Z^\prime} \leq 13.75$ TeV for 0.095 TeV $\leq M_N \leq 0.45$ TeV at $\sqrt{s}=1$ TeV which is shown in the top right panel of Fig.~\ref{MN-MZp-2-sigma}. Due to the change in the $U(1)_X$ charge of the heavy neutrinos in Case-II we find enhancement in the Majorana heavy neutrino pair production cross section which further allows us to obtain 2$-\sigma$ exclusion contours analyzing the signal and the backgrounds which are shown in the lower panel of Fig.~\ref{MN-MZp-2-sigma} for $x_H=-2$ (left) and 1 (right) respectively. In this case we can probe the $M_N-M_{Z^\prime}$ plane at 250 GeV for 6.1 TeV $\leq M_{Z^\prime} \leq 9.0$ TeV and 0.095 TeV $\leq M_N \leq 0.165$ TeV in case of $x_H=-2$ and that goes to 6.4 TeV $\leq M_{Z^\prime} \leq 11.2$ TeV and 0.095 TeV $\leq M_N \leq 0.12$ TeV in case of $x_H=1$. However, the reaches in $M_{Z^\prime}$ and $M_N$ increase with the increase in $\sqrt{s}$.
\section{Heavy neutrino pair production from BSM scalar at $e^-e^+$ colliders}
\label{RHNhiggs}
There is another interesting aspect of the Majorana heavy neutrino pair production from scalar involved in a general $U(1)_X$ scenario. In Case-I for simplicity,
applying the stationary conditions on the scalar potential in Eq.~\ref{pot} we find that the mass matrix of the scalars as
\bea
M^2=\begin{pmatrix}
2\lambda_h v_h^2 &  \lambda^\prime v_h  v_\Phi \\
\lambda^\prime v_h v_\Phi  &  2\lambda_{\Phi}v_\Phi^2 \\
\end{pmatrix}
\label{mm}
\eea
Hence diagonalizing the mass matrix in Eq.~\ref{mm} we obtain the mass eigenvalues of the physical scalars as 
\bea
m_{h_{1}}^{2}=\lambda_{h}v_h^{2}+\lambda_{\Phi}v_\Phi^{2}-\sqrt{(\lambda_{h}v_h^{2}-\lambda_{\Phi}v_\Phi^{2})^{2}+(\lambda^\prime v_h v_\Phi)^{2}},\,\,
m_{h_{2}}^{2}=\lambda_{h}v_h^{2}+\lambda_{\Phi}v_\Phi^{2}+\sqrt{(\lambda_{h}v_h^{2}-\lambda_{\Phi}v_\Phi^{2})^{2}+(\lambda^\prime v_h v_\Phi)^{2}}~~~~
\eea
and the scalar quartic couplings in Eq.~\ref{pot} of Case-I can be written as 
\bea
&\lambda_{h}=\frac{m_{h_{2}}^{2}}{4v_h^{2}}(1-\text{cos}~2\alpha)+\frac{m_{h_{1}}^{2}}{4v_h^{2}}(1+\text{cos}~2\alpha),\, \, \lambda_{\Phi}=\frac{m_{h_{1}}^{2}}{4v_\Phi^{2}}(1-\text{cos}~2\alpha)+\frac{m_{h_{2}}^{2}}{4v_\Phi^{2}}(1+\text{cos}~2\alpha),\\
&\lambda^\prime=\text{sin}~2\alpha\left(\frac{m_{h_{1}}^{2}-m_{h_{2}}^{2}}{2v_\Phi v_h}\right)
 \label{coup}
\eea
where $\alpha$ is the mixing angle between the two scalars required to diagonalize the mass matrix given in Eq.~\ref{mm}.
The BSM scalar $h_2$ can decay into SM particles or gauge bosons through the scalar mixing and the partial decay width of $h_2$ can be written as 
\bea
\Gamma (h_2 \to X_{\rm SM} X_{\rm SM}) = \sin^2 \alpha~\Gamma_{h \to X_{\rm SM} X_{\rm SM} } (m_{h} \to m_{h_2})
\eea
where $\Gamma_{h \to X_{SM} X_{SM}}$ is the partial decay widths of the SM Higgs. The BSM scalar can decay into SM Higgs $(m_{h_1}= 125~\rm GeV)$ through the interaction
\bea
\mathcal{L} =  C_{h_2 h_1 h_1} h_2 h_1 h_1
\eea
where
$C_{h_2 h_1 h_1} \equiv \frac{1}{2} [ (- s_\alpha^3 v_h +   c_\alpha^3 v_\Phi  + 2 c_\alpha^2 s_\alpha v_h - 2 c_\alpha s_\alpha^2 v_\Phi) \lambda^\prime
 - 6 c_\alpha^2 s_\alpha v_h \lambda_h + 6 c_\alpha s^2_\alpha v_\Phi \lambda_\Phi ]$ and $s_\alpha(c_\alpha) =\sin \alpha (\cos \alpha)$.
The decay width for $h_2 \to h_1 h_1$ is given by
\begin{equation}
\Gamma(h_2 \to h_1 h_1) = \frac{C_{h_2 h_1 h_1}^2}{8 \pi m_{h_2}} \sqrt{1 - 4 \frac{m_{h_1}^2}{m_{h_2}^2}}.
\end{equation}
The BSM scalar interacts with the pair of heavy neutrinos through the Yukawa interaction in Eq.~\ref{U1xy}. The partial decay process of $h_{2,1} \to N N$ can be written as
\bea
\Gamma(h_2 \to N N)= \frac{ 3Y_2^2 \cos^2\alpha}{16 \pi} m_{h_2} \left( 1 - 4 \frac{M_{N_i}^2}{m_{h_2}^2} \right)^{\frac32}, \, \,\Gamma(h_1 \to N N)= \frac{ 3Y_2^2 \sin^2\alpha}{16 \pi} m_{h_1} \left( 1 - 4 \frac{M_{N_i}^2}{m_{h_2}^2} \right)^{\frac32}
\eea
where $Y_2 = \frac{M_{N_i}\sqrt{2}}{ v_\Phi}$ with $v_\Phi = \frac{M_{Z^\prime}}{2 g^\prime}$. Hence we find a relation between the $Z^\prime$ and heavy neutrino mass as 
$g^\prime= \frac{1}{2\sqrt{2}} \frac{Y_2 M_{Z^\prime}}{M_{N_i}}$.

\begin{figure}[h]
\includegraphics[width=50mm]{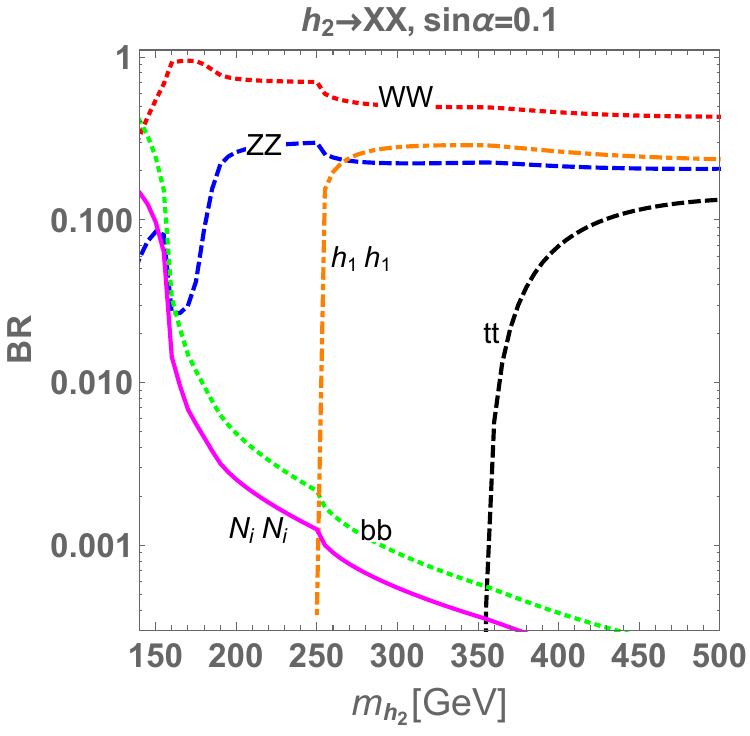}
\includegraphics[width=50mm]{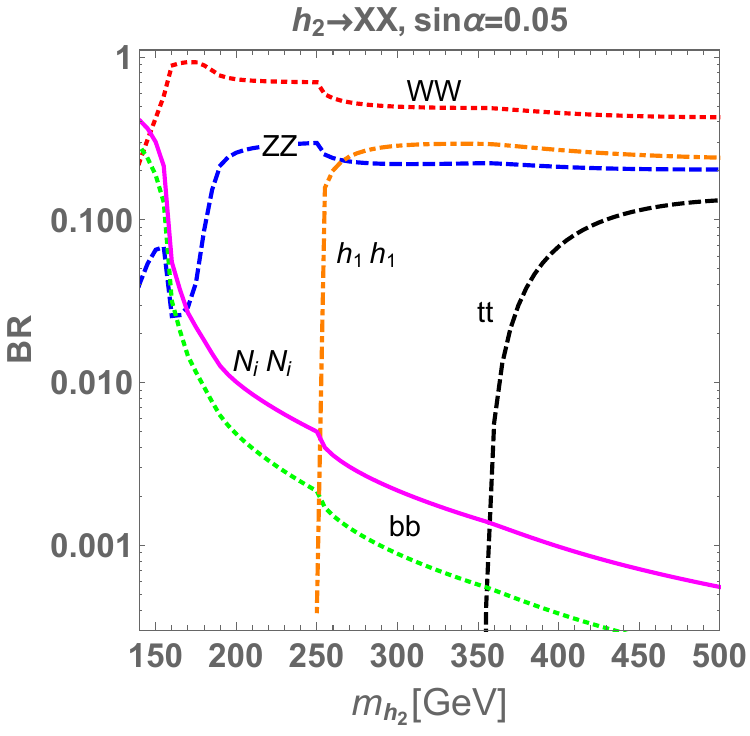}
\includegraphics[width=50mm]{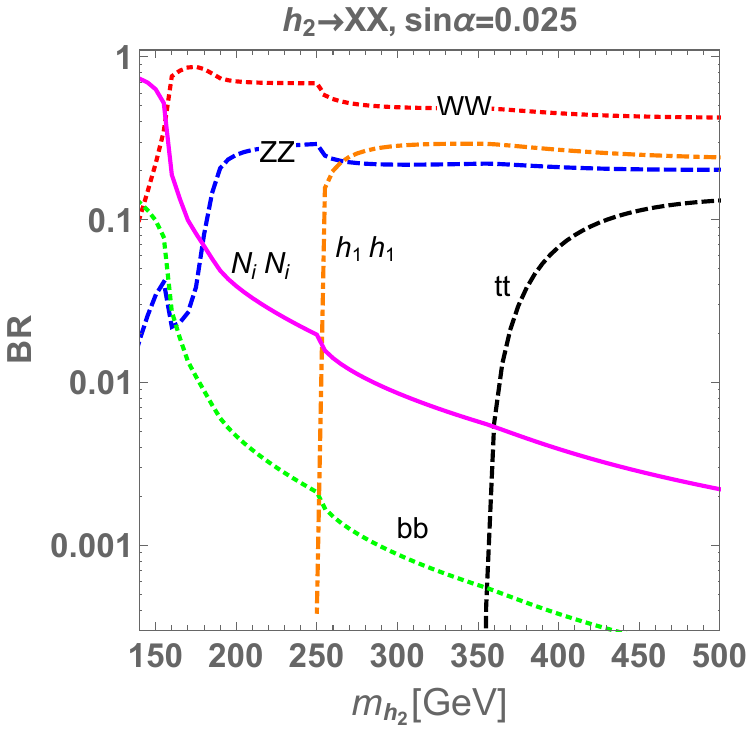}
\caption{Branching ratios of $h_2$ as functions of $m_{h_2}$ where $m_{Z'} = 5$ TeV, $g' =0.1$ and $M_{N_i} = 50$ GeV.}
\label{fig:BR}
\end{figure}

The complete form of the partial decay widths of $h_1$ and $h_2$ are given in the Appendix. We show the branching ratios of $h_2$ into different modes including a pair of RHNs in Fig.~\ref{fig:BR} as a function of $m_{h_2}$ considering $M_{Z^\prime} = 5$ TeV, $g^\prime =0.1$ and $M_{N_i}=$ 50 GeV. The allowed benchmark value of $g^\prime$ is taken from \cite{Das:2021esm} satisfying LHC dilepton, dijet and LEP-II constraints. We consider three benchmarks of $\sin\alpha=0.1$, 0.05 and 0.025 respectively which is taken from \cite{Das:2022oyx} which are allowed benchmark points after the application of LEP and LHC limits. In this case we consider $x_H=0$ which is the B$-$L case. 

\begin{figure}[h]
\includegraphics[width=50mm]{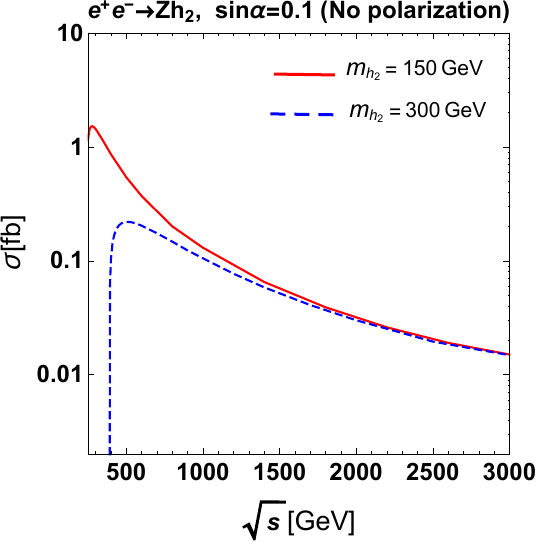}
\includegraphics[width=50mm]{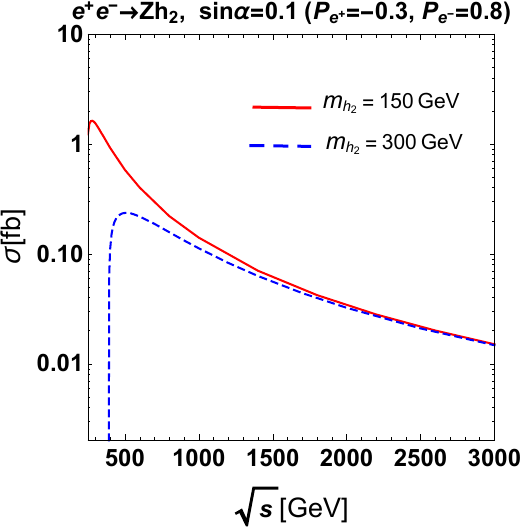}
\includegraphics[width=50mm]{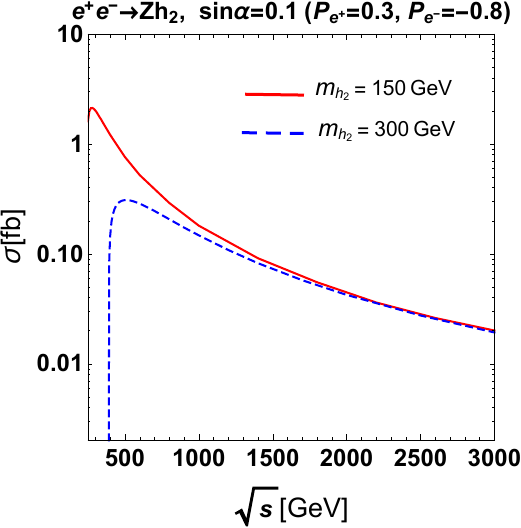}\\
\includegraphics[width=50mm]{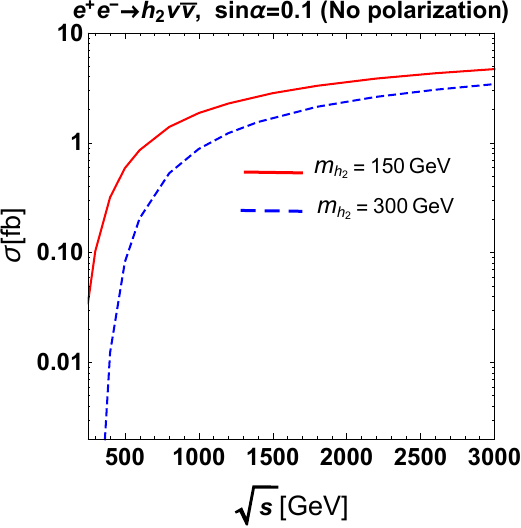}
\includegraphics[width=50mm]{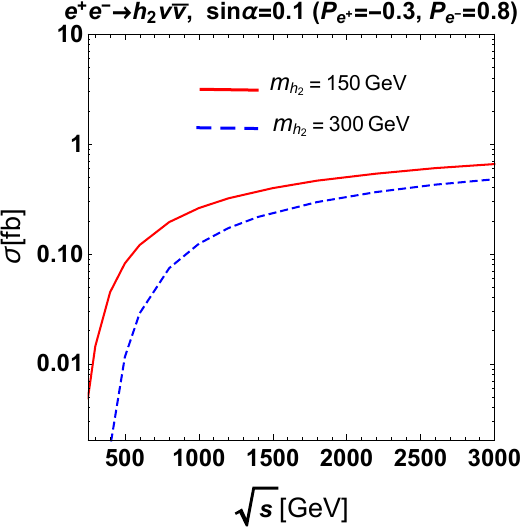}
\includegraphics[width=50mm]{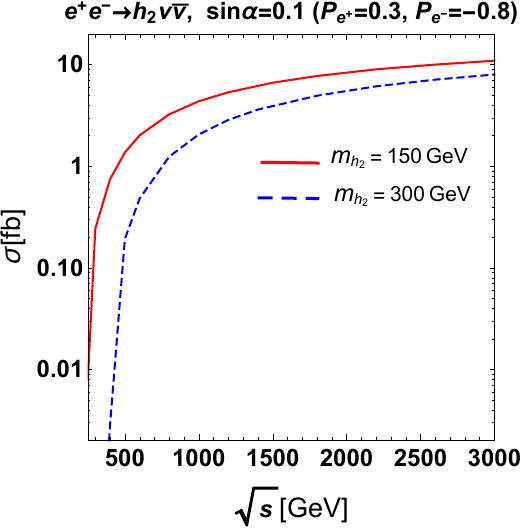}
\caption{Cross section for $e^+ e^- \to Z h_2$ (upper panel) and $e^+ e^- \to h_2 \nu \overline{\nu}$ (lower panel) process as functions of $\sqrt{s}$ for different values of $m_{h_2}$ and different electron and positron polarizations.}
\label{fig:CX}
\end{figure}

We consider two different modes of the $h_2$ production one at the electron positron collider.  One is $e^-e^+ \to Z h_2$ and the other is $e^+ e^- \to h_2 \nu \overline{\nu}$.
In Fig.~\ref{fig:CX} we show the cross sections of $Z h_2$ (upper panel) and $h_2 \nu \overline{\nu}$ (lower panel) respectively considering $\sin\alpha=0.1$ and $m_{h_2}=150$ GeV 
and 300 GeV respectively as a function of $\sqrt{s}$. In this study we consider three choices of the polarizations $P_{e^-}=P_{e^+}=0$, $P_{e^-}=-0.3, P_{e^+}=0.8$ and 
$P_{e^-}=0.3, P_{e^+}=-0.8$ respectively.
producing $h_2$ in the above modes we consider the heavy neutrino pair production followed by the decay $h_2 \to N_i N_i$. 
Note that $h_2$ can be produced via $e^+ e^- \to e^+ e^- h_2$ via $Z$ boson fusion but its production cross section is smaller compared to the other modes. 
As a result we did not consider it in this analysis.
In our following analysis masses of extra scalar and heavy neutrinos are fixed to be $m_{h_2} = 150$ GeV, $M_{N_1} = M_{N_2} = 50$ GeV for illustration.
We also choose $M_{Z^\prime} = 6.5$ TeV and $g^\prime=1$ to enhance $h_2 \to N_i N_i$ decay mode.
We consider two generations of the heavy neutrinos where $N_1$ dominantly decays into electron and $N_2$ dominantly decays into muon. 
Here we consider the dominant decay modes of the heavy neutrinos where $N_i \to \ell j j$ where two jets come from the hadronic decay mode of ${W^\pm}^\ast$.
\subsection{Signal from $e^+e^- \to Z h_2$}
Here we consider associated production process $e^+e^- \to Z h_2$ for $\sqrt{s} = 250$ GeV since the process provides the largest $h_2$ production cross section at the center of mass energy.
In this case our signal process is $ e^+ e^- \to h_2 Z \to N N j j \to \ell^\pm \ell^\pm + 6 j$ where two of these jets are coming from the hadronic decay of the $Z$ boson. We assume that the Yukawa coupling involved in this process is diagonal.
The signal indicates lepton number violation and number of SM background (BG) events can be reduced by same sign charged lepton tagging.
For the BG from SM processes at the LHC 250 GeV, we consider following processes $\ell^+ \ell^- W^+ W^-$ (BG1), $\ell^+ \ell^- Z Z$ (BG2), $\ell^\pm \nu W^\pm Z$ (BG3, BG4) and the cross sections 
of these processes are 0.33 fb, 0.094 fb and 2.5 fb respectively at $\sqrt{s}=250$ GeV. These backgrounds can mimic our signals providing same sign charged lepton events. We consider $M_N=50$ GeV so that heavy neutrinos can decay off-shell. 

We generate signal and SM background events using MadGraph \cite{Alwall:2014hca}. Then PYTHIA8 \cite{Sjostrand:2014zea}  is applied to deal with hadronization effects, the initial-state radiation (ISR) and final-state radiation (FSR) effects and the decays of SM particles, and Delphes \cite{deFavereau:2013fsa} is used for detector level simulation using $\{ P_{e^-},  P_{e^+} \} = \{0, 0\}$. At the detector level, event selection is imposed with kinematical cuts demanding an SSDL pair for electrons and muons as: (i) transverse momenta of leptons: $p_T^{\ell^\pm} > 10 $\ {\rm GeV} and those of the jets are: $p_T^{j} > 20$  \ {\rm GeV}, (ii) pseudo-rapidity: $\eta^{\ell, j} < 2.5$ and (iii) missing energy: $\slashed{E}_T < 20 \ {\rm GeV}$. We evaluate number of signals and SM baground events imposing selection cuts where $N_{\rm event} = \mathcal{L}_{\rm int} \sigma \frac{N_{\rm Selected}}{N_{\rm Generated}}$, $N_{\rm Select}$ is number of events after selection, $N_{\rm Generated}$ is number of generated events, $\sigma$ is a cross section for each process and $\mathcal{L}_{\rm int}=$2 ab$^{-1}$is integrated luminosity. We combine the electron and muon events claiming at least 3 jets $n_j \geq 3$ in the events. After the application of the selection cuts we evaluate the significance if the process using 
\begin{equation}
S = \sqrt{2 \left[ (N_S + N_{BG}  ) \ln \left( 1 + \frac{N_S}{N_{BG}} \right) - N_S\right]},
\label{signi1}
\end{equation}
In Table~\ref{tab:Event}, we summarize number of signal and SM background events after selection choosing $\sin \alpha = 0.1$ and $0.2$ as benchmark values. Hence we notice that a sizable significance can be observed at the 250 GeV electron positron collider.
\begin{center} 
\begin{table}[t]
\begin{tabular}{|c|c | c| c| c | c| c|}\hline
 signal/BGs  & signal [$s_\alpha =0.1$] & signal [$s_\alpha =0.2$] & BG1 & BG2 & BG3 & BG4  \\ \hline 
$N_{\rm ev}$ (without selection) & $6.6 \times 10^2 $ & $1.2 \times 10^3 $ &$ 6.6 \times 10^2$ & $1.9 \times 10^2$ & $5.0 \times 10^3$ & $5.0 \times 10^3$  \\ \hline
$N_{\rm ev}$ (with selection) & 28 & 51 & 0.81 & 0.30 & 16 & 16  \\ \hline
$S_{} [s_\alpha = 0.1]$ & \multicolumn{6}{c|}{4.4} \\ \hline
$S_{} [s_\alpha = 0.2]$ & \multicolumn{6}{c|}{7.4} \\ \hline
\end{tabular}
\caption{Number of signal and SM background events for $\sqrt{s} = 250$ GeV and integrated luminosity 2 ab$^{-1}$ before and after selection for SSDL mode.}
\label{tab:Event}
\end{table}
\end{center}
\subsection{Signal from $e^+e^- \to  h_2 \bar \nu \nu$}
Here we consider $W$ boson fusion for $h_2$ production, $e^+ e^- \to h_2 \bar \nu \nu$, with $\sqrt{s} = 1$ TeV since the cross section for $h_2$ production increases with $\sqrt{s}$. We use polarization of the $e^+$ and $e^-$ beam as  $\{ P_{e^-},  P_{e^+} \} = \{- 0.8, 0.3\}$, since this gives the largest cross section as shown in the lower panel of Fig.~\ref{fig:CX}. Our signal process is $e^+ e^- \to h_2 \nu \bar \nu \to  N N \nu \bar \nu \to \ell^\pm \ell^\pm + 4 j + \nu \bar \nu$. We simulate the signal and the SM backgrounds events using MadGraph  \cite{Alwall:2014hca} followed by hadronization using PYTHIA8 \cite{Sjostrand:2014zea} and detector simulation using Delphes \cite{deFavereau:2013fsa}. The SM backgrounds $\ell^+ \ell^- W^+ W^-$ (BG1), $\ell^+ \ell^- Z Z$ (BG2), $\ell^\pm \nu W^\pm Z$ (BG3, BG4) have the cross sections 44.0 fb, 0.48 fb and 39.0 fb at $\sqrt{s}=1$ TeV respectively. We use the selection cuts for the signal and SM background events as: $ p_T(\ell^\pm) > 10 \ {\rm GeV}$, $\eta^{\ell,j} < 2.5$, $p_T^j > 20 \ {\rm GeV}$ and $\slashed{E}_T > 30 \ {\rm GeV}$ respectively. As in the previous case we summarize number of events after the selection cuts and significance in Table~\ref{tab:Event2} where we write the events combining the signals with electron and muon. We find that the significance can be sizable at 1 TeV electron positron collider using 2 ab$^{-1}$ luminosity. 
\begin{center} 
\begin{table}[t]
\begin{tabular}{|c|c | c| c| c | c| c|}\hline
 signal/BGs  & signal [$s_\alpha =0.1$] & signal [$s_\alpha =0.2$] & BG1 & BG2 & BG3 & BG4  \\ \hline 
  $N_{\rm ev}$ (no selection) & $2.6 \times 10^3$ & $4.4 \times 10^3$ & $8.8 \times 10^4$  & $9.6 \times 10^2$ & $7.8 \times 10^4$ & $7.8 \times 10^4$  \\ \hline
$N_{\rm ev}$ (with selection)  &$1.1 \times 10^2$ & $2.0 \times 10^2$ & $5.7 \times 10^2$ & 7.4 & $4.6 \times 10^2$ & $4.5 \times 10^2$  \\ \hline
$S_{} [s_\alpha = 0.1]$ & \multicolumn{6}{c|}{3.0} \\ \hline
$S_{} [s_\alpha = 0.2]$ & \multicolumn{6}{c|}{5.1} \\ \hline
\end{tabular}
\caption{Number of signal and SM background events after selection cuts at $\sqrt{s} = 1$ TeV with $\{ P_{e^-},  P_{e^+} \} = \{- 0.8, 0.3\}$ and integrated luminosity 2 ab$^{-1}$ for SSDL plus four jet process.}
\label{tab:Event2}
\end{table}
\end{center}
\section{Conclusions}
\label{Sec8}
We have considered two general U$(1)$ extensions of the SM where neutrino mass can be generated by the seesaw mechanism. To cancel the gauge and mixed gauge-gravity anomalies, these models include three generations of the Majorana type heavy neutrinos which could be produced at the high energy colliders. Studying the existing constraints on the U$(1)_X$ gauge coupling as a function of the $M_{Z^\prime}$ for different $M_N$, we produce the RHNs in pair from $Z^\prime$ in hadron colliders at different center of mass energies. We study the SSDL and trilepton modes manifesting the Majorana nature of the RHNs considering the leading decay mode of the RHNs. Considering sufficiently heavy RHNs we find that the $W$ boson from the RHN decay can be sufficiently boosted. Selecting the signal and backgrounds and passing through advanced cuts we find that these signals can be obtained with reasonable significance for different benchmark scenarios of $M_N$ and $M_{Z^\prime}$ at hadron colliders. Hence scaning over a range of $M_{Z^\prime}$ and $M_N$ we estimate $2-\sigma$ exclusion contours at hadron colliders with different luminosities for the SSDL signal. Finally we conclude that the SSDL signal with two fat-jets can be probed at hadron colliders with different luminosities in the near future. Majorana heavy neutrinos can be produced in pair at the electron positron colliders where heavy $Z^\prime$ can be probed from the SSDL and trilepton signature. To do this we estimate the limits on the scale of the vacuum expectation value of the U$(1)$ breaking from LHC and LEP. Using those limits we simulate the SSDL and trilepton events.  Applying the kinematic cuts on the signal and SM backgrounds on the combined electron and muon events with jets, we estimate a 2$-\sigma$ contours on the $M_N-M_{Z^\prime}$ plane at 250 GeV, 500 GeV and 1 TeV respectively depending on the choice of the U$(1)$ charges of the particles. In addition to the $Z^\prime$ induced Majorana type heavy neutrino pair production we consider the production of a BSM scalar under the general U$(1)_X$ scenario considering the current bounds on the scalar mixing angle form LHC and LEP at different center of mass energies and polarizations. This BSM scalar can decay into a pair of Majorana type heavy neutrinos through the Yukawa interaction which can further decay into SSDL modes in association with jets and missing energy. Studying the signals and corresponding SM backgrounds we find that such process can also be probed at the electron positron colliders with sizable significance in the near future.
Furthermore studying the neutrino mass generation mechanism in the context of an inverse seesaw mechanism we can probe the pseudo-Dirac heavy neutrinos. The lepton number violating and conserving modes are different in the Majorana and pseudo-Dirac cases in colliders which will be studied in detail in our upcoming work (in progress) introducing heavy neutrino pair production mechanism from $Z^\prime$ to distinguish between Majorana and Dirac nature. Currently it is beyond the scope of this article. 
\section*{Appendix}
\label{app}
The partial decay widths of the SM like Higgs boson of mass $m_{h_1}$ into various modes are given below from \cite{Gunion:1989we}:
\begin{itemize}
\item[(i)] SM fermions (f):
\bea
\Gamma_{h_1 \to f{\bar f}} = \cos^2{\alpha} \times \frac{3 N_c m_{h_1} m_f^2}{8\pi v_{h}^2} \left( 1- \frac{4m_f^2}{m_h^2} \right)^{\frac32}, 
\eea
$N_c = 1$ and $3$ for SM leptons and quarks respectively. \\

\item[(ii)] on-shell gauge bosons (V $= W^\pm$ or $Z$ ):
\bea
\Gamma_{h_1 \to V V} =  \cos^2{\alpha} \times \frac{C_V}{32\pi} \frac{m_{h_1}^3}{ v_{h}^2} \left(1-4 \frac{m_V^2}{m_{h_1}^2}\right)^{\frac12}\left(1-4 \frac{m_V^2}{m_{h_1}^2}+12 \frac{m_V^4}{m_{h_1}^4}\right),
\eea
where $C_V = 1$ and $2$ for $V = Z$ or $W^\pm$ gauge boson, respectively. \\

\item[(iii)] gluon (g) via top-quark loop:
\bea
\Gamma_{h_1 \to gg} =  \cos^2{\alpha} \times \frac{\alpha_s^2 m_{h_1}^3 }{128\pi^3 v_h^2} (F_{1/2}(m_{h_1}))^2,
\eea
where 
\bea
F_{1/2}(m_{h_1}) = -2 \frac{4m_t^2}{m_{h_1}^2} \left[ 1- \left( 1 - \frac{4m_t^2}{m_{h_1}^2} 
\left({\rm sin^{-1}} \left( \frac{m_{h_1}}{2m_t}\right) \right)^2 \right) \right].
\eea 

\item[(iv)] one off-shell gauge boson :
\bea
\Gamma_{h_1 \to W^\pm {W^\pm}^\ast} = \cos^2{\alpha} \times \frac{3 m_W^4 m_{h_1} }{32\pi^3  v_{h}^4} G\left( \frac{m_W^2}{m_{h_1}^2} \right)
\eea
\bea
\Gamma_{h_1 \to Z Z^*} = \cos^2{\alpha} \times \frac{3m_Z^4 m_{h_1} }{32\pi^3  v_{h_1}^4} \left( \frac{7}{12} -\frac{10}{9} \sin^2 \theta_W + \frac{40}{9} \sin^4 \theta_W \right) 
G\left( \frac{m_W^2}{m_{h_1}^2} \right),
\eea
where $\sin^2\theta_W = 0.231$ and the loop functions can be represented as
\bea
G(x)= 3 \frac{1-8x+20x^2}{\sqrt{4x-1}} { \cos^{-1}}\left( \frac{3x-1}{2x^{3/2}}\right)-\frac{|1-x|}{2x}(2-13x+47x^2) -\frac{3}{2} (1-6x+4x^2)\log (\sqrt x), 
\eea
where $1/4 < x < 1$, for energetically allowed decays. For the $U(1)_X$ scalar $h_2$, $\cos{\alpha}$ will be replaced by $\sin\alpha$. 
\end{itemize}
\begin{acknowledgments} 
The work of S.M. is supported by KIAS Individual Grants (PG086001) at Korea Institute for Advanced Study.
S.S acknowledge the support of the SAMKHYA: High Performance Computing Facility provided by IOPB.
\end{acknowledgments}
\bibliography{bibliography}

\providecommand{\href}[2]{#2}\begingroup\raggedright\begin{thebibliography}{100}

\bibitem{Patrignani:2016xqp}
{\bfseries Particle Data Group} Collaboration, C.~Patrignani {\em et~al.},
  ``{Review of Particle Physics},''
\href{http://dx.doi.org/10.1088/1674-1137/40/10/100001}{{\em Chin. Phys.}
  {\bfseries C40} no.~10, (2016) 100001}.

\bibitem{Weinberg:1979sa}
S.~Weinberg, ``{Baryon and Lepton Nonconserving Processes},''
\href{http://dx.doi.org/10.1103/PhysRevLett.43.1566}{{\em Phys. Rev. Lett.}
  {\bfseries 43} (1979) 1566--1570}.

\bibitem{Minkowski:1977sc}
P.~Minkowski, ``{$\mu \to e\gamma$ at a Rate of One Out of $10^{9}$ Muon
  Decays?},''
\href{http://dx.doi.org/10.1016/0370-2693(77)90435-X}{{\em Phys. Lett.}
  {\bfseries 67B} (1977) 421--428}.

\bibitem{Mohapatra:1979ia}
R.~N. Mohapatra and G.~Senjanovic, ``{Neutrino Mass and Spontaneous Parity
  Nonconservation},'' \href{http://dx.doi.org/10.1103/PhysRevLett.44.912}{{\em
  Phys. Rev. Lett.} {\bfseries 44} (1980) 912}.
[,231(1979)].

\bibitem{Schechter:1980gr}
J.~Schechter and J.~W.~F. Valle, ``{Neutrino Masses in SU(2) x U(1)
  Theories},''
\href{http://dx.doi.org/10.1103/PhysRevD.22.2227}{{\em Phys. Rev.} {\bfseries
  D22} (1980) 2227}.

\bibitem{Yanagida:1979as}
T.~Yanagida, ``{Horizontal gauge symmetry and masses of neutrinos},''
{\em Conf. Proc.} {\bfseries C7902131} (1979) 95--99.

\bibitem{GellMann:1980vs}
M.~Gell-Mann, P.~Ramond, and R.~Slansky, ``{Complex Spinors and Unified
  Theories},'' {\em Conf. Proc.} {\bfseries C790927} (1979) 315--321,
\href{http://arxiv.org/abs/1306.4669}{{\ttfamily arXiv:1306.4669 [hep-th]}}.

\bibitem{Glashow:1979nm}
S.~L. Glashow, ``{The Future of Elementary Particle Physics},''
\href{http://dx.doi.org/10.1007/978-1-4684-7197-7_15}{{\em NATO Sci. Ser. B}
  {\bfseries 61} (1980) 687}.

\bibitem{Casas:2001sr}
J.~A. Casas and A.~Ibarra, ``{Oscillating neutrinos and muon $\to$ e, gamma},''
  \href{http://dx.doi.org/10.1016/S0550-3213(01)00475-8}{{\em Nucl. Phys.}
  {\bfseries B618} (2001) 171--204},
\href{http://arxiv.org/abs/hep-ph/0103065}{{\ttfamily arXiv:hep-ph/0103065
  [hep-ph]}}.

\bibitem{Das:2017nvm}
A.~Das and N.~Okada, ``{Bounds on heavy Majorana neutrinos in type-I seesaw and
  implications for collider searches},''
  \href{http://dx.doi.org/10.1016/j.physletb.2017.09.042}{{\em Phys. Lett. B}
  {\bfseries 774} (2017) 32--40},
  \href{http://arxiv.org/abs/1702.04668}{{\ttfamily arXiv:1702.04668
  [hep-ph]}}.

\bibitem{Asaka:2005pn}
T.~Asaka and M.~Shaposhnikov, ``{The $\nu$MSM, dark matter and baryon asymmetry
  of the universe},''
  \href{http://dx.doi.org/10.1016/j.physletb.2005.06.020}{{\em Phys. Lett. B}
  {\bfseries 620} (2005) 17--26},
  \href{http://arxiv.org/abs/hep-ph/0505013}{{\ttfamily arXiv:hep-ph/0505013}}.

\bibitem{Han:2006ip}
T.~Han and B.~Zhang, ``{Signatures for Majorana neutrinos at hadron
  colliders},'' \href{http://dx.doi.org/10.1103/PhysRevLett.97.171804}{{\em
  Phys. Rev. Lett.} {\bfseries 97} (2006) 171804},
  \href{http://arxiv.org/abs/hep-ph/0604064}{{\ttfamily arXiv:hep-ph/0604064}}.

\bibitem{Gorbunov:2007ak}
D.~Gorbunov and M.~Shaposhnikov, ``{How to find neutral leptons of the
  $\nu$MSM?},'' \href{http://dx.doi.org/10.1088/1126-6708/2007/10/015}{{\em
  JHEP} {\bfseries 10} (2007) 015},
  \href{http://arxiv.org/abs/0705.1729}{{\ttfamily arXiv:0705.1729 [hep-ph]}}.
  [Erratum: JHEP 11, 101 (2013)].

\bibitem{delAguila:2008cj}
F.~del Aguila and J.~A. Aguilar-Saavedra, ``{Distinguishing seesaw models at
  LHC with multi-lepton signals},''
  \href{http://dx.doi.org/10.1016/j.nuclphysb.2008.12.029}{{\em Nucl. Phys. B}
  {\bfseries 813} (2009) 22--90},
  \href{http://arxiv.org/abs/0808.2468}{{\ttfamily arXiv:0808.2468 [hep-ph]}}.

\bibitem{Mitra:2011qr}
M.~Mitra, G.~Senjanovic, and F.~Vissani, ``{Neutrinoless Double Beta Decay and
  Heavy Sterile Neutrinos},''
  \href{http://dx.doi.org/10.1016/j.nuclphysb.2011.10.035}{{\em Nucl. Phys. B}
  {\bfseries 856} (2012) 26--73},
  \href{http://arxiv.org/abs/1108.0004}{{\ttfamily arXiv:1108.0004 [hep-ph]}}.

\bibitem{BhupalDev:2012zg}
P.~S. Bhupal~Dev, R.~Franceschini, and R.~N. Mohapatra, ``{Bounds on TeV Seesaw
  Models from LHC Higgs Data},''
  \href{http://dx.doi.org/10.1103/PhysRevD.86.093010}{{\em Phys. Rev. D}
  {\bfseries 86} (2012) 093010},
  \href{http://arxiv.org/abs/1207.2756}{{\ttfamily arXiv:1207.2756 [hep-ph]}}.

\bibitem{Dev:2013wba}
P.~S.~B. Dev, A.~Pilaftsis, and U.-k. Yang, ``{New Production Mechanism for
  Heavy Neutrinos at the LHC},''
  \href{http://dx.doi.org/10.1103/PhysRevLett.112.081801}{{\em Phys. Rev.
  Lett.} {\bfseries 112} no.~8, (2014) 081801},
  \href{http://arxiv.org/abs/1308.2209}{{\ttfamily arXiv:1308.2209 [hep-ph]}}.

\bibitem{Das:2014jxa}
A.~Das, P.~S. Bhupal~Dev, and N.~Okada, ``{Direct bounds on electroweak scale
  pseudo-Dirac neutrinos from $\sqrt s=8$ TeV LHC data},''
  \href{http://dx.doi.org/10.1016/j.physletb.2014.06.058}{{\em Phys. Lett. B}
  {\bfseries 735} (2014) 364--370},
  \href{http://arxiv.org/abs/1405.0177}{{\ttfamily arXiv:1405.0177 [hep-ph]}}.

\bibitem{Das:2015toa}
A.~Das and N.~Okada, ``{Improved bounds on the heavy neutrino productions at
  the LHC},'' \href{http://dx.doi.org/10.1103/PhysRevD.93.033003}{{\em Phys.
  Rev. D} {\bfseries 93} no.~3, (2016) 033003},
  \href{http://arxiv.org/abs/1510.04790}{{\ttfamily arXiv:1510.04790
  [hep-ph]}}.

\bibitem{Antusch:2015gjw}
S.~Antusch, E.~Cazzato, and O.~Fischer, ``{Higgs production from sterile
  neutrinos at future lepton colliders},''
  \href{http://dx.doi.org/10.1007/JHEP04(2016)189}{{\em JHEP} {\bfseries 04}
  (2016) 189}, \href{http://arxiv.org/abs/1512.06035}{{\ttfamily
  arXiv:1512.06035 [hep-ph]}}.

\bibitem{Das:2016hof}
A.~Das, P.~Konar, and S.~Majhi, ``{Production of Heavy neutrino in
  next-to-leading order QCD at the LHC and beyond},''
  \href{http://dx.doi.org/10.1007/JHEP06(2016)019}{{\em JHEP} {\bfseries 06}
  (2016) 019}, \href{http://arxiv.org/abs/1604.00608}{{\ttfamily
  arXiv:1604.00608 [hep-ph]}}.

\bibitem{Das:2017gke}
A.~Das, P.~Konar, and A.~Thalapillil, ``{Jet substructure shedding light on
  heavy Majorana neutrinos at the LHC},''
  \href{http://dx.doi.org/10.1007/JHEP02(2018)083}{{\em JHEP} {\bfseries 02}
  (2018) 083}, \href{http://arxiv.org/abs/1709.09712}{{\ttfamily
  arXiv:1709.09712 [hep-ph]}}.

\bibitem{Bhardwaj:2018lma}
A.~Bhardwaj, A.~Das, P.~Konar, and A.~Thalapillil, ``{Looking for Minimal
  Inverse Seesaw scenarios at the LHC with Jet Substructure Techniques},''
  \href{http://dx.doi.org/10.1088/1361-6471/ab7769}{{\em J. Phys. G} {\bfseries
  47} no.~7, (2020) 075002}, \href{http://arxiv.org/abs/1801.00797}{{\ttfamily
  arXiv:1801.00797 [hep-ph]}}.

\bibitem{Antusch:2018bgr}
S.~Antusch, E.~Cazzato, O.~Fischer, A.~Hammad, and K.~Wang, ``{Lepton Flavor
  Violating Dilepton Dijet Signatures from Sterile Neutrinos at Proton
  Colliders},'' \href{http://dx.doi.org/10.1007/JHEP10(2018)067}{{\em JHEP}
  {\bfseries 10} (2018) 067}, \href{http://arxiv.org/abs/1805.11400}{{\ttfamily
  arXiv:1805.11400 [hep-ph]}}.

\bibitem{Das:2017pvt}
A.~Das, ``{Pair production of heavy neutrinos in next-to-leading order QCD at
  the hadron colliders in the inverse seesaw framework},''
  \href{http://dx.doi.org/10.1142/S0217751X21500123}{{\em Int. J. Mod. Phys. A}
  {\bfseries 36} no.~04, (2021) 2150012},
  \href{http://arxiv.org/abs/1701.04946}{{\ttfamily arXiv:1701.04946
  [hep-ph]}}.

\bibitem{Arganda:2015ija}
E.~Arganda, M.~J. Herrero, X.~Marcano, and C.~Weiland, ``{Exotic
  \ensuremath{\mu}\ensuremath{\tau}jj events from heavy ISS neutrinos at the
  LHC},'' \href{http://dx.doi.org/10.1016/j.physletb.2015.11.013}{{\em Phys.
  Lett. B} {\bfseries 752} (2016) 46--50},
  \href{http://arxiv.org/abs/1508.05074}{{\ttfamily arXiv:1508.05074
  [hep-ph]}}.

\bibitem{Dib:2016wge}
C.~O. Dib, C.~S. Kim, K.~Wang, and J.~Zhang, ``{Distinguishing Dirac/Majorana
  Sterile Neutrinos at the LHC},''
  \href{http://dx.doi.org/10.1103/PhysRevD.94.013005}{{\em Phys. Rev. D}
  {\bfseries 94} no.~1, (2016) 013005},
  \href{http://arxiv.org/abs/1605.01123}{{\ttfamily arXiv:1605.01123
  [hep-ph]}}.

\bibitem{Dib:2017vux}
C.~O. Dib, C.~S. Kim, and K.~Wang, ``{Search for Heavy Sterile Neutrinos in
  Trileptons at the LHC},''
  \href{http://dx.doi.org/10.1088/1674-1137/41/10/103103}{{\em Chin. Phys. C}
  {\bfseries 41} no.~10, (2017) 103103},
  \href{http://arxiv.org/abs/1703.01936}{{\ttfamily arXiv:1703.01936
  [hep-ph]}}.

\bibitem{Chakraborty:2018khw}
S.~Chakraborty, M.~Mitra, and S.~Shil, ``{Fat Jet Signature of a Heavy Neutrino
  at Lepton Collider},''
  \href{http://dx.doi.org/10.1103/PhysRevD.100.015012}{{\em Phys. Rev. D}
  {\bfseries 100} no.~1, (2019) 015012},
  \href{http://arxiv.org/abs/1810.08970}{{\ttfamily arXiv:1810.08970
  [hep-ph]}}.

\bibitem{Das:2018usr}
A.~Das, S.~Jana, S.~Mandal, and S.~Nandi, ``{Probing right handed neutrinos at
  the LHeC and lepton colliders using fat jet signatures},''
  \href{http://dx.doi.org/10.1103/PhysRevD.99.055030}{{\em Phys. Rev. D}
  {\bfseries 99} no.~5, (2019) 055030},
  \href{http://arxiv.org/abs/1811.04291}{{\ttfamily arXiv:1811.04291
  [hep-ph]}}.

\bibitem{Liu:2020vur}
J.~Liu, Z.~Liu, L.-T. Wang, and X.-P. Wang, ``{Enhancing Sensitivities to
  Long-lived Particles with High Granularity Calorimeters at the LHC},''
  \href{http://dx.doi.org/10.1007/JHEP11(2020)066}{{\em JHEP} {\bfseries 11}
  (2020) 066}, \href{http://arxiv.org/abs/2005.10836}{{\ttfamily
  arXiv:2005.10836 [hep-ph]}}.

\bibitem{Liu:2019ayx}
J.~Liu, Z.~Liu, L.-T. Wang, and X.-P. Wang, ``{Seeking for sterile neutrinos
  with displaced leptons at the LHC},''
  \href{http://dx.doi.org/10.1007/JHEP07(2019)159}{{\em JHEP} {\bfseries 07}
  (2019) 159}, \href{http://arxiv.org/abs/1904.01020}{{\ttfamily
  arXiv:1904.01020 [hep-ph]}}.

\bibitem{Mondal:2016kof}
S.~Mondal and S.~K. Rai, ``{Probing the Heavy Neutrinos of Inverse Seesaw Model
  at the LHeC},'' \href{http://dx.doi.org/10.1103/PhysRevD.94.033008}{{\em
  Phys. Rev. D} {\bfseries 94} no.~3, (2016) 033008},
  \href{http://arxiv.org/abs/1605.04508}{{\ttfamily arXiv:1605.04508
  [hep-ph]}}.

\bibitem{Helo:2018qej}
J.~C. Helo, M.~Hirsch, and Z.~S. Wang, ``{Heavy neutral fermions at the
  high-luminosity LHC},'' \href{http://dx.doi.org/10.1007/JHEP07(2018)056}{{\em
  JHEP} {\bfseries 07} (2018) 056},
  \href{http://arxiv.org/abs/1803.02212}{{\ttfamily arXiv:1803.02212
  [hep-ph]}}.

\bibitem{Drewes:2019fou}
M.~Drewes and J.~Hajer, ``{Heavy Neutrinos in displaced vertex searches at the
  LHC and HL-LHC},'' \href{http://dx.doi.org/10.1007/JHEP02(2020)070}{{\em
  JHEP} {\bfseries 02} (2020) 070},
  \href{http://arxiv.org/abs/1903.06100}{{\ttfamily arXiv:1903.06100
  [hep-ph]}}.

\bibitem{Drewes:2019vjy}
M.~Drewes, A.~Giammanco, J.~Hajer, and M.~Lucente, ``{New long-lived particle
  searches in heavy-ion collisions at the LHC},''
  \href{http://dx.doi.org/10.1103/PhysRevD.101.055002}{{\em Phys. Rev. D}
  {\bfseries 101} no.~5, (2020) 055002},
  \href{http://arxiv.org/abs/1905.09828}{{\ttfamily arXiv:1905.09828
  [hep-ph]}}.

\bibitem{Hirsch:2020klk}
M.~Hirsch and Z.~S. Wang, ``{Heavy neutral leptons at ANUBIS},''
  \href{http://dx.doi.org/10.1103/PhysRevD.101.055034}{{\em Phys. Rev. D}
  {\bfseries 101} no.~5, (2020) 055034},
  \href{http://arxiv.org/abs/2001.04750}{{\ttfamily arXiv:2001.04750
  [hep-ph]}}.

\bibitem{Gao:2021one}
Y.~Gao and K.~Wang, ``{Heavy Neutrino Searches via Same-sign Lepton Pairs at
  the Higgs Factory},'' \href{http://arxiv.org/abs/2102.12826}{{\ttfamily
  arXiv:2102.12826 [hep-ph]}}.

\bibitem{Deppisch:2019kvs}
F.~Deppisch, S.~Kulkarni, and W.~Liu, ``{Heavy neutrino production via $Z'$ at
  the lifetime frontier},''
  \href{http://dx.doi.org/10.1103/PhysRevD.100.035005}{{\em Phys. Rev. D}
  {\bfseries 100} no.~3, (2019) 035005},
  \href{http://arxiv.org/abs/1905.11889}{{\ttfamily arXiv:1905.11889
  [hep-ph]}}.

\bibitem{Cvetic:2020lyh}
G.~Cvetic, C.~S. Kim, S.~Mendizabal, and J.~Zamora-Saa, ``{Exploring
  CP-violation, via heavy neutrino oscillations, in rare B meson decays at
  Belle II},'' \href{http://dx.doi.org/10.1140/epjc/s10052-020-08625-0}{{\em
  Eur. Phys. J. C} {\bfseries 80} no.~11, (2020) 1052},
  \href{http://arxiv.org/abs/2007.04115}{{\ttfamily arXiv:2007.04115
  [hep-ph]}}.

\bibitem{Cvetic:2014nla}
G.~Cveti\v{c}, C.~S. Kim, and J.~Zamora-Sa\'a, ``{CP violation in lepton number
  violating semihadronic decays of $K,D,D_s,B,B_c$},''
  \href{http://dx.doi.org/10.1103/PhysRevD.89.093012}{{\em Phys. Rev. D}
  {\bfseries 89} no.~9, (2014) 093012},
  \href{http://arxiv.org/abs/1403.2555}{{\ttfamily arXiv:1403.2555 [hep-ph]}}.

\bibitem{Tastet:2021vwp}
J.-L. Tastet, O.~Ruchayskiy, and I.~Timiryasov, ``{Reinterpreting the ATLAS
  bounds on heavy neutral leptons in a realistic neutrino oscillation model},''
  \href{http://arxiv.org/abs/2107.12980}{{\ttfamily arXiv:2107.12980
  [hep-ph]}}.

\bibitem{Tastet:2020tzh}
J.-L. Tastet, E.~Goudzovski, I.~Timiryasov, and O.~Ruchayskiy, ``{Projected
  NA62 sensitivity to heavy neutral lepton production in
  K+\textrightarrow{}\ensuremath{\pi}0e+N decays},''
  \href{http://dx.doi.org/10.1103/PhysRevD.104.055005}{{\em Phys. Rev. D}
  {\bfseries 104} no.~5, (2021) 055005},
  \href{http://arxiv.org/abs/2008.11654}{{\ttfamily arXiv:2008.11654
  [hep-ph]}}.

\bibitem{Mekala:2022cmm}
K.~Mekala, J.~Reuter, and A.~F. Zarnecki, ``{Heavy neutrinos at future linear
  e$^+$e$^-$ colliders},'' \href{http://arxiv.org/abs/2202.06703}{{\ttfamily
  arXiv:2202.06703 [hep-ph]}}.

\bibitem{Liu:2022kid}
W.~Liu, S.~Kulkarni, and F.~F. Deppisch, ``{Heavy Neutrinos at the FCC-hh in
  the $U(1)_{B-L}$ Model},'' \href{http://arxiv.org/abs/2202.07310}{{\ttfamily
  arXiv:2202.07310 [hep-ph]}}.

\bibitem{Abada:2018sfh}
A.~Abada, N.~Bernal, M.~Losada, and X.~Marcano, ``{Inclusive Displaced Vertex
  Searches for Heavy Neutral Leptons at the LHC},''
  \href{http://dx.doi.org/10.1007/JHEP01(2019)093}{{\em JHEP} {\bfseries 01}
  (2019) 093}, \href{http://arxiv.org/abs/1807.10024}{{\ttfamily
  arXiv:1807.10024 [hep-ph]}}.

\bibitem{Choudhury:2020cpm}
D.~Choudhury, K.~Deka, T.~Mandal, and S.~Sadhukhan, ``{Neutrino and $Z'$
  phenomenology in an anomaly-free $\mathbf{U}(1)$ extension: role of
  higher-dimensional operators},''
  \href{http://dx.doi.org/10.1007/JHEP06(2020)111}{{\em JHEP} {\bfseries 06}
  (2020) 111}, \href{http://arxiv.org/abs/2002.02349}{{\ttfamily
  arXiv:2002.02349 [hep-ph]}}.

\bibitem{Deka:2021koh}
K.~Deka, T.~Mandal, A.~Mukherjee, and S.~Sadhukhan, ``{Leptogenesis in an
  anomaly-free $\mathrm{U}(1)$ extension with higher-dimensional operators},''
  \href{http://arxiv.org/abs/2105.15088}{{\ttfamily arXiv:2105.15088
  [hep-ph]}}.

\bibitem{Jana:2018rdf}
S.~Jana, N.~Okada, and D.~Raut, ``{Displaced vertex signature of type-I seesaw
  model},'' \href{http://dx.doi.org/10.1103/PhysRevD.98.035023}{{\em Phys. Rev.
  D} {\bfseries 98} no.~3, (2018) 035023},
  \href{http://arxiv.org/abs/1804.06828}{{\ttfamily arXiv:1804.06828
  [hep-ph]}}.

\bibitem{Abada:2021yot}
A.~Abada, N.~Bernal, A.~E.~C. Hern\'andez, X.~Marcano, and G.~Piazza, ``{Gauged
  inverse seesaw from dark matter},''
  \href{http://dx.doi.org/10.1140/epjc/s10052-021-09535-5}{{\em Eur. Phys. J.
  C} {\bfseries 81} no.~8, (2021) 758},
  \href{http://arxiv.org/abs/2107.02803}{{\ttfamily arXiv:2107.02803
  [hep-ph]}}.

\bibitem{FileviezPerez:2020cgn}
P.~Fileviez~P\'erez and A.~D. Plascencia, ``{Probing the Nature of Neutrinos
  with a New Force},''
  \href{http://dx.doi.org/10.1103/PhysRevD.102.015010}{{\em Phys. Rev. D}
  {\bfseries 102} no.~1, (2020) 015010},
  \href{http://arxiv.org/abs/2005.04235}{{\ttfamily arXiv:2005.04235
  [hep-ph]}}.

\bibitem{Davidson:1979wr}
A.~Davidson, M.~Koca, and K.~C. Wali, ``{U(1) as the Minimal Horizontal Gauge
  Symmetry},''
\href{http://dx.doi.org/10.1103/PhysRevLett.43.92}{{\em Phys. Rev. Lett.}
  {\bfseries 43} (1979) 92}.

\bibitem{Mohapatra:1980qe}
R.~N. Mohapatra and R.~E. Marshak, ``{Local B-L Symmetry of Electroweak
  Interactions, Majorana Neutrinos and Neutron Oscillations},''
  \href{http://dx.doi.org/10.1103/PhysRevLett.44.1644.2,
  10.1103/PhysRevLett.44.1316}{{\em Phys. Rev. Lett.} {\bfseries 44} (1980)
  1316--1319}.
[Erratum: Phys. Rev. Lett.44,1643(1980)].

\bibitem{Marshak:1979fm}
R.~E. Marshak and R.~N. Mohapatra, ``{Quark - Lepton Symmetry and B-L as the
  U(1) Generator of the Electroweak Symmetry Group},''
\href{http://dx.doi.org/10.1016/0370-2693(80)90436-0}{{\em Phys. Lett.}
  {\bfseries 91B} (1980) 222--224}.

\bibitem{Davidson:1978pm}
A.~Davidson, ``{$B-L$ as the fourth color within an $\mathrm{SU}(2)_L \times
  \mathrm{U}(1)_R \times \mathrm{U}(1)$ model},''
  \href{http://dx.doi.org/10.1103/PhysRevD.20.776}{{\em Phys. Rev. D}
  {\bfseries 20} (1979) 776}.

\bibitem{Davidson:1987mh}
A.~Davidson and K.~C. Wali, ``{Universal Seesaw Mechanism?},''
  \href{http://dx.doi.org/10.1103/PhysRevLett.59.393}{{\em Phys. Rev. Lett.}
  {\bfseries 59} (1987) 393}.

\bibitem{Wetterich:1981bx}
C.~Wetterich, ``{Neutrino Masses and the Scale of B-L Violation},''
\href{http://dx.doi.org/10.1016/0550-3213(81)90279-0}{{\em Nucl. Phys.}
  {\bfseries B187} (1981) 343--375}.

\bibitem{Masiero:1982fi}
A.~Masiero, J.~F. Nieves, and T.~Yanagida, ``{$B^-$l Violating Proton Decay and
  Late Cosmological Baryon Production},''
\href{http://dx.doi.org/10.1016/0370-2693(82)90024-7}{{\em Phys. Lett.}
  {\bfseries 116B} (1982) 11--15}.

\bibitem{Mohapatra:1982xz}
R.~N. Mohapatra and G.~Senjanovic, ``{Spontaneous Breaking of Global $B^-$l
  Symmetry and Matter - Antimatter Oscillations in Grand Unified Theories},''
\href{http://dx.doi.org/10.1103/PhysRevD.27.254}{{\em Phys. Rev.} {\bfseries
  D27} (1983) 254}.

\bibitem{Buchmuller:1991ce}
W.~Buchmuller, C.~Greub, and P.~Minkowski, ``{Neutrino masses, neutral vector
  bosons and the scale of B-L breaking},''
\href{http://dx.doi.org/10.1016/0370-2693(91)90952-M}{{\em Phys. Lett.}
  {\bfseries B267} (1991) 395--399}.

\bibitem{Amrith:2018yfb}
S.~Amrith, J.~M. Butterworth, F.~F. Deppisch, W.~Liu, A.~Varma, and D.~Yallup,
  ``{LHC Constraints on a $B-L$ Gauge Model using Contur},''
\href{http://arxiv.org/abs/1811.11452}{{\ttfamily arXiv:1811.11452 [hep-ph]}}.

\bibitem{Basso:2008iv}
L.~Basso, A.~Belyaev, S.~Moretti, and C.~H. Shepherd-Themistocleous,
  ``{Phenomenology of the minimal B-L extension of the Standard model: Z' and
  neutrinos},'' \href{http://dx.doi.org/10.1103/PhysRevD.80.055030}{{\em Phys.
  Rev.} {\bfseries D80} (2009) 055030},
\href{http://arxiv.org/abs/0812.4313}{{\ttfamily arXiv:0812.4313 [hep-ph]}}.

\bibitem{FileviezPerez:2009hdc}
P.~Fileviez~Perez, T.~Han, and T.~Li, ``{Testability of Type I Seesaw at the
  CERN LHC: Revealing the Existence of the B-L Symmetry},''
  \href{http://dx.doi.org/10.1103/PhysRevD.80.073015}{{\em Phys. Rev. D}
  {\bfseries 80} (2009) 073015},
  \href{http://arxiv.org/abs/0907.4186}{{\ttfamily arXiv:0907.4186 [hep-ph]}}.

\bibitem{Deppisch:2013cya}
F.~F. Deppisch, N.~Desai, and J.~W.~F. Valle, ``{Is charged lepton flavor
  violation a high energy phenomenon?},''
  \href{http://dx.doi.org/10.1103/PhysRevD.89.051302}{{\em Phys. Rev.}
  {\bfseries D89} no.~5, (2014) 051302},
\href{http://arxiv.org/abs/1308.6789}{{\ttfamily arXiv:1308.6789 [hep-ph]}}.

\bibitem{Kang:2015uoc}
Z.~Kang, P.~Ko, and J.~Li, ``{New Avenues to Heavy Right-handed Neutrinos with
  Pair Production at Hadronic Colliders},''
  \href{http://dx.doi.org/10.1103/PhysRevD.93.075037}{{\em Phys. Rev.}
  {\bfseries D93} no.~7, (2016) 075037},
\href{http://arxiv.org/abs/1512.08373}{{\ttfamily arXiv:1512.08373 [hep-ph]}}.

\bibitem{Cox:2017eme}
P.~Cox, C.~Han, and T.~T. Yanagida, ``{LHC Search for Right-handed Neutrinos in
  $Z^\prime$ Models},'' \href{http://dx.doi.org/10.1007/JHEP01(2018)037}{{\em
  JHEP} {\bfseries 01} (2018) 037},
\href{http://arxiv.org/abs/1707.04532}{{\ttfamily arXiv:1707.04532 [hep-ph]}}.

\bibitem{Accomando:2017qcs}
E.~Accomando, L.~Delle~Rose, S.~Moretti, E.~Olaiya, and C.~H.
  Shepherd-Themistocleous, ``{Extra Higgs boson and Z$^{'}$ as portals to
  signatures of heavy neutrinos at the LHC},''
  \href{http://dx.doi.org/10.1007/JHEP02(2018)109}{{\em JHEP} {\bfseries 02}
  (2018) 109}, \href{http://arxiv.org/abs/1708.03650}{{\ttfamily
  arXiv:1708.03650 [hep-ph]}}.

\bibitem{Appelquist:2002mw}
T.~Appelquist, B.~A. Dobrescu, and A.~R. Hopper, ``{Nonexotic Neutral Gauge
  Bosons},'' \href{http://dx.doi.org/10.1103/PhysRevD.68.035012}{{\em Phys.
  Rev.} {\bfseries D68} (2003) 035012},
\href{http://arxiv.org/abs/hep-ph/0212073}{{\ttfamily arXiv:hep-ph/0212073
  [hep-ph]}}.

\bibitem{Das:2017flq}
A.~Das, N.~Okada, and D.~Raut, ``{Enhanced pair production of heavy Majorana
  neutrinos at the LHC},''
  \href{http://dx.doi.org/10.1103/PhysRevD.97.115023}{{\em Phys. Rev.}
  {\bfseries D97} no.~11, (2018) 115023},
\href{http://arxiv.org/abs/1710.03377}{{\ttfamily arXiv:1710.03377 [hep-ph]}}.

\bibitem{Das:2018tbd}
A.~Das, N.~Okada, S.~Okada, and D.~Raut, ``{Probing the seesaw mechanism at the
  250 GeV ILC},''
\href{http://arxiv.org/abs/1812.11931}{{\ttfamily arXiv:1812.11931 [hep-ph]}}.

\bibitem{Das:2017deo}
A.~Das, N.~Okada, and D.~Raut, ``{Heavy Majorana neutrino pair productions at
  the LHC in minimal U(1) extended Standard Model},''
  \href{http://dx.doi.org/10.1140/epjc/s10052-018-6171-8}{{\em Eur. Phys. J.}
  {\bfseries C78} no.~9, (2018) 696},
\href{http://arxiv.org/abs/1711.09896}{{\ttfamily arXiv:1711.09896 [hep-ph]}}.

\bibitem{Das:2019fee}
A.~Das, P.~S.~B. Dev, and N.~Okada, ``{Long-Lived TeV-Scale Right-Handed
  Neutrino Production at the LHC in Gauged $U(1)_X$ Model},''
\href{http://arxiv.org/abs/1906.04132}{{\ttfamily arXiv:1906.04132 [hep-ph]}}.

\bibitem{Chiang:2019ajm}
C.-W. Chiang, G.~Cottin, A.~Das, and S.~Mandal, ``{Displaced heavy neutrinos
  from $Z'$ decays at the LHC},''
\href{http://arxiv.org/abs/1908.09838}{{\ttfamily arXiv:1908.09838 [hep-ph]}}.

\bibitem{Montero:2007cd}
J.~C. Montero and V.~Pleitez, ``{Gauging U(1) symmetries and the number of
  right-handed neutrinos},''
  \href{http://dx.doi.org/10.1016/j.physletb.2009.03.065}{{\em Phys. Lett.}
  {\bfseries B675} (2009) 64--68},
\href{http://arxiv.org/abs/0706.0473}{{\ttfamily arXiv:0706.0473 [hep-ph]}}.

\bibitem{Smirnov:1993af}
A.~Y. Smirnov, ``{Seesaw enhancement of lepton mixing},''
  \href{http://dx.doi.org/10.1103/PhysRevD.48.3264}{{\em Phys. Rev.} {\bfseries
  D48} (1993) 3264--3270},
\href{http://arxiv.org/abs/hep-ph/9304205}{{\ttfamily arXiv:hep-ph/9304205
  [hep-ph]}}.

\bibitem{King:1999mb}
S.~F. King, ``{Large mixing angle MSW and atmospheric neutrinos from single
  right-handed neutrino dominance and U(1) family symmetry},''
  \href{http://dx.doi.org/10.1016/S0550-3213(00)00109-7}{{\em Nucl. Phys.}
  {\bfseries B576} (2000) 85--105},
\href{http://arxiv.org/abs/hep-ph/9912492}{{\ttfamily arXiv:hep-ph/9912492
  [hep-ph]}}.

\bibitem{Frampton:2002qc}
P.~H. Frampton, S.~L. Glashow, and T.~Yanagida, ``{Cosmological sign of
  neutrino CP violation},''
  \href{http://dx.doi.org/10.1016/S0370-2693(02)02853-8}{{\em Phys. Lett.}
  {\bfseries B548} (2002) 119--121},
\href{http://arxiv.org/abs/hep-ph/0208157}{{\ttfamily arXiv:hep-ph/0208157
  [hep-ph]}}.

\bibitem{Ibarra:2003up}
A.~Ibarra and G.~G. Ross, ``{Neutrino phenomenology: The Case of two
  right-handed neutrinos},''
  \href{http://dx.doi.org/10.1016/j.physletb.2004.04.037}{{\em Phys. Lett.}
  {\bfseries B591} (2004) 285--296},
\href{http://arxiv.org/abs/hep-ph/0312138}{{\ttfamily arXiv:hep-ph/0312138
  [hep-ph]}}.

\bibitem{Ma:2000cc}
E.~Ma, ``{Naturally small seesaw neutrino mass with no new physics beyond the
  TeV scale},'' \href{http://dx.doi.org/10.1103/PhysRevLett.86.2502}{{\em Phys.
  Rev. Lett.} {\bfseries 86} (2001) 2502--2504},
\href{http://arxiv.org/abs/hep-ph/0011121}{{\ttfamily arXiv:hep-ph/0011121
  [hep-ph]}}.

\bibitem{Wang:2006jy}
F.~Wang, W.~Wang, and J.~M. Yang, ``{Split two-Higgs-doublet model and neutrino
  condensation},'' \href{http://dx.doi.org/10.1209/epl/i2006-10293-3}{{\em
  Europhys. Lett.} {\bfseries 76} (2006) 388--394},
\href{http://arxiv.org/abs/hep-ph/0601018}{{\ttfamily arXiv:hep-ph/0601018
  [hep-ph]}}.

\bibitem{Gabriel:2006ns}
S.~Gabriel and S.~Nandi, ``{A New two Higgs doublet model},''
  \href{http://dx.doi.org/10.1016/j.physletb.2007.04.062}{{\em Phys. Lett.}
  {\bfseries B655} (2007) 141--147},
\href{http://arxiv.org/abs/hep-ph/0610253}{{\ttfamily arXiv:hep-ph/0610253
  [hep-ph]}}.

\bibitem{Davidson:2009ha}
S.~M. Davidson and H.~E. Logan, ``{Dirac neutrinos from a second Higgs
  doublet},'' \href{http://dx.doi.org/10.1103/PhysRevD.80.095008}{{\em Phys.
  Rev.} {\bfseries D80} (2009) 095008},
\href{http://arxiv.org/abs/0906.3335}{{\ttfamily arXiv:0906.3335 [hep-ph]}}.

\bibitem{Haba:2010zi}
N.~Haba and M.~Hirotsu, ``{TeV-scale seesaw from a multi-Higgs model},''
  \href{http://dx.doi.org/10.1140/epjc/s10052-010-1414-3}{{\em Eur. Phys. J.}
  {\bfseries C69} (2010) 481--492},
\href{http://arxiv.org/abs/1005.1372}{{\ttfamily arXiv:1005.1372 [hep-ph]}}.

\bibitem{Aad:2019fac}
{\bfseries ATLAS} Collaboration, G.~Aad {\em et~al.}, ``{Search for high-mass
  dilepton resonances using 139 fb$^{-1}$ of $pp$ collision data collected at
  $\sqrt{s}=$13 TeV with the ATLAS detector},''
\href{http://arxiv.org/abs/1903.06248}{{\ttfamily arXiv:1903.06248 [hep-ex]}}.

\bibitem{CMS:2021ctt}
{\bfseries CMS} Collaboration, A.~M. Sirunyan {\em et~al.}, ``{Search for
  resonant and nonresonant new phenomena in high-mass dilepton final states at
  $ \sqrt{s} $ = 13 TeV},''
  \href{http://dx.doi.org/10.1007/JHEP07(2021)208}{{\em JHEP} {\bfseries 07}
  (2021) 208}, \href{http://arxiv.org/abs/2103.02708}{{\ttfamily
  arXiv:2103.02708 [hep-ex]}}.

\bibitem{CERN-LHCC-2017-018}
{\bfseries ATLAS Collaboration} Collaboration, ``{Technical Design Report for
  the Phase-II Upgrade of the ATLAS LAr Calorimeter},'' Tech. Rep.
  CERN-LHCC-2017-018. ATLAS-TDR-027, CERN, Geneva, Sep, 2017.
\newblock \url{https://cds.cern.ch/record/2285582}.

\bibitem{ATLAS:2019bov}
{\bfseries ATLAS} Collaboration, ``{Search for New Phenomena in Dijet Events
  using $139 \,\text{fb}^{-1}$ of $pp$ collisions at $\sqrt{s}=$ 13TeV
  collected with the ATLAS Detector},''.

\bibitem{Sirunyan:2018xlo}
{\bfseries CMS} Collaboration, A.~M. Sirunyan {\em et~al.}, ``{Search for
  narrow and broad dijet resonances in proton-proton collisions at $
  \sqrt{s}=13 $ TeV and constraints on dark matter mediators and other new
  particles},'' \href{http://dx.doi.org/10.1007/JHEP08(2018)130}{{\em JHEP}
  {\bfseries 08} (2018) 130}, \href{http://arxiv.org/abs/1806.00843}{{\ttfamily
  arXiv:1806.00843 [hep-ex]}}.

\bibitem{LEP:2003aa}
{\bfseries LEP, ALEPH, DELPHI, L3, OPAL, LEP Electroweak Working Group, SLD
  Electroweak Group, SLD Heavy Flavor Group} Collaboration, t.~S. Electroweak,
  ``{A Combination of preliminary electroweak measurements and constraints on
  the standard model},''
\href{http://arxiv.org/abs/hep-ex/0312023}{{\ttfamily arXiv:hep-ex/0312023
  [hep-ex]}}.

\bibitem{Carena:2004xs}
M.~Carena, A.~Daleo, B.~A. Dobrescu, and T.~M.~P. Tait, ``{$Z^\prime$ gauge
  bosons at the Tevatron},''
  \href{http://dx.doi.org/10.1103/PhysRevD.70.093009}{{\em Phys. Rev.}
  {\bfseries D70} (2004) 093009},
\href{http://arxiv.org/abs/hep-ph/0408098}{{\ttfamily arXiv:hep-ph/0408098
  [hep-ph]}}.

\bibitem{Schael:2013ita}
{\bfseries ALEPH, DELPHI, L3, OPAL, LEP Electroweak} Collaboration, S.~Schael
  {\em et~al.}, ``{Electroweak Measurements in Electron-Positron Collisions at
  W-Boson-Pair Energies at LEP},''
  \href{http://dx.doi.org/10.1016/j.physrep.2013.07.004}{{\em Phys. Rept.}
  {\bfseries 532} (2013) 119--244},
  \href{http://arxiv.org/abs/1302.3415}{{\ttfamily arXiv:1302.3415 [hep-ex]}}.

\bibitem{Das:2021esm}
A.~Das, P.~S.~B. Dev, Y.~Hosotani, and S.~Mandal, ``{Probing the minimal
  $U(1)_X$ model at future electron-positron colliders via the fermion
  pair-production channel},'' \href{http://arxiv.org/abs/2104.10902}{{\ttfamily
  arXiv:2104.10902 [hep-ph]}}.

\bibitem{Langacker:2008yv}
P.~Langacker, ``{The Physics of Heavy $Z^\prime$ Gauge Bosons},''
  \href{http://dx.doi.org/10.1103/RevModPhys.81.1199}{{\em Rev. Mod. Phys.}
  {\bfseries 81} (2009) 1199--1228},
\href{http://arxiv.org/abs/0801.1345}{{\ttfamily arXiv:0801.1345 [hep-ph]}}.

\bibitem{Pumplin:2002vw}
J.~Pumplin, D.~R. Stump, J.~Huston, H.~L. Lai, P.~M. Nadolsky, and W.~K. Tung,
  ``{New generation of parton distributions with uncertainties from global QCD
  analysis},'' \href{http://dx.doi.org/10.1088/1126-6708/2002/07/012}{{\em
  JHEP} {\bfseries 07} (2002) 012},
  \href{http://arxiv.org/abs/hep-ph/0201195}{{\ttfamily arXiv:hep-ph/0201195}}.

\bibitem{Sirunyan:2021khd}
{\bfseries CMS} Collaboration, A.~M. Sirunyan {\em et~al.}, ``{Search for
  resonant and nonresonant new phenomena in high-mass dilepton final states at
  $\sqrt{s} = $ 13 TeV},'' \href{http://arxiv.org/abs/2103.02708}{{\ttfamily
  arXiv:2103.02708 [hep-ex]}}.

\bibitem{Chakraborty:2021apc}
K.~Chakraborty, A.~Das, S.~Goswami, and S.~Roy, ``{Constraining general U(1)
  interactions from neutrino-electron scattering measurements at DUNE near
  detector},'' \href{http://dx.doi.org/10.1007/JHEP04(2022)008}{{\em JHEP}
  {\bfseries 04} (2022) 008}, \href{http://arxiv.org/abs/2111.08767}{{\ttfamily
  arXiv:2111.08767 [hep-ph]}}.

\bibitem{Christensen:2008py}
N.~D. Christensen and C.~Duhr, ``{FeynRules - Feynman rules made easy},''
  \href{http://dx.doi.org/10.1016/j.cpc.2009.02.018}{{\em Comput. Phys.
  Commun.} {\bfseries 180} (2009) 1614--1641},
  \href{http://arxiv.org/abs/0806.4194}{{\ttfamily arXiv:0806.4194 [hep-ph]}}.

\bibitem{Alloul:2013bka}
A.~Alloul, N.~D. Christensen, C.~Degrande, C.~Duhr, and B.~Fuks, ``{FeynRules
  2.0 - A complete toolbox for tree-level phenomenology},''
  \href{http://dx.doi.org/10.1016/j.cpc.2014.04.012}{{\em Comput. Phys.
  Commun.} {\bfseries 185} (2014) 2250--2300},
  \href{http://arxiv.org/abs/1310.1921}{{\ttfamily arXiv:1310.1921 [hep-ph]}}.

\bibitem{Alwall:2011uj}
J.~Alwall, M.~Herquet, F.~Maltoni, O.~Mattelaer, and T.~Stelzer, ``{MadGraph 5
  : Going Beyond},'' \href{http://dx.doi.org/10.1007/JHEP06(2011)128}{{\em
  JHEP} {\bfseries 06} (2011) 128},
  \href{http://arxiv.org/abs/1106.0522}{{\ttfamily arXiv:1106.0522 [hep-ph]}}.

\bibitem{Alwall:2014hca}
J.~Alwall, R.~Frederix, S.~Frixione, V.~Hirschi, F.~Maltoni, O.~Mattelaer,
  H.~S. Shao, T.~Stelzer, P.~Torrielli, and M.~Zaro, ``{The automated
  computation of tree-level and next-to-leading order differential cross
  sections, and their matching to parton shower simulations},''
  \href{http://dx.doi.org/10.1007/JHEP07(2014)079}{{\em JHEP} {\bfseries 07}
  (2014) 079}, \href{http://arxiv.org/abs/1405.0301}{{\ttfamily arXiv:1405.0301
  [hep-ph]}}.

\bibitem{Sjostrand:2007gs}
T.~Sjostrand, S.~Mrenna, and P.~Z. Skands, ``{A Brief Introduction to PYTHIA
  8.1},'' \href{http://dx.doi.org/10.1016/j.cpc.2008.01.036}{{\em Comput. Phys.
  Commun.} {\bfseries 178} (2008) 852--867},
  \href{http://arxiv.org/abs/0710.3820}{{\ttfamily arXiv:0710.3820 [hep-ph]}}.

\bibitem{deFavereau:2013fsa}
{\bfseries DELPHES 3} Collaboration, J.~de~Favereau, C.~Delaere, P.~Demin,
  A.~Giammanco, V.~Lema\^\i{}tre, A.~Mertens, and M.~Selvaggi, ``{DELPHES 3, A
  modular framework for fast simulation of a generic collider experiment},''
  \href{http://dx.doi.org/10.1007/JHEP02(2014)057}{{\em JHEP} {\bfseries 02}
  (2014) 057}, \href{http://arxiv.org/abs/1307.6346}{{\ttfamily arXiv:1307.6346
  [hep-ex]}}.

\bibitem{Dokshitzer:1997in}
Y.~L. Dokshitzer, G.~D. Leder, S.~Moretti, and B.~R. Webber, ``{Better jet
  clustering algorithms},''
  \href{http://dx.doi.org/10.1088/1126-6708/1997/08/001}{{\em JHEP} {\bfseries
  08} (1997) 001}, \href{http://arxiv.org/abs/hep-ph/9707323}{{\ttfamily
  arXiv:hep-ph/9707323}}.

\bibitem{Wobisch:1998wt}
M.~Wobisch and T.~Wengler, ``{Hadronization corrections to jet cross-sections
  in deep inelastic scattering},'' in {\em {Workshop on Monte Carlo Generators
  for HERA Physics (Plenary Starting Meeting)}}.
\newblock 4, 1998.
\newblock \href{http://arxiv.org/abs/hep-ph/9907280}{{\ttfamily
  arXiv:hep-ph/9907280}}.

\bibitem{Larkoski:2014wba}
A.~J. Larkoski, S.~Marzani, G.~Soyez, and J.~Thaler, ``{Soft Drop},''
  \href{http://dx.doi.org/10.1007/JHEP05(2014)146}{{\em JHEP} {\bfseries 05}
  (2014) 146}, \href{http://arxiv.org/abs/1402.2657}{{\ttfamily arXiv:1402.2657
  [hep-ph]}}.

\bibitem{Dasgupta:2013ihk}
M.~Dasgupta, A.~Fregoso, S.~Marzani, and G.~P. Salam, ``{Towards an
  understanding of jet substructure},''
  \href{http://dx.doi.org/10.1007/JHEP09(2013)029}{{\em JHEP} {\bfseries 09}
  (2013) 029}, \href{http://arxiv.org/abs/1307.0007}{{\ttfamily arXiv:1307.0007
  [hep-ph]}}.

\bibitem{Butterworth:2008iy}
J.~M. Butterworth, A.~R. Davison, M.~Rubin, and G.~P. Salam, ``{Jet
  substructure as a new Higgs search channel at the LHC},''
  \href{http://dx.doi.org/10.1103/PhysRevLett.100.242001}{{\em Phys. Rev.
  Lett.} {\bfseries 100} (2008) 242001},
  \href{http://arxiv.org/abs/0802.2470}{{\ttfamily arXiv:0802.2470 [hep-ph]}}.

\bibitem{LCCPhysicsWorkingGroup:2019fvj}
{\bfseries LCC Physics Working Group} Collaboration, K.~Fujii {\em et~al.},
  ``{Tests of the Standard Model at the International Linear Collider},''
  \href{http://arxiv.org/abs/1908.11299}{{\ttfamily arXiv:1908.11299
  [hep-ex]}}.

\bibitem{Barklow:2015tja}
T.~Barklow, J.~Brau, K.~Fujii, J.~Gao, J.~List, N.~Walker, and K.~Yokoya,
  ``{ILC Operating Scenarios},''
  \href{http://arxiv.org/abs/1506.07830}{{\ttfamily arXiv:1506.07830
  [hep-ex]}}.

\bibitem{Das:2022oyx}
A.~Das, S.~Gola, S.~Mandal, and N.~Sinha, ``{Two-component scalar and fermionic
  dark matter candidates in a generic U$(1)_X$ model},''
  \href{http://arxiv.org/abs/2202.01443}{{\ttfamily arXiv:2202.01443
  [hep-ph]}}.

\bibitem{Sjostrand:2014zea}
T.~SjÃ¶strand, S.~Ask, J.~R. Christiansen, R.~Corke, N.~Desai, P.~Ilten,
  S.~Mrenna, S.~Prestel, C.~O. Rasmussen, and P.~Z. Skands, ``{An Introduction
  to PYTHIA 8.2},'' \href{http://dx.doi.org/10.1016/j.cpc.2015.01.024}{{\em
  Comput. Phys. Commun.} {\bfseries 191} (2015) 159--177},
\href{http://arxiv.org/abs/1410.3012}{{\ttfamily arXiv:1410.3012 [hep-ph]}}.

\bibitem{Gunion:1989we}
J.~F. Gunion, H.~E. Haber, G.~L. Kane, and S.~Dawson, {\em {The Higgs Hunter's
  Guide}}, vol.~80.
\newblock 2000.

\end{thebibliography}\endgroup
\bibliographystyle{utphys}
\end{document}